\documentclass{aa}  
\usepackage{natbib}
\bibpunct{(}{)}{;}{a}{}{,} 

\usepackage[english]{babel}           
\usepackage{graphicx,epsfig}
\usepackage{txfonts}
\usepackage{lscape}
\usepackage{eucal}
\usepackage{longtable}

\def \cc{\ifmmode{\,{\rm cm}^{-3}}\else{$\,{\rm cm}^{-3}$}\fi}
\def \cq{\ifmmode{\,{\rm cm}^{-2}}\else{$\,{\rm cm}^{-2}$}\fi}
\def \mic{\ifmmode{\,\mu{\rm m}}\else{$\mu$m}\fi}
\def \eccs{\ifmmode{\,{\rm erg}\,{\rm cm}^{-3} {\rm s}^{-1}}\else{$\,{\rm
erg}\,{\rm cm}^{-3} {\rm s}^{-1}$}\fi}
\def \ecqs{\ifmmode{\,{\rm erg}\,{\rm cm}^{-2}\,{\rm s}^{-1}\,{\rm
sr}^{-1}}\else{$\,{\rm erg}\,{\rm cm}^{-2}\,{\rm s}^{-1}\,{\rm sr}^{-1}$}\fi}
\def \deg{\ifmmode{^{\circ}}\else{$^{\circ}$}\fi} 
\def \pc{\ifmmode{\,{\rm pc}}\else{$\,{\rm pc}$}\fi} 
\def \kms{\ifmmode{\,{\rm km}\,{\rm s}^{-1}}\else{km s$^{-1}$}\fi} 
\def \kmspc{\ifmmode{\,{\rm km}\,{\rm s}^{-1}\,{\rm pc}^{-1}}\else{km
s$^{-1}$ pc$^{-1}$}\fi} 
\def \MJysr{\ifmmode{\,{\rm MJy\,sr}^{-1}}\else{$\,{\rm MJy\,sr}^{-1}$}\fi} 
\def \Kkms{\ifmmode{\,{\rm K\,km\,s}^{-1}}\else{$\,{\rm K\,km\,s}^{-1}$}\fi} 
\def \twCO{\ifmmode{\rm ^{12}CO}\else{$\rm^{12}CO$}\fi} 
\def \thCO{\ifmmode{\rm ^{13}CO}\else{$\rm^{13}CO$}\fi} 
\def \twCN{\ifmmode{\rm ^{12}CN}\else{$\rm^{12}CN$}\fi} 
\def \thCN{\ifmmode{\rm ^{13}CN}\else{$\rm^{13}CN$}\fi} 
\def \HdCO{\ifmmode{\rm H_{2}CO}\else{$\rm H_{2}CO$}\fi} 
\def \twHdCO{\ifmmode{\rm H_{2}^{12}CO}\else{$\rm H_{2}^{12}CO$}\fi} 
\def \thHdCO{\ifmmode{\rm H_{2}^{13}CO}\else{$\rm H_{2}^{13}CO$}\fi} 
\def \twC{\ifmmode{\rm ^{12}C}\else{$\rm^{12}C$}\fi} 
\def \thC{\ifmmode{\rm ^{13}C}\else{$\rm^{13}C$}\fi} 
\def \Cp{\ifmmode{\rm C^+}\else{$\rm C^+$}\fi} 
\def \Cpp{\ifmmode{\rm C^{++}}\else{$\rm C^{++}$}\fi} 
\def \Sp{\ifmmode{\rm C^+}\else{$\rm S^+$}\fi} 
\def \Spp{\ifmmode{\rm S^{++}}\else{$\rm S^{++}$}\fi} 
\def \CFp{\ifmmode{\rm CF^+}\else{$\rm CF^+$}\fi}
\def \CHp{\ifmmode{\rm CH^+}\else{$\rm CH^+$}\fi}
\def \CHdp{\ifmmode{\rm CH_2^+}\else{$\rm CH_2^+$}\fi}
\def \CHtp{\ifmmode{\rm CH_3^+}\else{$\rm CH_3^+$}\fi} 
\def \SHp{\ifmmode{\rm SH^+}\else{$\rm SH^+$}\fi}
\def \SH2p{\ifmmode{\rm SH_2^+}\else{$\rm SH_2^+$}\fi}
\def \twCHp{\ifmmode{\rm ^{12}CH^+}\else{$\rm^{12}CH^+$}\fi}
\def \thCHp{\ifmmode{\rm ^{13}CH^+}\else{$\rm^{13}CH^+$}\fi}
\def \CtH{\ifmmode{\rm C_2H}\else{$\rm C_2H$}\fi} 
\def \CthHt{\ifmmode{\rm C_3H_2}\else{$\rm C_3H_2$}\fi} 
\def \Htp{\ifmmode{\rm H_3^+}\else{$\rm H_3^+$}\fi} 
\def \HCOp{\ifmmode{\rm HCO^+}\else{$\rm HCO^+$}\fi} 
\def \HtOp{\ifmmode{\rm H_3O^+}\else{$\rm H_3O^+$}\fi} 
\def \HCfiN{\ifmmode{\rm HC_5N}\else{$\rm HC_5N$}\fi} 
\def \wat{\ifmmode{\rm H_2O}\else{$\rm H_2O$}\fi} 
\def \HdO{\ifmmode{\rm H_2O}\else{$\rm H_2O$}\fi} 
\def \OHp{\ifmmode{\rm OH^+}\else{$\rm OH^+$}\fi} 
\def \HdOp{\ifmmode{\rm H_2O^+}\else{$\rm H_2O^+$}\fi} 
\def \NHd{\ifmmode{\rm NH_2}\else{$\rm NH_2$}\fi} 
\def \NHtrois{\ifmmode{\rm NH_3}\else{$\rm NH_3$}\fi} 
\def \oxy{\ifmmode{\rm O_2}\else{$\rm O_2$}\fi} 
\def \HH{\ifmmode{\rm H_2}\else{$\rm H_2$}\fi}
\def \Jone{\ifmmode{\rm {(J=1--0)}}\else{{(J=1--0)}}\fi} 
\def \Jtwo{\ifmmode{\rm {(J=2--1)}}\else{{(J=2--1)}}\fi} 
\def \Jthr{\ifmmode{\rm {(J=3--2)}}\else{{(J=3--2)}}\fi} 
\def \Jfou{\ifmmode{\rm {(J=4--3)}}\else{{(J=4--3)}}\fi} 
\def \Jfiv{\ifmmode{\rm {J=4--3}}\else{{J=4--3}}\fi} 
\def \Ta{\ifmmode{\rm T_A}\else{$\rm T_A$}\fi} 
\def \Tas{\ifmmode{\rm T_A^*}\else{$\rm T_A^*$}\fi} 
\def \Tmb{\ifmmode{\rm T_{mb}}\else{$\rm T_{mb}$}\fi} 
\def \Tr{\ifmmode{\rm T_r}\else{$\rm T_r$}\fi} 
\def \Trs{\ifmmode{\rm T_r^*}\else{$\rm T_r^*$}\fi}
\def \NHt{\ifmmode{N_{\rm H}}\else{$N_{\rm H}$}\fi}
\def \NH{\ifmmode{N({\rm H})}\else{$N({\rm H})$}\fi}
\def \NH2{\ifmmode{N({\rm H}_2)}\else{$N({\rm H}_2)$}\fi}
\def \NCH{\ifmmode{N({\rm CH})}\else{$N({\rm CH})$}\fi}
\def \NHF{\ifmmode{N({\rm HF})}\else{$N({\rm HF})$}\fi}
\def \dens{\ifmmode{n_{\rm H}}\else{$n_{\rm H}$}\fi}
\def \nCO{\ifmmode{n({\rm CO})}\else{$n({\rm CO})$}\fi}
\def \nHF{\ifmmode{n({\rm HF})}\else{$n({\rm HF})$}\fi}
\def \nH2{\ifmmode{n({\rm H}_2)}\else{$n({\rm H}_2)$}\fi}

\begin{document}

\title{Comparative study of \CHp\ and \SHp\ absorption lines observed towards distant 
star-forming regions\thanks{Based on observations obtained with the HIFI instrument 
onboard the \emph{Herschel} space telescope in the framework of the 
key programmes PRISMAS and HEXOS.}$^{,}$\thanks{\emph{Herschel} is an ESA space 
observatory with science instruments provided by European-led Principal Investigator 
consortia and with important participation from NASA.}}

\author{
  B. Godard              \inst{1},
  E. Falgarone           \inst{2},
  M. Gerin               \inst{2},
  D. C. Lis              \inst{3},
  M. De Luca             \inst{2},
  J. H. Black            \inst{4}, 
  J. R. Goicoechea       \inst{1},
  J. Cernicharo          \inst{1}, 
  D. A. Neufeld          \inst{5}, 
  K. M. Menten           \inst{6},
  M. Emprechtinger       \inst{3}
}

\institute{
  Departamento de Astrof\'isica, Centro de Astrobiolog\'ia, CSIC-INTA, Torrej\'on de Ardoz, Madrid, Spain
  \and
  LERMA, CNRS UMR 8112, \'Ecole Normale Sup\'erieure \& Observatoire de Paris, Paris, France
  \and
  California Institute of Technology, Pasadena, CA 91125, USA
  \and
  Department of Earth and Space Sciences, Chalmers University of Technology, Onsala Space Observatory, 43992 Onsala, Sweden.
  \and
  The Johns Hopkins University, Baltimore, MD 21218, USA
  \and
  MPI f\"ur Radioastronomie, Bonn, Germany. 
  }

 \date{Received 8 July 2011 / Accepted 19 January 2012}

\abstract{} 
{The HIFI instrument onboard \emph{Herschel} has allowed high spectral resolution 
and sensitive observations of ground-state transitions of three molecular ions: 
the methylidyne cation \CHp, its isotopologue \thCHp, and sulfanylium \SHp.  
Because of their unique chemical properties, a comparative analysis of these cations
provides essential clues to the link between the chemistry and dynamics of the diffuse 
interstellar medium.}
{The \CHp, \thCHp, and \SHp\ lines are observed in absorption towards the distant 
high-mass star-forming regions (SFRs) DR21(OH), G34.3+0.1, W31C, W33A, W49N, and W51, 
and towards two sources close to the Galactic centre, SgrB2(N) and SgrA*+50. All sight 
lines sample the diffuse interstellar matter along pathlengths of several kiloparsecs 
across the Galactic Plane. In order to compare the velocity structure of each species, 
the observed line profiles were deconvolved from the hyperfine structure of the \SHp\ 
transition and the \CHp, \thCHp, and \SHp\ spectra were independently decomposed into
Gaussian velocity components. To analyse the chemical composition of the foreground gas, 
all spectra were divided, in a second step, into velocity intervals over which the 
\CHp, \thCHp, and \SHp\ column densities and abundances were derived.}  
{\SHp\ is detected along all observed lines of sight, with a velocity structure close 
to that of \CHp\ and \thCHp. The linewidth distributions of the \CHp, \SHp, and \thCHp\ 
Gaussian components are found to be similar.  These distributions have the same mean 
($\left< \Delta \upsilon \right> \sim 4.2$ \kms) and standard deviation 
($\sigma (\Delta \upsilon) \sim 1.5$ \kms). This mean value is also close to that of 
the linewidth distribution of the \CHp\ visible transitions detected in the solar 
neighbourhood. We show that the lack of absorption components narrower than 2 \kms\ 
is not an artefact caused by noise: the \CHp, \thCHp, and \SHp\ line profiles are 
therefore statistically broader than those of most species detected in absorption 
in diffuse interstellar gas (e. g. HCO$^+$, CH, or CN). 
The $\SHp/\CHp$ column density ratio observed in the components located away from 
the Galactic centre spans two orders of magnitude and correlates with the \CHp\ 
abundance. Conversely, the ratio observed in the components close to the Galactic 
centre varies over less than one order of magnitude with no apparent correlation
with the \CHp\ abundance.
The observed dynamical and chemical properties of \SHp\ and \CHp\ 
are proposed to trace the ubiquitous process of turbulent dissipation, 
in shocks or shears, in the diffuse ISM and the specific environment of the Galactic 
centre regions.} 
{}


   \authorrunning{B. Godard et al.}
   \titlerunning{Comparative study of \CHp\ and \SHp\ absorption lines.}
   \maketitle
%

\section{Introduction}

Studying the diffuse phases of the interstellar medium (ISM) is
essential, not only because they contain a large part of the total
gas mass  of the cold ISM and are the precursors of dense clouds, but also because they
harbour the first steps of interstellar chemistry. Since the detection 
of the first diatomic molecules
CN, CH, and \CHp\ (see references in the review of \citealt{Snow2006})
through their narrow visible absorption lines, the improvement of the
observational techniques and instruments has
provided a more comprehensive view of the diffuse ISM and led to a
better understanding of its dynamical, thermal, and chemical
evolution.  A wide variety of diatomic and triatomic molecular
species has already been observed in the diffuse medium. Moreover, its
chemical composition has now been probed in the solar
neighbourhood through UV
(e.g. \citealt{Shull1982,Crawford1997,Snow2000,Rachford2002,Gry2002,Lacour2005}),
visible
(e.g. \citealt{Crane1995,Gredel1997,Thorburn2003,Weselak2008,Maier2001}),
and radio (e.g. \citealt{Haud2007,Liszt2006} and references therein)
spectroscopy, and the inner Galaxy material through submillimetre and
radio (millimetre, centimetre) spectroscopy
(e.g. \citealt{Koo1997,Fish2003,Nyman1989,Greaves1994,Neufeld2002,Plume2004}).

The \emph{Herschel} space mission has broadened this investigation,
giving access to the full submillimetre domain, which has allowed 
the detection of many molecular species that could not be
detected from the ground before because of the high opacity of the
atmosphere. In the framework of the HIFI key programme PRISMAS (PRobing
InterStellar Molecules with Absorption line Studies) many
hydrides such as HF \citep{Neufeld2010,Sonnentrucker2010}, \OHp, \HdOp
\citep{Neufeld2010a,Gerin2010a}, CH \citep{Gerin2010}, NH, \NHd,
\NHtrois\ \citep{Persson2010}, and \CHp\ \citep{Falgarone2010} were
detected in absorption against the strong continuum emission of distant
star-forming regions, providing for the first time a good census of
these light hydrides in the inner Galaxy.

Of all molecules targeted by PRISMAS, the methylidyne cation
\CHp\ is particularly interesting because its presence in the diffuse
ISM remains one of the most intriguing puzzle in astrophysics. The
\CHp\ abundances predicted by steady-state, UV-dominated, PDR-type
(PhotoDissociation Regions) chemical models are several orders of
magnitude lower than the observed values, because all the
\CHp\ formation pathways that are alternative to the highly endothermic $\Cp +
\HH \rightarrow \CHp + {\rm H}$ reaction, are inefficient in balancing
its fast destruction by hydrogenation. \citet{Indriolo2010} recently
showed that the upper limit on the $\CHtp / \CHp$ abundance ratio
observed towards Cyg OB2 can only be reproduced in diffuse molecular
clouds with extreme conditions (i.e. $f_{{\rm H}_2} < 0.2$, or $T >
1000$ K ). So far, the solution to this puzzle has been to invoke
regions of the diffuse gas where a warm chemistry is activated by the
release of supra-thermal energy in the cold ISM induced by low-velocity
magnetohydrodynamic (MHD) shocks \citep{Draine1986a,Pineau-des-Forets1986}, 
Alfv\'en waves \citep{Federman1996}, turbulent mixing
\citep{Xie1995,Lesaffre2007}, or turbulent dissipation
\citep{Falgarone1995,Joulain1998,Godard2009}. \citet{Indriolo2010}
claimed that observations of the excited levels of \CHtp\ are able to
provide the clues necessary to favour one theory over the other.

A potentially related problem is the existence of sulfanylium
(\SHp) in the diffuse gas. Because the hydrogenation reaction of
\Sp\ has an endothermicity twice as high as that of \Cp, a
measurement of the $\SHp / \CHp$ abundance ratio should provide 
valuable insights about the regions where \CHp\ is formed. 
Sought for without success in the local diffuse ISM in absorption against nearby 
stars in the UV domain since 1988 \citep{Millar1988,Magnani1989,Magnani1991}, this 
molecular ion was recently detected by \citet{Menten2011} through its ground-state 
rotational transition in the submillimetre range observed in absorption towards the 
Galactic centre line of sight SgrB2(M) with the APEX telescope.

In this paper, we report the detection of \SHp, \CHp, and
\thCHp\ towards the six distant star-forming regions DR21(OH), G34.3+0.1, 
W31C, W33A, W49N, and W51, and the Galactic centre sight lines\footnote{The SgrA*+50
sight line (GCM-0.02-0.07) corresponds to the 50 \kms\ cloud located 
in the vicinity of SgrA*, which is known to be a bright submillimetre source 
\citep{Dowell1999}.} SgrA*+50 and SgrB2(N), 
and we perform a cross analysis of the observed properties of those three species. 
The observations, obtained in the framework of the key programmes PRISMAS and HEXOS,
are presented in Sect. \ref{SectObs}. The methods used for the
analysis and the results obtained are shown in Sects. \ref{SectAna}
and \ref{SectRes}, respectively. We conclude this work in Sect. \ref{SectDisc} 
with a discussion on the chemical and dynamical properties of the gas seen in absorption. 
The comparison with the model predictions will be the subject 
of a forthcoming paper (Godard et al., in prep.).

\section{Observations and data reduction} \label{SectObs}

\begin{table*}[!!!ht]
\begin{center}
\caption{Properties of background sources.}
\begin{tabular}{l @{\hspace{0.3cm}} l @{\hspace{0.3cm}} l @{\hspace{0.3cm}} r
    @{\hspace{0.3cm}} r @{\hspace{0.3cm}} r @{\hspace{0.3cm}} r @{\hspace{0.3cm}} r
    @{\hspace{0.3cm}} r}
\hline
Source  & RA(J2000)   & Dec (J2000) & l & b & $D$$^{a}$ & $R_G$ & $T_c ({\rm K})$$^{b}$ & $T_c ({\rm K})$$^{b}$ \\
        & (h) (m) (s) & ($^{\circ}$) ($'$) ($''$) &  ($^{\circ}$) & ($^{\circ}$) & (kpc) & (kpc) & (830 GHz) & (530 GHz) \\
\hline
SgrA*+50 & 17 45 50.2 & -28 59 53 & 359.97 & -0.07 &  8.4 & 0.1 & 0.4 & 0.2 \\
SgrB2(N) & 17 47 19.9 & -28 22 18 &   0.68 & -0.03 &  8.4 & 0.1 & 8.0 & 2.9 \\
G34.3+0.1& 18 53 18.7 &  01 14 58 &  34.26 &  0.15 &  3.8 & 5.8 & 2.3 & 0.7 \\
W31C     & 18 10 28.7 & -19 55 50 &  10.62 & -0.38 &  4.8 & 3.9 & 1.9 & 0.5 \\
W33A     & 18 14 39.4 & -17 52 00 &  12.91 & -0.26 &  4.0 & 4.7 & 0.6 & 0.2 \\
W49N     & 19 10 13.2 & +09 06 12 &  43.17 & +0.01 & 11.5 & 7.9 & 2.4 & 0.8 \\
W51      & 19 23 43.9 & +14 30 31 &  49.49 & -0.39 &  7.0 & 6.6 & 2.7 & 1.0 \\
DR21(OH) & 20 39 01.0 &  42 22 48 &  81.72 &  0.57 &  1.0 & 8.4 & 1.3 & 0.4 \\
\hline
\end{tabular}
\begin{list}{}{}
\item[$^{a}$] For G34.3+0.1, W31C, W33A, and W49N, the source distances $D$, and the subsequent Galactocentric 
distances $R_G$, are taken from \citet{Fish2003} and \citet{Pandian2008}, who resolved the kinematic distance ambiguity. 
Distance uncertainties are of the order of 0.5 kpc.
\item[$^{b}$] Uncertainties on the continnuum temperatures $T_c$ are of the order of 10\% \citep{Roelfsema2012}.
\end{list}
\label{TabSources}
\end{center}
\end{table*}

\begin{table*}
\begin{center}
\caption{\CHp $X^{1}\Sigma^{+}$, \thCHp $X^{1}\Sigma^{+}$, and \SHp\ $X^{3}\Sigma^{-}$ spectroscopic 
parameters for the observed pure rotational transitions. Numbers in parenthesis are power of 10.}
\begin{tabular}{r @{\hspace{0.2cm}} c @{\hspace{0.1cm}} c @{\hspace{0.1cm}} r @{\hspace{0.2cm}} r @{\hspace{0.2cm}} r
@{\hspace{0.2cm}} c @{\hspace{0.3cm}} c @{\hspace{0.1cm}} c @{\hspace{0.1cm}} c @{\hspace{0.1cm}} c
@{\hspace{0.1cm}} c @{\hspace{0.1cm}} c @{\hspace{0.1cm}} c @{\hspace{0.1cm}} c}
\hline
\multicolumn{2}{c}{Transition}  & \multicolumn{1}{l}{$g_u$} & \multicolumn{1}{c}{$\nu_0$$^a$} & \multicolumn{1}{c}{$A$}        & \multicolumn{1}{c}{$\Delta \upsilon_h$$^b$} & I$_h$$^c$ & \multicolumn{8}{c}{$\sigma_l/T_{\rm c}$$^d$}\\
                  &             &                           & \multicolumn{1}{c}{(GHz)}       & \multicolumn{1}{c}{(s$^{-1}$)} & \multicolumn{1}{c}{(\kms)}                  &           & SgrA*+50 & SgrB2(N) & DR21(OH) & G34 & W31C & W33A & W49N & W51 \\
\hline
\CHp   & 1 - 0             & 3 & 835.137504 & 5.83 (-3) &        &      &  1.1 (-2) & 9.8 (-3) & 1.5 (-2) & 1.9 (-2) & 5.2 (-2) & 2.0 (-2) & 1.7 (-2) & 1.7 (-2) \\
\thCHp & 1,1/2 - 0,1/2     & 2 & 830.215004 & 5.83 (-3) &  0.59  & 0.50 &   \\
\thCHp & 1,3/2 - 0,1/2     & 4 & 830.216640 & 5.83 (-3) &  0.00  & 1.00 &  5.5 (-2) & 5.5 (-3) & 1.5 (-2) & 2.0 (-2) & 1.5 (-2) & 2.9 (-2) & 1.4 (-2) & 2.0 (-2) \\
\SHp   & 1,2,3/2 - 0,1,1/2 & 4 & 526.038722 & 7.99 (-4) &  +5.26 & 0.56 &   \\
\SHp   & 1,2,5/2 - 0,1,3/2 & 6 & 526.047947 & 9.59 (-4) &  0.00  & 1.00 &  5.0 (-2) & 1.5 (-2) & 2.9 (-2) & 2.5 (-2) & 9.0 (-3) & 2.4 (-2) & 1.1 (-2) & 1.5 (-2) \\
\SHp   & 1,2,3/2 - 0,1,3/2 & 4 & 526.124976 & 1.60 (-4) & -43.93 & 0.11 &   \\
\hline
\end{tabular}
\begin{list}{}{}
\item[$^{a}$] \CHp\, \thCHp\, and \SHp\ line frequencies are from \citet{Amano2010} and \citet{Savage2004}, respectively.
\item[$^{b}$] Velocity shifts of the hyperfine components of the \SHp\ $(1,2 \gets 0,1)$ line, relative to the
$F = 5/2 \gets 3/2$ transition, the reference hyperfine transition in the text.
\item[$^{c}$] Intensities of the hyperfine components computed (at $T_{\rm ex} \rightarrow \infty$) as 
$g_u A_{ul} / g_{u0}A_{ul0}$, relative to a chosen reference hyperfine transition (designated by the index 0 
in the previous formula).
\item[$^{d}$] $\sigma_l/T_{\rm c}$ is the rms noise level divided by the continuum intensities of the spectra
at the resolution of 1.1 MHz.
\end{list}
\label{TabCondObs}
\end{center}
\end{table*}

\subsection{Observing conditions}

\begin{figure}[!h]
\begin{center}
\includegraphics[width=9.001cm,angle=0]{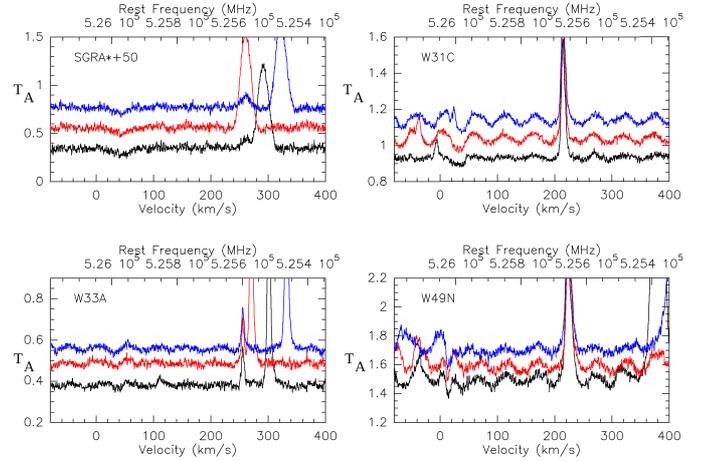} \\
\caption{Original \SHp\ spectra (double-sideband antenna temperature $T_A$ before removing the standing waves) 
observed towards SgrA*+50, W31C, W33A, and W49N in the horizontal polarization and for the three 
different LO settings (in black, red, and blue). For more clarity the red and blue curves have 
been shifted from the black curve by 0.2 and 0.4 K (for the SgrA*+50 data), and by 0.1 
and 0.2 K (for the W31C, W33A, and W49N data).}
\label{FigOriginal}
\end{center}
\end{figure}

The observations were carried out from March to October 2010
and in April 2011 towards the eight submillimetre background 
continuum sources listed in Table \ref{TabSources} (with their Galactic
coordinates, their distance from the Sun, and their measured single-sideband 
continuum temperature $T_c$ in K at $\sim 830$ and $\sim 530$ GHz). Using the
dual-beam switch (DBS) mode (with a throw at 3' from the source) and the 
wide-band spectrometer (WBS) of Herschel/HIFI (see \citealt{Roelfsema2012} for 
a detailed description of the properties and performances of HIFI), we observed
\begin{itemize}
\item[$\bullet$] the $J=1 \gets 0$ absorption lines of \CHp\ and \thCHp,
in the upper and lower sidebands of band 3a, 
\item[$\bullet$] and the $F=3/2 \gets 1/2$, $5/2 \gets 3/2$, and $3/2 \gets 3/2$ 
hyperfine components of the
  $N,J=1,2 \gets 0,1$ absorption line of \SHp, in the lower sideband of band 1a.
\end{itemize}

The data obtained towards SgrB2(N) were part of the full HIFI spectral 
scan performed by the HEXOS key programme; the corresponding double-sideband 
spectra were deconvolved into single-sideband spectra, including the 
continuum \citep{Comito2002}. Towards the other sources, the observations
were performed in the framework of the PRISMAS key programme;
to separate spectral features from upper and lower sidebands of
the WBS spectrometer, each transition was observed using three slightly
different settings of the local oscillator (LO) frequency, adapted to
induce a relative velocity shift of $\sim 30$ \kms\ between the two sidebands. 
The spectroscopic parameters of the observed lines are listed in Table
\ref{TabCondObs}, along with the rms noise levels relative to the
single-sideband continuum intensities obtained with an on-source integration 
time ranging from 1 to 20 min. In these frequency ranges, the WBS resolution 
of 1.1 MHz corresponds to velocity resolutions of $\sim 0.36$ \kms\ for 
the \CHp\ and \thCHp\ transitions, and of $\sim 0.57$ \kms\ for the
\SHp\ transitions, and the {\it Herschel} HPBW is 26'' at 835 GHz and
41'' at 526 GHz.

The data were calibrated with hot and cold blackbodies \citep{Roelfsema2012}, 
reduced using the standard Herschel pipeline to Level 2,
and subsequently analysed using the Herschel Interactive Processing
Environment\footnote{See
  http://herschel.esac.esa.int/HIPE\_download.shtml for more
  information about HIPE.}  (HIPE v5.1, \citealt{Ott2010}). The final analysis 
was performed with the
GILDAS-CLASS90 software\footnote{See http://www.iram.fr/IRAMFR/GILDAS
  for more information about GILDAS softwares.} \citep{Pety2005}, 
  and a set of Fortran95 numerical routines
that we developed.  While the signals measured in the two orthogonal
polarizations that were obtained with the three LO settings agreed 
excellently for both the \CHp\ and \thCHp\ line observations,
the \SHp\ spectra, displayed in Fig.  \ref{FigOriginal}, exhibit
standing waves (SW) of identified origin\footnote{Most of the observed SW
  have a period of $\sim 90 - 100$ MHz, identified by \citet{Roelfsema2012} 
  as reflections occuring between the mixer focus and the
  cold and hot black bodies. For two spectra observed towards W49N we
  had to remove an additional standing wave with a period $\sim 150$
  MHz.} and removed using the HIPE sine wave fitting task
FitHifiFringe (FHF). Since the inferred detected opacities of \SHp\ are low,
and since the period and the amplitude of the waves are similar to the
size and the depth of the \SHp\ absorption features (see
Fig. \ref{FigOriginal}), the resulting spectra were sensitive to the
FHF input parameters: several plausible solutions were found
depending on the number of sub-bands taken into account for the fit,
the number of sine waves to remove, and the use of a frequency
mask. We estimated a maximal error of 50 \% on the
deepest absorption features of G34.3+0.1 and W51, 30 \% on those 
of W33A, and W31C and 10 \% on the others. The latter value is  
comparable to that due to the uncertainties on 
the beam efficiency and the sideband gain ratio \citep{Roelfsema2012}.

\begin{figure*}[!hb]
\begin{center}
\includegraphics[width=18cm,angle=0]{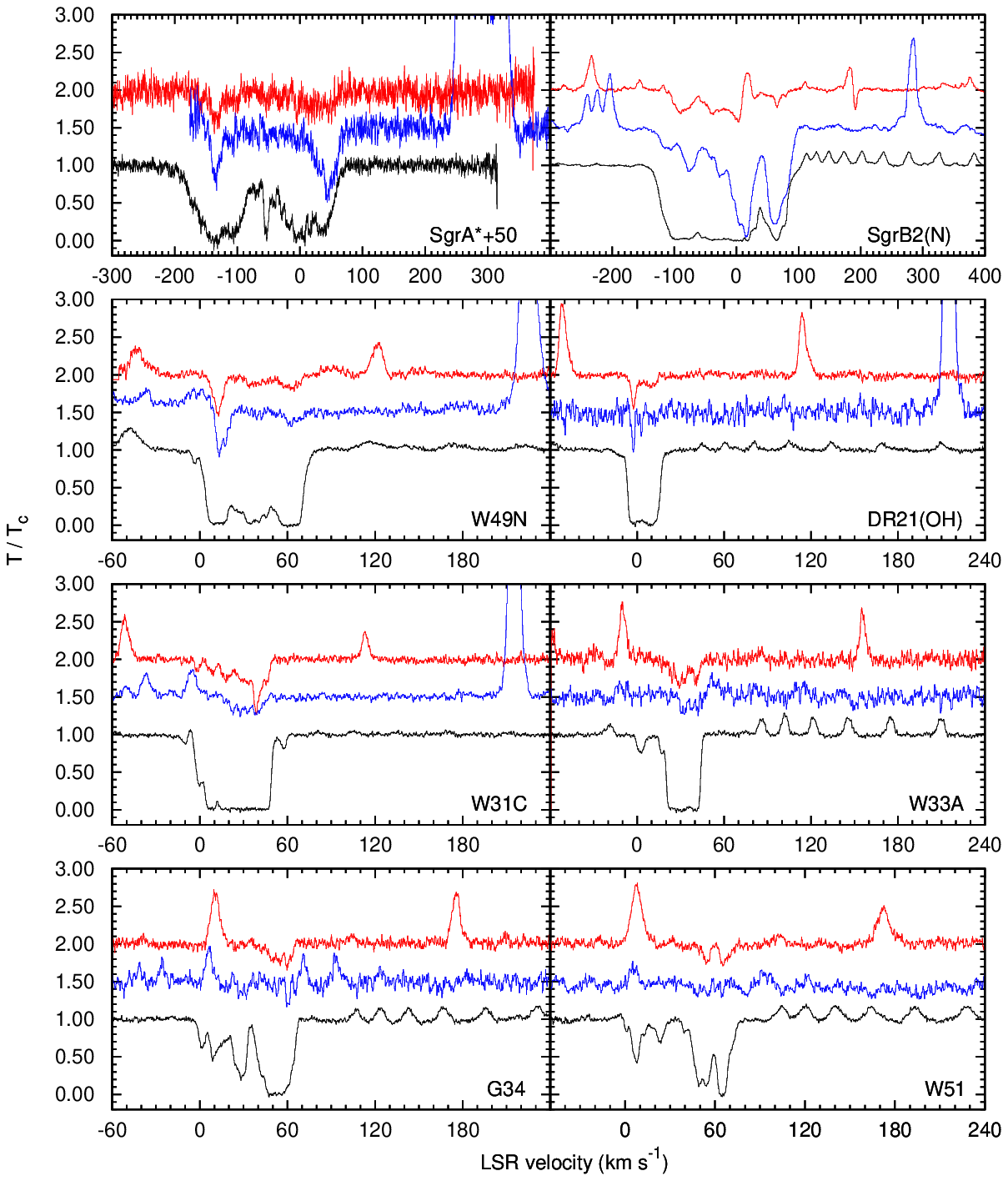}
\caption{Spectra observed at the frequencies of the ground-state transitions of \CHp\ (black), 
         \thCHp\ (red), and \SHp\ (blue) in the direction of DR21(OH), G34.3+0.1, W31C, W33A, W49N, 
         W51, SgrA*+50, and SgrB2(N). All spectra have been normalized to the continuum 
         temperatures. To make the absorption features discernible, the \thCHp\ and \SHp\ 
         signals were mutiplied by a factor of 2 towards SgrA*+50, and SgrB2(N), and by factors of 2 and
         3, respectively, towards the other sources. For more clarity the \thCHp\ and \SHp\ spectra were 
         shifted from the \CHp\ spectra by 0.5 and 1.0.}
\label{FigSpectres}
\end{center}
\end{figure*}

For each transition and both polarizations, we obtained an average
spectrum by combining the data from the three observations with
different LO settings. Because we are interested in the velocity
structure and the properties of the absorbing gas, the spectra in both
polarizations were normalized to their respective continuum
temperature.  As in \citet{Falgarone2010}, we used the saturated shape
of the \CHp\ absorption line profiles to measure the sideband gain
ratios $R$ at 835.1375 GHz, defined as the ratio of the continuum
temperatures in the lower and upper sidebands. For all spectra
with saturated absorption lines, we found $R \sim 0.9-1$ and $R \sim
0.7-0.8$ in the horizontal and vertical polarizations, respectively. We
finally used these values at 835 GHz, and $R=1$ at 830 GHz and 526
GHz, to combine the data from both polarizations, and obtain the final
average spectra (with rms noise levels given in Table \ref{TabCondObs}) 
shown in Fig. \ref{FigSpectres} as functions of the LSR (local standard 
of rest) velocity. This figure illustrates the quality of the baselines 
over the large bandwidth for most of the spectra.
The strong emission lines detected in the spectra shown in Figs. 
\ref{FigOriginal} and \ref{FigSpectres} were identified as the 
\HCOp\ $(6 \rightarrow 5)$ and \HdCO\ $(6 \rightarrow 5)$ transitions near 535.1 GHz 
and 525.6 GHz, three methanol lines near 829.9 GHz and 830.3 GHz, and 
the SO$_2$ emission band near 835 GHz, emitted by the SFRs themselves.

\subsection{Deconvolution of the \SHp\ hyperfine structure}

\begin{figure*}[!hb]
\begin{center}
\includegraphics[width=18cm,angle=0]{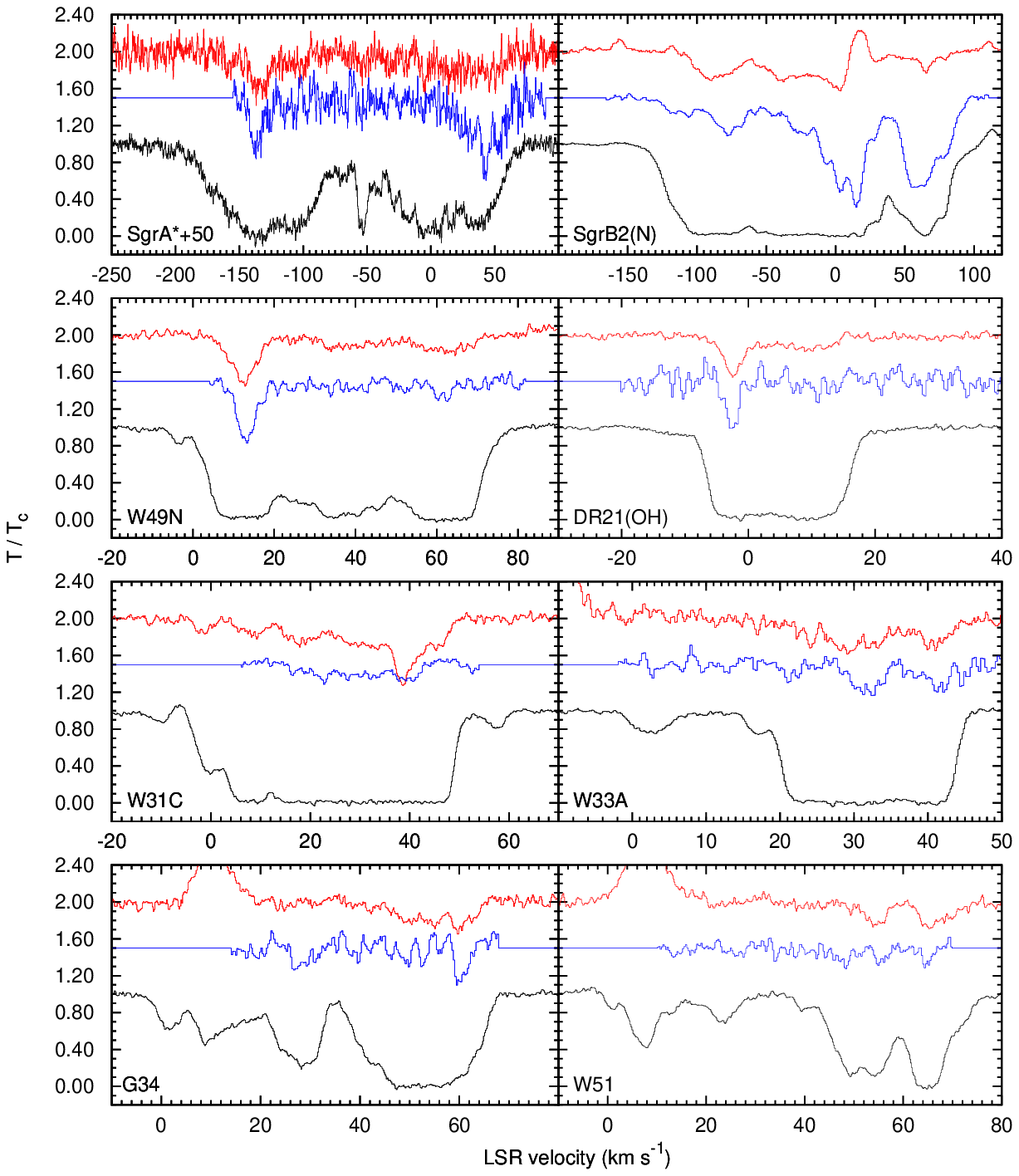}
\caption{Absorption profiles observed in the direction of DR21(OH), G34.3+0.1, W31C, W33A, W49N, 
  W51, SgrA*+50, and SgrB2(N) in the ground-state transitions of \CHp\ (black) and \thCHp\ (red), and 
  in the main hyperfine component (see Tab. \ref{TabCondObs}) of the ground-state transition of \SHp\ (blue). 
  All spectra have been normalized to the continuum temperatures. To make the 
  absorption features discernible, the \thCHp\ and \SHp\ signals were mutiplied by a factor of 2 towards 
  SgrA*+50, and SgrB2(N), and by factors of 2 and 3, respectively, towards the other sources. For more 
  clarity the \thCHp\ and \SHp\ spectra were shifted from the \CHp\ spectra by 0.5 and 1.0. 
  The x- and y- scales were chosen to display the velocity structure of the absorption features 
  in detail.}
\label{FigSpectres2}
\end{center}
\end{figure*}

\begin{figure*}[!hb]
\begin{center}
\includegraphics[width=18cm,angle=0]{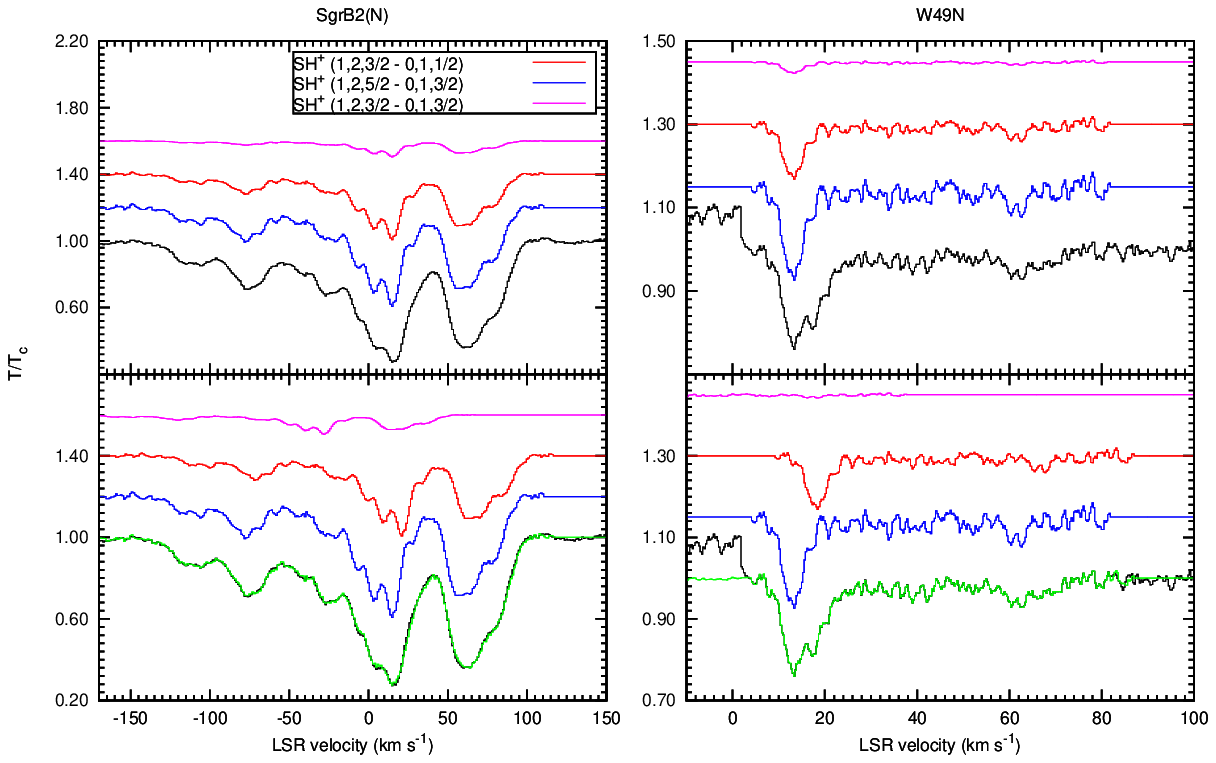}
\caption{Result of the hyperfine structure deconvolution code applied on the \SHp\ $(1,2 \gets 0,1)$ 
absorption spectrum observed towards SgrB2(N) and W49N. The top panel displays the original data and 
the resulting decomposition in three hyperfine components, aligned to the same velocity scale. 
The hyperfine components are shifted by their velocity shift $\Delta \upsilon_h$ (see Table \ref{TabCondObs}) 
in the bottom panels, to show how they can be combined (green line) to reproduce the original data 
(black line). For more clarity, the hyperfine components were vertically shifted from
the original data.}
\label{FigHyper}
\end{center}
\end{figure*}

Unfortunately (see Table \ref{TabCondObs}), the velocity shifts $\Delta \upsilon_h$ associated 
with the \SHp\ hyperfine transitions are smaller than the observed velocity ranges of the \SHp\ absorption 
spectra, and prevent us from performing a direct cross comparison of the velocity profiles of the \SHp, \CHp, 
and \thCHp\ lines. To solve this problem, 
we developed a numerical procedure to extract the signal associated with each hyperfine transition, 
solving the following set of equations for $\tau_r (\upsilon)$, 
over the entire absorption velocity domain,
\begin{equation}
\sum_{k=1}^{N_{h}} I_{h}(k) \, \tau_r (\upsilon-\Delta \upsilon_h(k)) = -{\rm ln} \left[ T(\upsilon)/T_{\rm c} \right],
\end{equation}
where $N_h$, $I_h$, $\tau_r$, and $T(\upsilon)/T_c$ are the
number of hyperfine transitions, the intensities relative to the
strongest hyperfine component, the opacity of the reference hyperfine
transition, and the normalized line profile (line/continuum), 
respectively. The resulting spectra are shown in Fig. \ref{FigSpectres2}, 
and, as an example, the outcome of the
hyperfine decomposition code, applied to the absorption lines observed
towards SgrB2(N) and W49N, is shown in Fig. \ref{FigHyper}. 
This figure illustrates the excellent agreement between 
the original data (in black) and the spectra rebuilt after decomposition 
(in green).


The \thCHp\ $J=1 \gets 0$ line also exhibits a spin-rotation splitting
\citep{Amano2010}, although the associated $F=3/2 \gets 1/2$ and
$F=1/2 \gets 1/2$ transitions are separated by only 1.636 MHz ($\sim
0.59$ \kms) and are therefore too close to be individually resolved
given the significant velocity dispersion of the gas. This hyperfine
structure induces a systematic broadening of the absorption velocity
components that depends on their $FWHM$ $\Delta \upsilon_{\rm real}$ :
for $\Delta \upsilon_{\rm real}$ varying between 2 and 10 \kms\ the
broadening ranges\footnote{This result on the line profile broadening
  is derived from the analysis of 1760 synthetic spectra taking into
  account the hyperfine structure of \thCHp\ (line strength and
  velocity structure, \citealt{Amano2010})} between 4\% and 0.1\%.
Since this error is far smaller than that imputable to the rms
noise levels of the \thCHp\ spectra (see Sect. \ref{SectAna}), we
chose to ignore the hyperfine structure of \thCHp\ in the following
analysis.


The spectra shown in Fig. \ref{FigSpectres2} are highly structured and
have the following remarkable properties: (1) thanks to the high
sensitivity of the HIFI receiver, the \SHp\ ion is seen in absorption 
along every line of sight; 
(2) all hydride lines are detected in absorption, and 
within the limits imposed by the signal/noise (S/N) ratio, \CHp\, \thCHp\, and \SHp\ absorptions 
are detected over the whole velocity range of the foreground matter along each line of sight;
(3) although the
opacity ratios vary from one line of sight to another and from one
velocity range to another, the velocity structure of \SHp\ is similar
to those of \CHp\ and \thCHp. It is the similarity and differences of these 
absorption line profiles that are the focus of the present study. 

\section{Analysis of the line profiles} \label{SectAna}

\subsection{Multi-Gaussian decomposition}

As in \citet{Godard2010}, the decomposition of the spectra in velocity components
was deduced through a multi-Gaussian fitting procedure that we developed, based on
the Levenberg-Marquardt algorithm, which takes advantage of the information carried 
by the hyperfine structure of a given transition. This algorithm is applied  
to adjust, in the least-squares sense, the {\it minimal} number of Gaussians required to 
describe the data within the observational errors without introducing any systematic 
effect. The number of Gaussians was increased, for instance, when we visually spotted  
serpentine curves - characteristic of poor fits in the line wings - in the residuals.
Thus, for each transition, the observed normalized line profile (line/continuum) is written
\begin{equation}
  \frac{T(\upsilon)}{T_{\rm c}} = {\rm exp}\left[ - \sum_{j=1}^{N_{c}}
  \sum_{k=1}^{N_{h}} I_{h}(k) \, \tau_0(j) \, e^{-\frac{1}{2}\left[\frac{\upsilon -
  \upsilon_{0}(j) - \Delta \upsilon_h(k)}{\sigma_0 (j)}\right]^{2}}\right],
\end{equation}
where $\tau_0$, $\upsilon_0$, and $\sigma_0$ are the usual Gaussian
fit parameters. All spectra were decomposed independently from one 
another, without imposing any constraints on the Gaussian parameters.
The choice of the input parameters, namely the initial 
values of $\upsilon_0 (j)$ and $\sigma_0 (j)$ for each velocity components 
$j$, was guided by the comparison of the
different lines observed towards each source.  To correctly determine
the opacity of weak absorption features blended with saturated lines,
as observed in the \CHp\ spectra, we applied an empirical model
constrained by the wings of the saturated line profiles. The results 
of the multi-Gaussian decomposition procedure and the associated errors 
on the Gaussian parameters are given in Table \ref{TabFit} of Appendix 
\ref{AppendGauss}, and the resulting fits and models of the saturated line 
profiles are displayed in Figs. \ref{FigGaussDR21} - \ref{FigGaussSGRB2N}.
Because we aim to compare the kinematic signatures of the \CHp, \thCHp, 
and \SHp\ spectra, we discuss below the reliability
of the extracted Gaussian components in view of the errors on their 
parameters.


\subsection{Validity and self-consistency of the multi-Gaussian decompositions} \label{GaussVal}

Since the decomposition algorithm allows the
detection of components with very weak central opacities, 
the numerical procedure may converge upon Gaussian components whose 
reality is questionable.
To keep only the most reliable
velocity components for our subsequent analysis of the linewidths, we
applied the following detection criterion: any Gaussian component was
considered real if its Gaussian parameters simultaneously verify 
$\Delta \upsilon > 3 \sigma (\Delta \upsilon)$ and $\tau > 2.5 \sigma
(\tau)$, where $\sigma (\Delta \upsilon)$ and $\sigma (\tau)$ are the
errors on $\Delta \upsilon$ and $\tau$ respectively. The resulting
confirmed or uncertain Gaussian components (in the following
C-components and U-components) are indicated in Table \ref{TabFit} (in
columns 4, 8, and 12), and in Figs. \ref{FigGaussDR21} -
\ref{FigGaussSGRB2N} (solid red and dashed blue curves) of Appendix
\ref{AppendGauss}.

In total we found that 25 C-components are simultaneously observed in at least two molecular spectra 
towards DR21(OH), G34.3+0.1, W31C, W33A, W49N, and W51. When compared, the positions of these components 
are found to agree with one another within their respective error for 18 of them, within 0.5 \kms\ for 
6 of them, and within 1.3 \kms\ for 1 of them; similarly, out of the 17 common C-components observed 
towards SgrA*+50 and SgrB2(N), 10 are found to agree with one another within 1 \kms, 6 within 2 \kms, 
and 1 within 3.5 \kms. Except for the \SHp\ components at -126 \kms\ observed towards SgrA*+50 these
shifts in the central positions are at least four times smaller than the corresponding Gaussian linewidths.
Moreover, many of the U-components have corresponding C-components in other species, 
indicating that the fitting process is robust and that the selection method is severe.

These concordances combined with the strict selection on the Gaussian parameters suggest that all 
C-components are real detections and not artefacts caused by noise or the standing waves 
removing procedure.

\subsection{Comparison of the Gaussian linewidths}\label{SectResDyn}

\begin{figure*}[!hb]
\begin{center}
\includegraphics[width=12.0cm,angle=0]{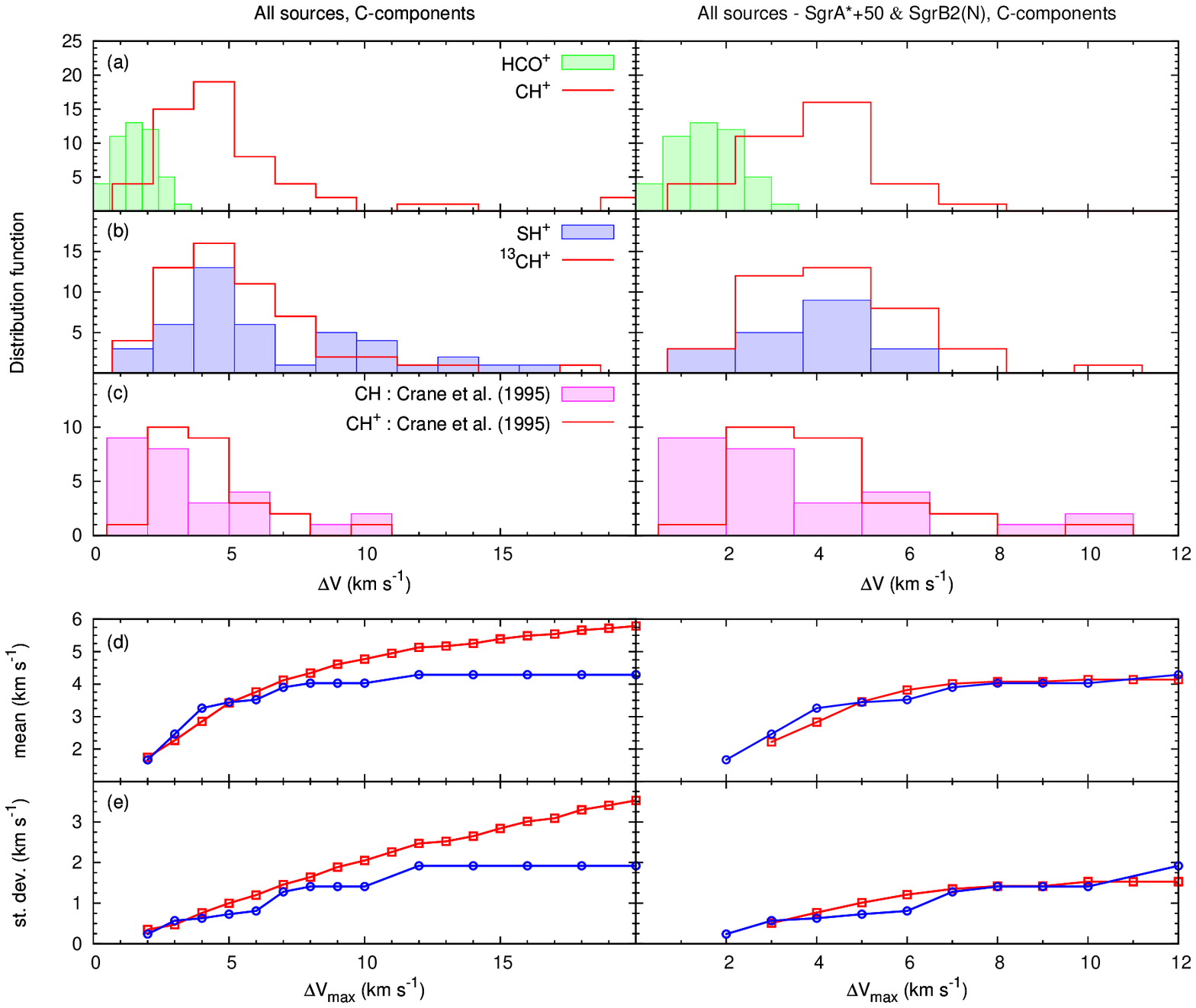}
\caption{Top: (a) and (b) histogram of the linewidths of the \CHp, \thCHp, \SHp\ (this work), and \HCOp\ 
\citep{Godard2010} velocity components obtained with the multi-Gaussian decomposition procedure (see 
Table \ref{TabFit}); (c) linewidth distributions of \CHp\ and CH observed by Crane et al. (1995) in the 
local diffuse medium. Bottom: first- (mean) and second- (standard deviation) order moments 
of the $\Delta \upsilon$ distributions issued from the combined \CHp, \thCHp, and \SHp\ data 
(red squares) and from the \CHp\ data observed in the solar neighbourhood (blue circles); 
the moments are computed over increasing $\Delta \upsilon$ intervals: 
mean = 
$\int_0^{\Delta \upsilon_{\rm max}} \Delta \upsilon N(\Delta \upsilon) d(\Delta \upsilon) /
\int_0^{\Delta \upsilon_{\rm max}} N(\Delta \upsilon) d(\Delta \upsilon)$; 
standard deviation = 
$[\int_0^{\Delta \upsilon_{\rm max}} (\Delta \upsilon - {\rm mean})^2 N(\Delta \upsilon) d(\Delta \upsilon) /
\int_0^{\Delta \upsilon_{\rm max}} N(\Delta \upsilon) d(\Delta \upsilon)]^{1/2}$, where $N(\Delta \upsilon)$
is the linewidth distribution.}
\label{FigDistrW}
\end{center}
\end{figure*}

In Fig. \ref{FigDistrW}, we display the distributions of linewidths
associated with the C-components extracted from the \CHp, \thCHp, and \SHp\ 
spectra. To emphasize the differences observed along the Galactic centre 
sight lines ($l \sim 0$), these distributions are computed for all sight lines 
(left panels), and for the $l \ne 0$ sight lines only (right panels).
As a comparison, the linewidth distributions of the \HCOp\ ground state 
radio transition observed by \citet{Godard2010} towards W31C, W49N, W51, 
and G34.6 are displayed in panels (a), and those of CH and \CHp\ visible 
transitions observed at high spectral resolution ($\sim 0.3$ \kms) in the 
local diffuse medium by \citet{Crane1995} are shown in panels (c) of Fig. \ref{FigDistrW}.
Lastly, panels (d) and (e) of Fig. \ref{FigDistrW} display the first- and second-order 
moments of the $\Delta \upsilon$-distributions issued from 
the combined \CHp, \thCHp, and \SHp\ data (red squares) and from the \CHp\ 
data observed in the solar neighbourhood (blue circles).
In panels (a), the histogram of \HCOp\ linewidths is narrower and
peaks at lower values than those of the submillimetre lines of \CHp,
\thCHp, and \SHp. Conversely, the histogram of the visible \CHp\ data,
characterising the local diffuse interstellar matter, is very similar
to that of the submillimetre data. We discuss the validity and
the significance of these comparisons below.

To demonstrate that the absence of narrow velocity components in 
the \CHp, \thCHp, and \SHp spectra is real and not due to a limitation 
of our extraction algorithm, we have derived the
minimum width of a Gaussian component of optical depth $\tau_0$ that
can be extracted from a given profile characterised by a noise
$\sigma_\tau$ at the velocity resolution $\delta \upsilon$.  Because
$\sigma_\tau$ scales inversely with the square root of the velocity
resolution, a component detected at a $3 \sigma_\tau$ level
necessarily verifies
\begin{equation}
\Delta \upsilon \geqslant \delta \upsilon \left[3 \sigma_\tau /\tau_0 \right]^2.
\end{equation}
While the noise level of the \SHp\ spectra towards SgrA*+50, DR21(OH), 
G34.3+0.1, and W33A
forbids us to extract components with linewidth smaller than 1 \kms,
that of all the other spectra is small enough to allow us to detect
narrow velocity structures down to two velocity channels (0.72
\kms\ for \CHp\ and \thCHp, and 1.14 \kms\ for \SHp). Therefore the
scarcity of components in the first bin [0 ; 2 \kms] of the \CHp,
\thCHp, and \SHp\ histograms is not a noise artefact.

Now, can we compare these linewidth distributions with one
another? Because the S/N ratios of \HCOp\ (see Table. 1 of
\citealt{Godard2010}) and \CHp\ (see Table. \ref{TabCondObs}) profiles
are high and comparable, those two sets of spectra are decomposed with
the same level of detection. The noise therefore affects the
statistics of the component linewidths at the same level.  The same is
true for the lines of \SHp\ and \thCHp, which have comparable, though
poorer, S/N ratios.  Finally, because \CHp\ and \thCHp\ necessarily
bear the same dynamical signatures, all distributions displayed in
Fig. \ref{FigDistrW} can be compared with one another.

We recall that all absorption spectra have been decomposed into
Gaussians independently from one another, without imposing the same
velocity centroids or width to the different velocity components
of a given line of sight. In addition, since the most intense
\SHp\ components are saturated in \CHp, and because the moderate S/N
ratio of the \thCHp\ and \SHp\ data prevents us from observing the
low-opacity structures detected in the \CHp\ spectra, the respective
distributions do not correspond to the same velocity components. It is
therefore remarkable that the distributions of the \CHp, \thCHp, and
\SHp\ widths of the Gaussian velocity components are so similar, 
even identical within the statistical uncertainty. 
They exhibit a common pattern: a peak (hereafter called P1), observed 
towards all the sources, 
defined by a first- (mean) and a second- (standard deviation) 
order moments\footnote{The uncertainties on the moments of the distributions
are the statistical standard errors computed for the first- and second-order 
moments as $\sigma/\sqrt{N}$ and $\sigma/\sqrt{2N}$, respectively,
where $\sigma$ is the standard deviation and $N$ the size of the sample.} 
of $4.2 \pm 0.2$ \kms\ and $1.5 \pm 0.1$ \kms, 
respectively, and an extended tail (hereafter called P2), observed 
only on the Galactic centre sight lines (left panels), with 
$\Delta \upsilon$-values up to 20 \kms.

Because the lines of sight sample kiloparsecs of interstellar material 
in the Galaxy, the patterns P1 and P2 result from the small-scale 
dynamics of the production processes of \CHp\ and \SHp,
the turbulent dynamics of the diffuse ISM, and the 
Galactic dynamics.
The comparison between the left and right (d) and (e) panels of Fig. 
\ref{FigDistrW} shows that the first- and second-order moments of 
the $\Delta \upsilon$ distributions are the same for the components of  
the Galactic ISM along the $l \ne 0$ sight lines and 
the visible \CHp\ lines sampling the solar neighbourhood
(defined by first- and second-order moments of
$4.3 \pm 0.4$ \kms\ and $1.85 \pm 0.3$ \kms, respectively). Because the 
latter is unaffected by the Galactic dynamics, this similarity suggests 
that the peak P1 results from the dynamics of the formation processes 
of \CHp\ and \SHp\ convolved with that of the turbulence of the diffuse gas. 
Because the tail of the
$\Delta \upsilon$ distributions is observed only on the Galactic 
centre sight lines, and because the high absorption dips 
($\Delta \upsilon > 8$ \kms) observed on the $l \ne 0$ sight lines 
can be decomposed into many narrow components (as performed by 
\citealt{Godard2010} with the \HCOp\ spectra), we conclude that P2 is 
caused by the Galactic dynamics only.

Lastly we note that while the first- and second-order moments of 
peak P1 both agree with those of the $\Delta \upsilon$ distribution obtained 
by \citet{Crane1995} within the statistical uncertainties, they also substantially differ
from those of the \CHp\ $\Delta \upsilon$ distribution derived in the solar 
neighbourhood by \citet{Pan2004}, who find first- and second-order moments 
of $3.3 \pm 0.04$ \kms\ and $0.4 \pm 0.03$ \kms.
As proposed by \citet{Pan2005}, these differences may originate from
their profile fitting method: (1) the number of \CHp\ components and
their position were constrained by the observations of other species,  
possibly unrelated chemically (e.g. CN, CO, Ca), and (2) a maximal value of 
$\Delta \upsilon$ of 5.8 \kms\ was set.

\section{Analysis of the integrated opacities} \label{SectRes}

Since the multi-Gaussian decomposition of \CHp, \thCHp, and \SHp\ absorption profiles 
provides sets of C-components that do not always strictly coincide in velocity and width, 
and sets of U-components that sometimes clearly correspond to real absorption but with 
width and depth poorly constrained, our subsequent analysis and comparison of the column 
densities and abundances of these species is based on integrals of opacities computed over 
broad velocity intervals corresponding to marked absorption features common to all lines. 
These broad intervals (typically 5 to 20 \kms) are given in the two first columns of 
Table \ref{TabIntTau}. The column densities 
given in columns 4, 5, and 6 of Table \ref{TabIntTau} are derived assuming a single excitation 
temperature of 4.4 K for \twCHp\ and \thCHp, and of 3.0 K for \SHp\ (see Appendix \ref{AppendCD}). 
Following the method set in Sect. \ref{GaussVal} to keep only the 
most reliable absorption features, a $3\sigma$ detection level was adopted.
Throughout, every measurement below this level is considered as a lower limit.
Conversely, we adopt a conservative lower limit of 2.3 on the optical depth 
for the saturated velocity intervals \citep{Neufeld2010}.

\subsection{Variation of the $^{12}$C/$^{13}$C 
isotopic ratio across the Galactic disk} \label{SecResIso}

\begin{figure}[!hb]
\begin{center}
\includegraphics[width=6.4cm,angle=-90]{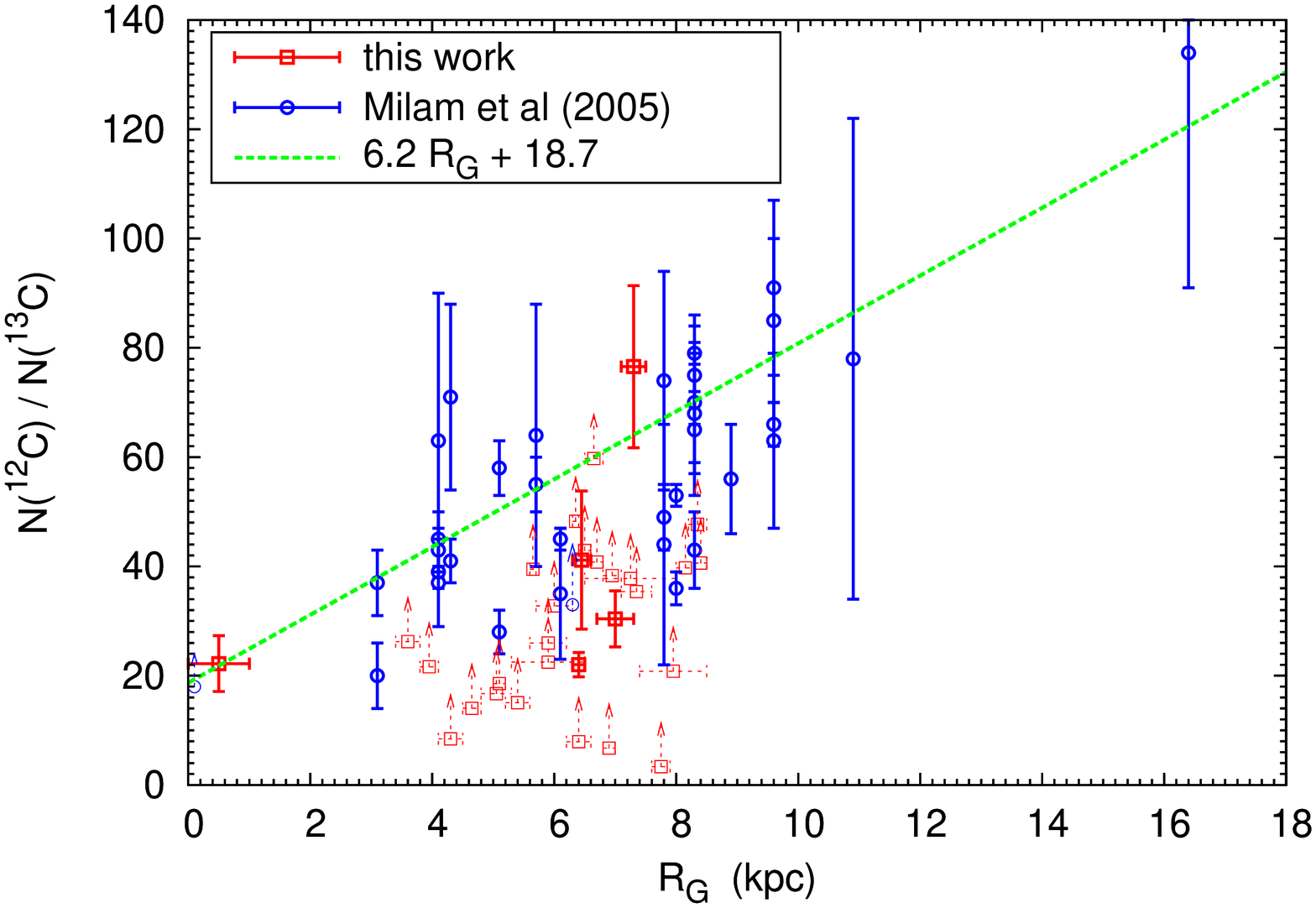}
\caption{$N(\twC) / N(\thC)$ column density ratio as a function of the
  galactocentric distance $R_G$.  The red squares are from this work,
  and therefore evaluated with measurements of the $N(\twCHp) /
  N(\thCHp)$ column density ratio. The blue circles are from previous
  measurements of the $N(\twCN) / N(\thCN)$, $N(\twCO) / N(\thCO)$,
  and $N(\twHdCO) / N(\thHdCO)$ column density ratios
  (\citealt{Milam2005} and references therein) while the green curve
  corresponds to a linear least-squares fit applied to these data.}
\label{Fig-Isotopic}
\end{center}
\end{figure}

In Fig. \ref{Fig-Isotopic}, we display the $\twC / \thC$ column
density ratio derived from the present \twCHp\ and \thCHp\ data
as a function of the Galactocentric distance $R_G$ (given in 
columns 7 and 8 of Table \ref{TabIntTau}) which was computed 
assuming a flat Galactic rotation curve. We compared
these results to those deduced from previous observations of CN, CO,
\HdCO\ and their respective isotopologues \citep{Milam2005}. 
We found that the five firm values (indicated in Table 3) 
derived from  the simultaneous detections of $\twCHp$ and $\thCHp$ 
absorption lines over the same velocity range
are consistent with those derived from the neutral
species. The 25 lower limits inferred from saturated \twCHp\ lines 
(dashed symbols) are also consistent with  \citet{Milam2005}.
This result not only suggests that the isotopic ratios
measured with ions and neutrals are not substantially influenced by
chemical fractionation processes, but it also validates the use of the
empirical relation
\begin{equation} \label{Eq12C13C}
\twC / \thC = 6.2 (\pm 1.0) R_G + 18.7 (\pm 7.4)
\end{equation}
found by \citet{Milam2005} to infer the \twCHp\ column densities from those measured 
in the \thCHp\ spectra. 
The $\twC / \thC$ ratio and $N(\twCHp)$ computed from Eq. \ref{Eq12C13C}
for the velocity intervals towards DR21(OH), G34.3+0.1, W31C, W33A, W49N, and W51 are 
given in columns 9 and 10 of Table \ref{TabIntTau}. Towards SgrA*+50 and SgrB2(N), 
because most of the gas appears to be associated with the Galactic centre environment
\citep{Rodriguez-Fernandez2006}, a $\thC / \twC$ abundance ratio of 20 is assumed 
everywhere except for the velocity interval -20 to +30 \kms, a velocity domain where 
the absorption features are also associated with gas in the Galactic plane, and where 
we use two alternative values, 20 and 60, to bracket the result.

The uncertainties given in Eq. \ref{Eq12C13C} correspond to the standard deviation 
of the best least-squares fit performed by Milam et al. (2005) on the CN, CO, and \HdCO\ data. 
They do not take into account the uncertainty on $R_G$ due to the random motion of interstellar 
clouds. To do so, we relied on the analysis of the HI emission in the first Galactic quadrant by 
\citet{Elmegreen1987}: they found that a large part of the mass of the diffuse gas
is distributed into $\sim$ 200 pc superclouds, separated along spiral arms by 1.5 kpc.
Their one-dimensional internal velocity dispersion of 5.3 \kms\ is the main source of uncertainty 
on $R_G$. Taking into account the $(l,v)$ location of the absorbing gas observed towards each 
source (see the equation of $R_G$ in the caption of Table \ref{TabIntTau}), we obtain a maximal
uncertainty of about 30 \% on $R_G$. When combined with the errors given in Eq. \ref{Eq12C13C}, 
we find a total uncertainty of about 50\% on the $\twC / \thC$ ratio and the subsequent $N(\twCHp)$ 
given in columns 9 and 10 of Table \ref{TabIntTau}.

\subsection{Comparison of the column densities of \CHp\ and \SHp\ } \label{SectResComp}

\begin{figure}[!hb]
\begin{center}
\includegraphics[width=6.4cm,angle=-90]{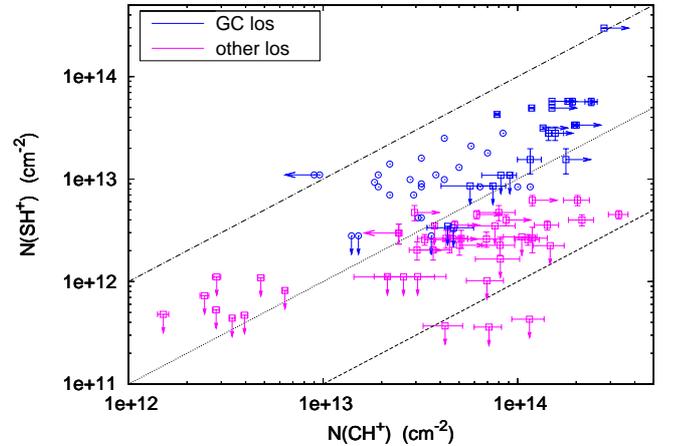}
\caption{\SHp\ column density as a function of the \CHp\ column density per broad velocity interval 
(see Table \ref{TabIntTau}). The open squares and open circles are from the present analysis and from 
\citet{Menten2011}, respectively. The blue and purple points are from the absorption lines observed along the 
Galactic centre sight lines (SgrA*+50, SgrB2(N), and SgrB2(M)) and along the other sight lines (DR21(OH),
G34.3+0.1, W31C, W33A, W49N, and W51), respectively. The black dashed, dotted, and dashed-dotted curves
indicate  $N(\SHp)/N(\CHp)$ ratios of 0.01, 0.1, and 1, respectively.}
\label{Fig-Correl-SHp-CHp}
\end{center}
\end{figure}

Fig. \ref{Fig-Correl-SHp-CHp} displays the \SHp\ and \CHp\ column densities
inferred for each broad velocity interval. The \CHp\ column densities were derived 
either from the \twCHp\ profile, where unsaturated, or from \thCHp\ and the above 
isotopic ratio in the other case. The data set includes the results of the present 
study and those obtained by \citet{Menten2011} towards SgrB2(M). While 
\citet{Menten2011} reported the detection of two absorption components 
at $N(\CHp) \sim 3 \times 10^{12}$ and $N(\SHp) \sim 10^{13}$ \cq, we 
removed those points from Fig. \ref{Fig-Correl-SHp-CHp} because the \thCHp\ 
spectrum in the corresponding velocity range clearly exhibits contamination 
of the absorption features by a strong and broad emission line. 
Last, we note that the few points at $N(\CHp)<10^{13}$ \cq\ correspond to
the faintest \twCHp\ absorption features. Since they are not detected in 
\SHp, these points weakly constrain our subsequent analysis of the 
\SHp/\CHp ratio.

Taking into account the upper and lower limits shown in Fig. 
\ref{Fig-Correl-SHp-CHp}, we find that
$N(\CHp)$ and $N(\SHp)$ span more than two orders of magnitude and that the 
$\SHp / \CHp$ column density ratio varies from less than 0.01 to 1.
Interestingly, the $N(\SHp)/N(\CHp)$ ratios observed towards SgrB2(M), 
SgrB2(N), and SgrA*+50 are very similar, with a mean value of 
$0.28 \pm 0.02$, $\sim$ 9 times higher than that towards DR21(OH), 
G34.3+0.1, W31C, W33A, W49N, and W51, $\left<N(\SHp)/N(\CHp) \right> 
\sim 0.03 \pm 0.007$. 
The above difference in the mean $\SHp / \CHp$ ratio is 
much larger than the 50 \% uncertainty on the computed $\twC / \thC$ 
ratio. Moreover, we obtain the same difference of values if we use only the 
few points detected in \twCHp. It is therefore unlikely that this difference
can be ascribed to an uncertainty on the $\twCHp /\thCHp$ ratio.
Last, \citet{Daflon2004} found that carbon and sulfur have similar abundance
gradients across the Galactic disk (with slopes of -0.037 and -0.040
dex kpc$^{-1}$, respectively).  Consequently, the difference observed
in the $\SHp / \CHp$ column density ratios measured on the $l\sim 0$ and 
$l\ne 0$ sight lines is most likely tracing variations of both the physical and
chemical conditions of the diffuse gas sampled in each case.  

Finally, with a correlation coefficient of 0.1, no evident chemical
relationship seems to stand out from Fig. \ref{Fig-Correl-SHp-CHp}, a
surprising finding in view of the fact that \CHp\ and \SHp\ are
clearly linked by their dynamics (see Sect. \ref{SectResDyn}).
However, this lack of correlation applies to the column densities.
In the following, we discuss the properties of the abundances relative to hydrogen, 
a discussion that requires the knowledge of $N_{\rm H}$.


\subsection{Estimation of the hydrogen column densities in the broad velocity intervals} 
\label{EstimationOfHydrogen}

To (1) estimate the mean molecular abundances that are to be compared
with the chemical model predictions (Godard et al., in preparation), 
and (2) to
establish a possible relation between the \CHp\ and \SHp\ abundances,
it is essential to estimate the hydrogen column density $\NHt$ in the broad 
velocity intervals over which $N(\CHp)$ and $N(\SHp)$ are measured. 
Since the molecular fraction of the gas where \CHp\ and \SHp\ are detected
is low (0.4 on average, see Appendix \ref{AppendTracerHH}), both atomic and 
molecular hydrogen HI and \HH\ are needed to estimate $\NHt$.

The method consists in separately evaluating the column densities of 
HI and \HH, using the VLA observations of the $\lambda 21$ cm absorption line 
of H \citep{Koo1997,Fish2003,Dwarakanath2004,Pandian2008,Lang2010} and a
relevant tracer for the molecular hydrogen because \HH\ is not directly
observable. Then, $\NHt=N(HI)+2 N(\HH)$.
In Appendix \ref{AppendTracerHH} we discuss the validity of using the
HIFI observations of CH and HF to compute the \HH\ column
densities. If available, HF is preferentially used in the
following section to infer $N(\HH)$ and the ensuing \CHp, \thCHp,
and \SHp\ mean abundances. If not, $N(\HH)$ is derived from CH,
assuming a HF/CH mean abundance ratio of 0.4 (a value defined 
with a large standard deviation of 0.25), deduced from columns
5 and 6 of Table \ref{TabHI}. Last, we compare these values of $\NHt$ 
to those inferred, as in Godard et al. (2010),
from the analysis of the 2MASS survey
\citep{Cutri2003}. \citet{Marshall2006} have measured the near
infrared colour excess in large areas of the inner Galaxy
($|l|<100^{\circ}$, $|b|<10^{\circ}$) to obtain the visible
extinctions ($A_V \sim 10A_K$), providing an estimate of the total
hydrogen column density along the lines of sight.  However, because of
the low resolution of the 2MASS extinction analysis ($\sim 15$
arcmin), the uncertainty on $\NHt$ (computed as the standard deviation
of the extinction measured along the four closest lines of sight
surrounding a given source) can be important, and is as high as 50\% 
for DR21(OH), W31C, and W33A.
The last columns of Table \ref{TabHI} show that these two independent 
measurements of $\NHt$ differ by less than 25 \%.

\section{Results and discussion} \label{SectDisc}

\subsection{Chemical properties of the gas seen in absorption}

\begin{figure}[!ht]
\begin{center}
\includegraphics[width=6.4cm,angle=-90]{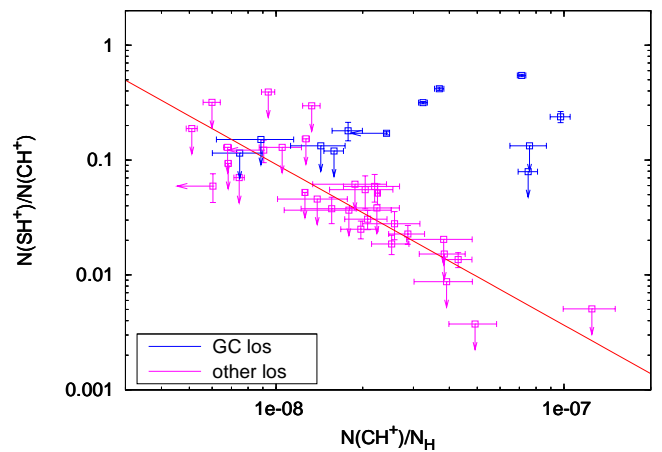}
\caption{$N(\SHp) / N(\CHp)$ column density ratio as a function of the \CHp\ mean abundance. 
The \CHp, \SHp, and H column densities are computed from \CHp, \thCHp, \SHp, HI, HF, and 
CH opacities integrated over the velocity intervals given in Tables \ref{TabIntTau} and \ref{TabHI}.
The blue and purple points are from the absorption lines observed along the Galactic centre 
sight lines (SgrA*+50 and SgrB2(N)) and along the other sight lines (DR21(OH), G34.3+0.1, W31C, 
W33A, W49N, and W51), respectively. The red line corresponds to a least-squares fit
of the latter data.}
\label{Fig-Correlation-Norm}
\end{center}
\end{figure}


In Fig. \ref{Fig-Correlation-Norm}, we display the $N(\SHp)/N(\CHp)$
column density ratio as a function of the \CHp\ mean abundance (with
respect to the total hydrogen column density $N_{\rm H}$) integrated
over the velocity intervals given in Table \ref{TabHI}. Because
  \citet{Menten2011} used a different method than we did to estimate the
  \HH\ column densities
\footnote{In \citet{Menten2011} the \HH\ column densities are deduced
from those of \HCOp, assuming $N(\HCOp) / N(\HH) = 5 \times 10^{-9}$
\citep{Lucas1996}.} the data obtained towards SgrB2(M) are not included 
in this plot. We found that both the mean abundances and the abundance
ratios vary by two orders of magnitude in the diffuse ISM sampled by
those lines of sight. In addition, Fig. \ref{Fig-Correlation-Norm}
reveals major differences between the results obtained on the $l\sim 0$ and 
the $l \ne 0$ sight lines. While the \CHp\ mean abundances span the same 
range of values in both cases, the $\SHp/ \CHp$ ratio measured towards 
SgrA*+50 and SgrB2(N) shows no correlation with $N(\CHp)/N_{\rm H}$. 
Conversely, the points corresponding to the observations
performed on the $l \ne 0$ sight lines exhibit a trend
\begin{equation}
N(\SHp)/N(\CHp) \sim  0.09 \,\, \big[ 10^{8} \times N(\CHp)/N_{\rm H} \big]^{-1.4},
\end{equation}
with a correlation coefficient of 0.8. These are the results that have 
to be compared with the predictions of chemical models applied to the 
diffuse ISM.

\subsection{Carbon and sulfur chemistries in the diffuse ISM} \label{Sect-C-S-Chemistry}

\subsubsection{UV-driven chemistry}

\begin{figure}[!ht]
\begin{center}
\includegraphics[width=6.4cm,angle=-90]{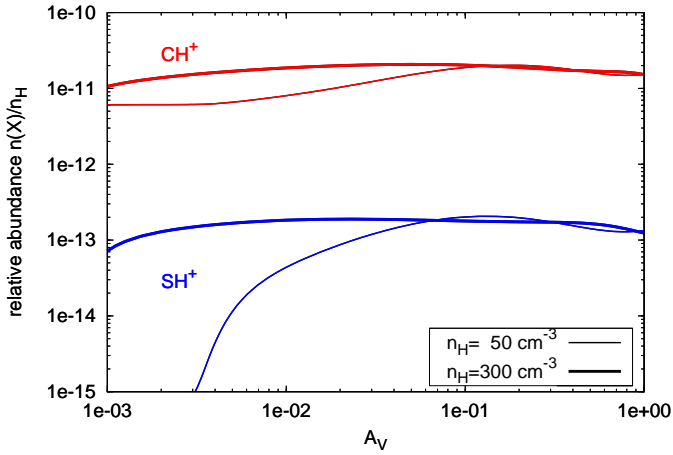}
\caption{Predictions of two PDR models for gas densities $\dens=50$ and $300$ \cc. The \CHp\ (in red)
and \SHp\ (in blue) relative abundances are displayed as functions of the shielding $A_V$ from the 
interstellar radiation field for a slab of gas illuminated on one side only.}
\label{Fig-PDR}
\end{center}
\end{figure}

In a chemistry entirely driven by the UV-radiation field and the cosmic 
ray particles, the hydrogenation chains of carbon and sulfur, and the subsequent 
productions of \CHp\ and \SHp\, are initiated by the radiative associations of \Cp\ and \Sp\ 
with molecular hydrogen,
\begin{equation}
\begin{array}{l l l l l l l}
{\rm C}^+ & + & {\rm H}_2 & \rightarrow & {\rm CH}_2^+ & + & \gamma \,\, {\rm and} \\[6.0pt]
{\rm S}^+ & + & {\rm H}_2 & \rightarrow & {\rm SH}_2^+ & + & \gamma, \\
\end{array}
\end{equation}
two reactions with long timescales:
$2 \times 10^{6} \, {\rm yr} \,\, f_{{\rm H}_2}^{-1} \left( 50 \, \cc / \dens \right)$ and 
$1 \times 10^{8} \, {\rm yr} \,\, f_{{\rm H}_2}^{-1} \left( 50 \, \cc / \dens \right)$, respectively
\citep{Herbst1985,Herbst1989},
where $f_{\HH}$	is the molecular fraction defined as $f_{\HH} = 2n(\HH)/\dens$.
In comparison, \CHp\ and \SHp\ are mainly destroyed by hydrogenation and dissociative 
recombination\footnote{Because the hydrogenation of \SHp\ is highly endo-energetic: 
$\Delta E / k = 6380$ K.}, 
\begin{equation}
\begin{array}{l l l l l l l}
{\rm CH}^+ & + & {\rm H}_2 & \rightarrow & {\rm CH}_2^+ & + & {\rm H} \,\, {\rm and} \\[6.0pt]
{\rm SH}^+ & + & e^-       & \rightarrow & {\rm S}      & + & {\rm H}, \\
\end{array}
\end{equation}
two processes with short timescales: $1 \, {\rm yr} \,\, f_{{\rm
    H}_2}^{-1} \left( 50 \, \cc / \dens \right)$ and $5 \, {\rm yr}
\,\, (T /100 \, {\rm K})^{0.72} ( 1.38 \times 10^{-4} / x_{e^-} ) ( 50
\, \cc / \dens )$, respectively. Because of the lack of efficient
production pathways to balance their rapid destruction, the \CHp\ and
\SHp\ abundances predicted by UV-dominated chemistry are very
low. Fig. \ref{Fig-PDR} displays the \CHp\ and \SHp\ relative
abundances computed with two models of 
PhotoDissociation Regions (PDR)\footnote{The Meudon PDR model employs a
one-dimensional chemical code in which a slab of gas with a given
density profile is illuminated by the ambient interstellar radiation
field \citep{Le-Petit2006}.} illuminated on one side as
functions of the shielding from the ISRF $A_V$. For the physical
conditions of the diffuse gas, we obtain $N(\CHp)/N_{\rm H} \sim 1.8
\times 10^{-11}$ and $N(\SHp)/N_{\rm H} \sim 1.6 \times 10^{-13}$, two
to four orders of magnitude lower than the observed values, and with
a corresponding abundance ratio never higher than 0.01.

\subsubsection{Alternative models}


In the diffuse interstellar medium sampled by the $l \ne 0$ sight lines, the only production pathways 
efficient enough to balance the fast destruction of \CHp\ and \SHp\ are
\begin{equation}\label{reac-CHp} 
{\rm C}^+  + {\rm H}_2 \rightarrow {\rm CH}^+ + {\rm H} \quad \Delta E / k = 4640 {\rm K}, \quad {\rm and}
\end{equation}
\begin{equation}\label{reac-SHp}
{\rm S}^+  + {\rm H}_2 \rightarrow {\rm SH}^+ + {\rm H} \quad \Delta E / k = 9860 {\rm K}.
\end{equation}
Since these reactions are highly endothermic, it has been proposed that
high \CHp\ and \SHp\ abundances are the signatures of shock waves
propagating through the ISM \citep{Draine1986,Millar1986}.  Comparing
the predictions of HD and MHD
shocks with the \CHp\ column density observed towards $\zeta$ Oph,
\citet{Draine1986} and \citet{Pineau-des-Forets1986} favoured the MHD
case in which \CHp\ and \SHp\ form mainly through ion-neutral
friction. Including the sulfur chemistry \citet{Millar1986} and
\citet{Pineau-des-Forets1986} predicted a $\SHp / \CHp$ abundance
ratio increasing from 0.01 to 0.4 for a shock speed increasing from 9 to
16 \kms, a transverse magnetic field of $5 \mu$G and a preshock
density of 20 \cc. These results agree excellently with our observations.

Another scenario, alternative to the shock waves, is the TDR
(turbulent dissipation regions) model.  In this model the turbulent
energy is dissipated in many small-scale magnetized vortices 
in which the ionized and neutral fluids decouple. Dissipation is caused by both 
ion-neutral friction and viscous dissipation at the edge of the vortices. 
Comparing the predictions of TDRs to observations of CH, \CHp, OH, and \HCOp\
in the local diffuse gas, \citet{Godard2009} also favoured models in which 
the dissipation is dominated by the ion-neutral friction and where the 
production of \CHp\ and \SHp\ via reactions \ref{reac-CHp} and \ref{reac-SHp} 
is triggered by the ambipolar diffusion. An analysis of the $\SHp / \CHp$ ratio 
obtained in the framework of the TDR model will be presented in a
forthcoming paper (Godard et al. in prep.).

While the scenarios of shocks and vortices could equally apply to gas
sampled by the $l \sim 0$ sight lines, an alternative chemical process
may be at work there. Large amounts of the diffuse gas detected along
SgrA*+50, SgrB2(N), and SgrB2(M) belong to the Central Molecular Zone
(CMZ).  Because the CMZ is pervaded by a strong X-ray radiation field,
high \CHp\ and \SHp\ abundances could be due to regions where
\Cpp\ and \Spp\ ions co-exist with \HH\ and form \CHp\ and \SHp\ by
the reactions
\begin{equation}\label{reac-CHp2} 
{\rm C}^{++}  + {\rm H}_2 \rightarrow {\rm CH}^+ + {\rm H}^+, \quad {\rm and}
\end{equation}
\begin{equation}\label{reac-SHp2}
{\rm S}^{++}  + {\rm H}_2 \rightarrow {\rm SH}^+ + {\rm H}^+,
\end{equation}
as proposed by \citet{Langer1978}. Using the high rate of reaction
(\ref{reac-SHp2}) obtained by \citet{Chen2003}, \citet{Abel2008} found
that the predicted \SHp\ column density for molecular gas surrounding
an active galactic nucleus is two orders of magnitude higher than that
predicted by UV-dominated chemistry. Since rate measurements of
reaction (\ref{reac-CHp2}) point to lower values, this chemical process
could account for the high mean $\SHp / \CHp$ abundance ratio derived
towards SgrA*+50, SgrB2(N), and SgrB2(M) (about nine times higher than
that obtained along the other Galactic sight lines, see
Sect. \ref{SectResComp}).  

The above results are in line with the detection of large amounts 
 of warm diffuse gas recently identified with the 3$\mu m$ absorption lines 
 of \Htp\ \citep{Oka2005,Geballe2010} in the CMZ. The temperature ($\sim 200 -
  300$ K) of the low-density gas phase is found to be considerably
  higher than that of typical diffuse clouds, and unique to the
  Galactic CMZ \citep{Goto2008}. They also corroborate 
the specific dynamics of molecular clouds
  associated with the Galactic centre regions
  (e.g. \citealt{Rodriguez-Fernandez2006}).

\section{Summary and perspectives}

We have presented the analysis of {\it Herschel}/HIFI observations of
the ground-state transitions of \CHp, \thCHp, and \SHp, all detected in absorption against 
the submillimetre dust continuum of 
distant star-forming regions and the 
Galactic centre sources, SgrA*+50 and SgrB2(N). The velocity range
over which the absorption features are detected corresponds to
diffuse or transluscent environments. The deconvolution of the
hyperfine structure embedded in the \SHp\ $1,2 - 0,1$ spectra, and the
independent decomposition of the absorption domains in Gaussian
velocity components allowed us to identify many velocity
components per sight line and to perform a cross comparison of the
dynamical and chemical signatures of those three species.

This study provides the following main results.
(1) The linewidth distributions of \CHp, \thCHp, and \SHp\ are found to be similar 
and likely trace the kinematics of the chemical production processes of these species
convolved with that of the turbulent and Galactic dynamics of the diffuse ISM.
(2) These lines are broad ($\sim 4.2$ \kms), similar to those found in
visible absorption lines in the solar neighbourhood, and broader than those of \HCOp\ and 
CN along the same lines of sight.
(3) The $\SHp / \CHp$ abundance ratio covers a broad range of values from 0.01 to more than 1, 
shows higher values in warmer environments (such as the Galactic centre clouds), and appears to 
be proportional to $(N(\CHp)/N_{\rm H})^{-1.4}$ in the diffuse gas 
sampled by the $l \ne 0$ sight lines.
(4) As for \CHp, the \SHp\ abundances cannot be reproduced by UV-driven chemistry in the diffuse gas.

The unique properties of the carbon and sulfur chemistries support the
framework of a warm chemistry triggered by turbulent dissipation
(either in shocks or intense velocity shears) that selectively
enhances the production of \SHp\ for which the formation
endothermicity is the highest.  A detailed comparison of the TDR model
predictions with these observational results will be given in a
forthcoming paper (Godard et al. in prep.).

\begin{acknowledgements}

We are most grateful to the referee for providing constructive comments 
and helping in improving the content of this paper.
HIFI has been designed and built by a consortium of institutes and university
departments from across Europe, Canada and the United States (NASA) under the
leadership of SRON, Netherlands Institute for Space Research, Groningen, The
Netherlands, and with major contributions from Germany, France and the
US. Consortium members are : Canada: CSA, U. Waterloo; France : CESR, LAB,
LERMA, IRAM; Germany : KOSMA, MPIfR, MPS; Ireland : NUI Maynooth; Italy : ASI,
IFSI-INAF, Osservatorio Astrofisico di Arcetri-INAF; Netherlands : SRON, TUD;
Poland : CAMK, CBK; Spain : Observatorio Astron\`omico Nacional (IGN), Centro
de Astrobiologia; Sweden : Chalmers University of Technology - MC2, RSS \&
GARD, Onsala Space Observatory, Swedish National Space Board, Stockholm
University - Stockholm Observatory; Switzerland : ETH Zurich, FHNW; USA :
CalTech, JPL, NHSC.  
BG, EF, MG, and MDL acknowledge the support from the Centre National de Recherche 
Spatiale (CNES), and from ANR through the SCHISM project (ANR-09-BLAN-231).
BG, JC, and JRG thank the Spanish MICINN for funding support through grants,
AYA2009-07304 and CSD2009-00038

\end{acknowledgements}



\longtab{3}{
\begin{longtable}{r @{\hspace{0.05cm}} r      @{\hspace{0.30cm}} c @{\hspace{0.45cm}} 
                  r @{\hspace{0.05cm}} r@{.}l @{\hspace{0.05cm}} l @{\hspace{0.45cm}}
                  r @{\hspace{0.05cm}} r@{.}l @{\hspace{0.05cm}} l @{\hspace{0.45cm}}
                  r @{\hspace{0.05cm}} r@{.}l @{\hspace{0.05cm}} l @{\hspace{0.45cm}}
                  c @{\hspace{0.1cm}}  c      @{\hspace{0.10cm}} c @{\hspace{0.45cm}}
                  r @{\hspace{0.05cm}} r@{.}l @{\hspace{0.05cm}} l @{\hspace{0.45cm}}
                  r @{\hspace{0.05cm}} r@{.}l @{\hspace{0.05cm}} l @{\hspace{0.45cm}}
                  r @{\hspace{0.05cm}} r@{.}l @{\hspace{0.05cm}} l}
\caption{Comparison between the column densities of \CHp, \thCHp, and \SHp obtained with the integrations of
the respective opacities over several velocity intervals. The two last columns contain the ratios 
$r_{12} = N(\SHp)/N(\CHp)$ and $r_{13} = N(\SHp)/N(\thCHp) \times \thC/\twC$. In column 10, \CHp\ column 
densities are inferred from those of \thCHp\ and the \twC/\thC\ Galactic gradient derived by \citet{Milam2005}:
$N(\CHp) = N(\thCHp) \times \twC/\thC$. The blank spaces correspond to unusable \thCHp\ or \SHp\ data owing
to the contamination of an emission line from the SFRs, or because the spectrum measured over the associated 
velocity interval could not be deconvolved from the \SHp\ hyperfine structure.} \\
\hline
\multicolumn{1}{c}{$\upsilon_{\rm min}$} & \multicolumn{1}{c}{$\upsilon_{\rm max}$} & $*$ & \multicolumn{4}{l}{$N$(CH$^{+}$)$^b$}     & \multicolumn{4}{l}{$N$(SH$^{+}$)}         & \multicolumn{4}{l}{$N$($^{13}$CH$^{+}$)}  & $R_{g,{\rm min}}^{a}$ & $R_{g,{\rm max}}^{a}$ & $\twC/\thC$ & \multicolumn{4}{l}{$N$(CH$^{+}$)$^b$}     & \multicolumn{4}{l}{$r_{12}$$^b$}& \multicolumn{4}{l}{$r_{13}$$^b$}\\
\multicolumn{1}{c}{(km s$^{-1}$)}        & \multicolumn{1}{c}{(km s$^{-1}$)}        &     & \multicolumn{4}{l}{($10^{13}$ cm$^{-2}$)} & \multicolumn{4}{l}{($10^{12}$ cm$^{-2}$)} & \multicolumn{4}{l}{($10^{12}$ cm$^{-2}$)} & (kpc)                 & (kpc)                 &             & \multicolumn{4}{l}{($10^{13}$ cm$^{-2}$)} & \multicolumn{4}{l}{$\times 10$} & \multicolumn{4}{l}{$\times 10$} \\
\hline
\endfirsthead
\caption{continued.} \\
\hline
\multicolumn{1}{c}{$\upsilon_{\rm min}$} & \multicolumn{1}{c}{$\upsilon_{\rm max}$} & $*$ & \multicolumn{4}{l}{$N$(CH$^{+}$)$^b$}     & \multicolumn{4}{l}{$N$(SH$^{+}$)}         & \multicolumn{4}{l}{$N$($^{13}$CH$^{+}$)}  & $R_{g,{\rm min}}^{a}$ & $R_{g,{\rm max}}^{a}$ & $\twC/\thC$ & \multicolumn{4}{l}{$N$(CH$^{+}$)$^b$}     & \multicolumn{4}{l}{$r_{12}$$^b$}& \multicolumn{4}{l}{$r_{13}$$^b$}\\
\multicolumn{1}{c}{(km s$^{-1}$)}        & \multicolumn{1}{c}{(km s$^{-1}$)}        &     & \multicolumn{4}{l}{($10^{13}$ cm$^{-2}$)} & \multicolumn{4}{l}{($10^{12}$ cm$^{-2}$)} & \multicolumn{4}{l}{($10^{12}$ cm$^{-2}$)} & (kpc)                 & (kpc)                 &             & \multicolumn{4}{l}{($10^{13}$ cm$^{-2}$)} & \multicolumn{4}{l}{$\times 10$} & \multicolumn{4}{l}{$\times 10$} \\
\hline
\endhead
\hline
\\
\multicolumn{30}{p{7.2in}}{ $*$ E = absorption line profile observed in the star-forming region; $T_{\rm ex}$ may be underestimated, 
hence the lower limit on $N(\CHp)$, $N(\thCHp)$, and $N(\SHp)$. 1,2,3,4,5 = velocity intervals where \twCHp\ and \thCHp\ are observed
simultaneously with a $3\sigma$ detection level.}\\
\multicolumn{30}{p{7.2in}}{ $^a$ Galactocentric distances derived as 
$R_G = R_0 \frac{\theta(R_G) \, {\rm sin}(l) {\rm cos}(b)}{V_{\rm lsr} + \theta_0 \, {\rm sin}(l) {\rm cos}(b)}$
with $R_0 = 8.5$ kpc and $\theta_0 = 220$ \kms\ as recommended by the IAU \citep{Kerr1986}, 
and assuming a flat Galactic rotation curve ($\theta(R_G) = \theta_0$).} \\
\multicolumn{30}{p{7.2in}}{$^b$ Towards SgrA*+50 and SgrB2(N), when the comparison is possible, 
$N(\CHp)$ derived from \twCHp\ and \thCHp\ and the ratios $r_{12}$ and $r_{13}$ agree within the 
combination of their respective errors and the uncertainty of 50 \% on the $\twC/\thC$ ratio (see main 
text).}
\endfoot
\multicolumn{30}{c}{\bf SgrA*+50} \\
\hline
-215	&	-198	&		&		&	0	&	3	&	$\pm$	0.1	&		&\multicolumn{2}{}{}&				&	$<$	&	0	&	6		&				&		&		&	20	&	$<$	&	1	&	2		&				&		&\multicolumn{2}{}{}&				&		&\multicolumn{2}{}{}&				\\
-198	&	-168	&		&		&	3	&	4	&	$\pm$	0.1	&		&\multicolumn{2}{}{}&				&	$<$	&	0	&	8		&				&		&		&	20	&	$<$	&	1	&	6		&				&		&\multicolumn{2}{}{}&				&		&\multicolumn{2}{}{}&				\\
-168	&	-150	&		&	$>$	&	6	&	9	&				&	$<$	&	3	&	4		&				&		&	2	&	3		&	$\pm$	0.6	&		&		&	20	&		&	4	&	7		&	$\pm$	1.3	&	$<$	&	0	&	5		&				&	$<$	&	0	&	7		&				\\
-150	&	-123	&		&	$>$	&	19	&	1	&				&		&	56	&	8		&	$\pm$	4.8	&		&	11	&	9		&	$\pm$	0.9	&		&		&	20	&		&	23	&	9		&	$\pm$	1.7	&	$<$	&	3	&	0		&				&		&	2	&	4		&	$\pm$	0.3	\\
-123	&	-95		&		&	$>$	&	17	&	7	&				&		&	15	&	5		&	$\pm$	4.3	&		&	5	&	8		&	$\pm$	0.8	&		&		&	20	&		&	11	&	6		&	$\pm$	1.6	&	$<$	&	0	&	9		&				&		&	1	&	3		&	$\pm$	0.4	\\
-95		&	-75		&		&		&	4	&	4	&	$\pm$	0.2	&	$<$	&	3	&	5		&				&	$<$	&	0	&	6		&				&		&		&	20	&	$<$	&	1	&	3		&				&	$<$	&	0	&	8		&				&		&\multicolumn{2}{}{}&				\\
-75		&	-45		&		&	$>$	&	7	&	9	&				&		&	8	&	6		&	$\pm$	4.3	&		&	2	&	9		&	$\pm$	0.8	&		&		&	20	&		&	5	&	7		&	$\pm$	1.7	&	$<$	&	1	&	1		&				&		&	1	&	5		&	$\pm$	0.9	\\
-45		&	-15		&	1	&		&	9	&	1	&	$\pm$	0.3	&		&	10	&	9		&	$\pm$	4.3	&		&	4	&	1		&	$\pm$	0.8	&		&		&	20	&		&	8	&	2		&	$\pm$	1.7	&		&	1	&	2		&	$\pm$	0.5	&		&	1	&	3		&	$\pm$	0.6	\\
-15		&	0		&		&	$>$	&	9	&	6	&				&		&	8	&	6		&	$\pm$	3.1	&		&	3	&	7		&	$\pm$	0.6	&		&		&	20	&		&	7	&	5		&	$\pm$	1.2	&	$<$	&	0	&	9		&				&		&	1	&	1		&	$\pm$	0.5	\\
		&			&		&		&\multicolumn{2}{}{}&			&		&\multicolumn{2}{}{}&				&		&\multicolumn{2}{}{}&				&		&		&	60	&		&	22	&	5		&	$\pm$	3.6	&		&\multicolumn{2}{}{}&				&		&	0	&	4		&	$\pm$	0.1	\\
0		&	26		&		&	$>$	&	14	&	4	&				&		&	28	&	0		&	$\pm$	4.2	&		&	7	&	8		&	$\pm$	0.8	&		&		&	20	&		&	15	&	6		&	$\pm$	1.6	&	$<$	&	2	&	0		&				&		&	1	&	8		&	$\pm$	0.3	\\
		&			&		&		&\multicolumn{2}{}{}&			&		&\multicolumn{2}{}{}&				&		&\multicolumn{2}{}{}&				&		&		&	60	&		&	46	&	8		&	$\pm$	4.8	&		&\multicolumn{2}{}{}&				&		&	0	&	6		&	$\pm$	0.1	\\
26		&	49		&	E	&	$>$	&	12	&	9	&				&	$>$	&	89	&	2		&				&	$>$	&	8	&	1		&				&		&		&	20	&	$>$	&	16	&	1		&				&		&\multicolumn{2}{}{}&				&		&\multicolumn{2}{}{}&				\\
49		&	67		&	E	&	$>$	&	2	&	8	&				&	$>$	&	41	&	0		&				&	$>$	&	3	&	2		&				&		&		&	20	&	$>$	&	6	&	5		&				&		&\multicolumn{2}{}{}&				&		&\multicolumn{2}{}{}&				\\
67		&	75		&	E	&	$>$	&	0	&	2	&				&		&\multicolumn{2}{}{}&				&		&\multicolumn{2}{}{}&				&		&		&	20	&		&\multicolumn{2}{}{}&				&		&\multicolumn{2}{}{}&				&		&\multicolumn{2}{}{}&				\\
\hline
\multicolumn{30}{c}{\bf SgrB2(N)} \\
\hline
-135	&	-100	&		&	$>$	&	13	&	5	&				&		&	31	&	5		&	$\pm$	0.9	&		&\multicolumn{2}{}{}&				&		&		&	20	&		&\multicolumn{2}{}{}&				&	$<$	&	2	&	3		&				&		&\multicolumn{2}{}{}&				\\
-82		&	-62		&		&	$>$	&	15	&	0	&				&		&	49	&	3		&	$\pm$	0.8	&		&	5	&	9		&	$\pm$	0.2	&		&		&	20	&		&	11	&	8		&	$\pm$	0.3	&	$<$	&	3	&	3		&				&		&	4	&	2		&	$\pm$	0.1	\\
-62		&	-35		&		&	$>$	&	20	&	0	&				&		&	33	&	6		&	$\pm$	0.8	&		&	9	&	8		&	$\pm$	0.2	&		&		&	20	&		&	19	&	6		&	$\pm$	0.4	&	$<$	&	1	&	7		&				&		&	1	&	7		&	$\pm$	0.1	\\
-35		&	-15		&		&	$>$	&	15	&	0	&				&		&	57	&	3		&	$\pm$	0.8	&		&	9	&	1		&	$\pm$	0.2	&		&		&	20	&		&	18	&	1		&	$\pm$	0.4	&	$<$	&	3	&	8		&				&		&	3	&	2		&	$\pm$	0.1	\\
-15		&	23		&		&	$>$	&	27	&	8	&				&		&	296	&	8		&	$\pm$	1.6	&		&\multicolumn{2}{}{}&				&		&		&	20	&		&\multicolumn{2}{}{}&				&	$<$	&	10	&	7		&				&		&\multicolumn{2}{}{}&				\\
		&			&		&		&\multicolumn{2}{}{}&			&		&\multicolumn{2}{}{}&				&		&\multicolumn{2}{}{}&				&		&		&	60	&		&\multicolumn{2}{}{}&				&		&\multicolumn{2}{}{}&				&		&\multicolumn{2}{}{}&				\\
23		&	38		&		&		&	7	&	8	&	$\pm$	0.1	&		&	42	&	7		&	$\pm$	0.7	&		&\multicolumn{2}{}{}&				&		&		&	20	&		&\multicolumn{2}{}{}&				&		&	5	&	5		&	$\pm$	0.1	&		&\multicolumn{2}{}{}&				\\
38		&	74		&	E	&	$>$	&	20	&	9	&				&	$>$	&	235	&	0		&				&	$>$	&	5	&	6		&				&		&		&	20	&	$>$	&	11	&	2		&				&		&\multicolumn{2}{}{}&				&		&\multicolumn{2}{}{}&				\\
74		&	100		&	E	&	$>$	&	4	&	9	&				&	$>$	&	68	&	4		&				&	$>$	&	0	&	8		&				&		&		&	20	&	$>$	&	1	&	6		&				&		&\multicolumn{2}{}{}&				&		&\multicolumn{2}{}{}&				\\
\hline
\multicolumn{30}{c}{\bf DR21(OH)} \\
\hline
-17		&	-9		&	E	&	$>$	&	0	&	2	&				&		&\multicolumn{2}{}{}&				&		&\multicolumn{2}{}{}&				&	8.9	&	9.2	&	75	&		&\multicolumn{2}{}{}&				&		&\multicolumn{2}{}{}&				&		&\multicolumn{2}{}{}&				\\
-8		&	1		&	E	&	$>$	&	5	&	4	&				&	$>$	&	5	&	4		&				&	$>$	&	3	&	1		&				&	8.5	&	8.8	&	72	&	$>$	&	22	&	1		&				&		&\multicolumn{2}{}{}&				&		&\multicolumn{2}{}{}&				\\
1		&	7		&		&	$>$	&	4	&	0	&				&	$<$	&	1	&	0		&				&		&	1	&	0		&	$\pm$	0.1	&	8.3	&	8.5	&	71	&		&	7	&	0		&	$\pm$	1.5	&	$<$	&	0	&	3		&				&	$<$	&	0	&	1		&				\\
7		&	15		&		&	$>$	&	6	&	0	&				&		&	2	&	7		&	$\pm$	1.3	&		&	1	&	5		&	$\pm$	0.1	&	8.0	&	8.3	&	69	&		&	10	&	5		&	$\pm$	2.1	&	$<$	&	0	&	5		&				&		&	0	&	3		&	$\pm$	0.1	\\
18		&	26		&	E	&	$>$	&	0	&	1	&				&		&\multicolumn{2}{}{}&				&	$>$	&	0	&	3		&				&	7.6	&	7.9	&	67	&	$>$	&	2	&	1		&				&		&\multicolumn{2}{}{}&				&		&\multicolumn{2}{}{}&				\\
\hline
\multicolumn{30}{c}{\bf G34.3+0.1} \\
\hline
-4		&	5		&		&		&	0	&	8	&	$\pm$	0.02&		&\multicolumn{2}{}{}&				&	$<$	&	0	&	2		&				&	8.2	&	8.5	&	71	&	$<$	&	1	&	1		&				&		&\multicolumn{2}{}{}&				&		&\multicolumn{2}{}{}&				\\
5		&	21		&		&		&	2	&	2	&	$\pm$	0.03&		&\multicolumn{2}{}{}&				&		&\multicolumn{2}{}{}&				&	7.3	&	8.2	&	67	&		&\multicolumn{2}{}{}&				&		&\multicolumn{2}{}{}&				&		&\multicolumn{2}{}{}&				\\
21		&	34		&	2	&		&	3	&	7	&	$\pm$	0.05&		&	3	&	5		&	$\pm$	1.4	&		&	1	&	2		&	$\pm$	0.2	&	6.7	&	7.3	&	62	&		&	7	&	6		&	$\pm$	2.1	&		&	0	&	9		&	$\pm$	0.4	&		&	0	&	5		&	$\pm$	0.2	\\
36		&	44		&	3	&		&	2	&	1	&	$\pm$	0.04&	$<$	&	1	&	1		&				&		&	0	&	5		&	$\pm$	0.2	&	6.3	&	6.6	&	59	&		&	3	&	1		&	$\pm$	1.2	&	$<$	&	0	&	5		&				&	$<$	&	0	&	4		&				\\
44		&	58		&	E	&	$>$	&	9	&	5	&				&	$>$	&	3	&	4		&				&	$>$	&	3	&	7		&				&	5.8	&	6.3	&	56	&	$>$	&	20	&	7		&				&		&\multicolumn{2}{}{}&				&		&\multicolumn{2}{}{}&				\\
58		&	66		&	E	&	$>$	&	4	&	2	&				&	$>$	&	3	&	5		&				&	$>$	&	2	&	2		&				&	5.5	&	5.8	&	54	&	$>$	&	11	&	7		&				&		&\multicolumn{2}{}{}&				&		&\multicolumn{2}{}{}&				\\
\hline
\multicolumn{30}{c}{\bf W31C} \\
\hline
-5		&	3		&	E	&	$>$	&	1	&	8	&				&		&\multicolumn{2}{}{}&				&	$>$	&	1	&	0		&				&	8.0	&	8.5	&	70	&	$>$	&	6	&	9		&				&		&\multicolumn{2}{}{}&				&		&\multicolumn{2}{}{}&				\\
3		&	13		&		&	$>$	&	6	&	8	&				&		&\multicolumn{2}{}{}&				&		&	1	&	8		&	$\pm$	0.1	&	6.5	&	8.0	&	64	&		&	11	&	5		&	$\pm$	2.2	&		&\multicolumn{2}{}{}&				&		&\multicolumn{2}{}{}&				\\
13		&	25		&		&	$>$	&	8	&	7	&				&		&	4	&	0		&	$\pm$	0.5	&		&	3	&	9		&	$\pm$	0.2	&	5.3	&	6.5	&	55	&		&	21	&	4		&	$\pm$	3.1	&	$<$	&	0	&	5		&				&		&	0	&	2		&	$\pm$	0.04	\\
25		&	31		&		&	$>$	&	4	&	8	&				&		&	3	&	6		&	$\pm$	0.4	&		&	2	&	8		&	$\pm$	0.1	&	4.8	&	5.3	&	50	&		&	14	&	2		&	$\pm$	2.0	&	$<$	&	0	&	7		&				&		&	0	&	2		&	$\pm$	0.04	\\
31		&	36		&		&	$>$	&	3	&	3	&				&		&	2	&	6		&	$\pm$	0.3	&		&	2	&	4		&	$\pm$	0.1	&	4.5	&	4.8	&	48	&		&	11	&	3		&	$\pm$	1.6	&	$<$	&	0	&	8		&				&		&	0	&	2		&	$\pm$	0.04	\\
36		&	44		&		&	$>$	&	6	&	2	&				&		&	4	&	5		&	$\pm$	0.4	&		&	7	&	3		&	$\pm$	0.2	&	4.1	&	4.5	&	45	&		&	33	&	1		&	$\pm$	3.8	&	$<$	&	0	&	7		&				&		&	0	&	1		&	$\pm$	0.03	\\
44		&	51		&		&	$>$	&	3	&	6	&				&	$<$	&	0	&	4		&				&		&	1	&	7		&	$\pm$	0.1	&	3.8	&	4.1	&	43	&		&	7	&	1		&	$\pm$	1.1	&	$<$	&	0	&	1		&				&	$<$	&	0	&	1		&				\\
51		&	61		&		&		&	0	&	3	&	$\pm$	0.01&	$<$	&	0	&	4		&				&	$<$	&	0	&	1		&				&	3.4	&	3.8	&	41	&	$<$	&	0	&	6		&				&	$<$	&	1	&	3		&				&		&\multicolumn{2}{}{}&				\\
\hline
\multicolumn{30}{c}{\bf W33A} \\
\hline
-2		&	7		&		&		&	0	&	5	&	$\pm$	0.02&	$<$	&	1	&	1		&				&	$<$	&	0	&	2		&				&	7.4	&	8.5	&	69	&	$<$	&	1	&	2		&				&	$<$	&	2	&	3		&				&		&\multicolumn{2}{}{}&				\\
14		&	18		&		&		&	0	&	3	&	$\pm$	0.01&	$<$	&	0	&	7		&				&		&	0	&	3		&	$\pm$	0.2	&	6.2	&	6.6	&	59	&		&	1	&	8		&	$\pm$	1.1	&	$<$	&	3	&	0		&				&	$<$	&	0	&	4		&				\\
18		&	25		&		&	$>$	&	3	&	8	&				&		&	1	&	7		&	$\pm$	1.0	&		&	1	&	5		&	$\pm$	0.2	&	5.6	&	6.2	&	56	&		&	8	&	1		&	$\pm$	2.1	&	$<$	&	0	&	4		&				&		&	0	&	2		&	$\pm$	0.1	\\
25		&	31		&		&	$>$	&	4	&	2	&				&		&	2	&	2		&	$\pm$	0.9	&		&	2	&	8		&	$\pm$	0.2	&	5.2	&	5.6	&	52	&		&	14	&	7		&	$\pm$	2.7	&	$<$	&	0	&	5		&				&		&	0	&	2		&	$\pm$	0.1	\\
31		&	35		&		&	$>$	&	3	&	0	&				&		&	4	&	7		&	$\pm$	0.8	&		&	1	&	6		&	$\pm$	0.2	&	5.0	&	5.2	&	50	&		&	8	&	0		&	$\pm$	1.7	&	$<$	&	1	&	6		&				&		&	0	&	6		&	$\pm$	0.2	\\
35		&	39		&	E	&	$>$	&	3	&	0	&				&	$>$	&	1	&	2		&				&	$>$	&	1	&	1		&				&	4.7	&	5.0	&	49	&	$>$	&	5	&	3		&				&		&\multicolumn{2}{}{}&				&		&\multicolumn{2}{}{}&				\\
39		&	47		&	E	&	$>$	&	3	&	4	&				&	$>$	&	5	&	3		&				&	$>$	&	1	&	9		&				&	4.3	&	4.7	&	47	&	$>$	&	8	&	8		&				&		&\multicolumn{2}{}{}&				&		&\multicolumn{2}{}{}&				\\
\hline
\multicolumn{30}{c}{\bf W49N} \\
\hline
-8		&	2		&	E	&	$>$	&	0	&	5	&				&		&\multicolumn{2}{}{}&				&		&\multicolumn{2}{}{}&				&	8.4	&	8.5	&	71	&		&\multicolumn{2}{}{}&				&		&\multicolumn{2}{}{}&				&		&\multicolumn{2}{}{}&				\\
5		&	20		&	E	&	$>$	&	10	&	4	&				&	$>$	&	19	&	6		&				&	$>$	&	6	&	6		&				&	7.5	&	8.2	&	68	&	$>$	&	44	&	8		&				&		&\multicolumn{2}{}{}&				&		&\multicolumn{2}{}{}&				\\
20		&	30		&	4	&		&	5	&	1	&	$\pm$	0.1	&		&	2	&	6		&	$\pm$	0.5	&		&	0	&	7		&	$\pm$	0.1	&	7.1	&	7.5	&	64	&		&	4	&	3		&	$\pm$	1.2	&		&	0	&	5		&	$\pm$	0.1	&		&	0	&	6		&	$\pm$	0.2	\\
30		&	37		&		&	$>$	&	5	&	0	&				&		&	2	&	3		&	$\pm$	0.5	&		&	1	&	3		&	$\pm$	0.1	&	6.8	&	7.1	&	62	&		&	8	&	1		&	$\pm$	1.5	&	$<$	&	0	&	4		&				&		&	0	&	3		&	$\pm$	0.1	\\
37		&	44		&		&	$>$	&	4	&	7	&				&		&	2	&	6		&	$\pm$	0.4	&		&	1	&	1		&	$\pm$	0.1	&	6.6	&	6.8	&	60	&		&	6	&	9		&	$\pm$	1.4	&	$<$	&	0	&	6		&				&		&	0	&	4		&	$\pm$	0.1	\\
44		&	49		&		&	$>$	&	3	&	1	&				&	$<$	&	0	&	4		&				&		&	0	&	7		&	$\pm$	0.1	&	6.4	&	6.6	&	59	&		&	4	&	2		&	$\pm$	1.0	&	$<$	&	0	&	1		&				&	$<$	&	0	&	1		&				\\
49		&	54		&		&	$>$	&	3	&	0	&				&		&	2	&	0		&	$\pm$	0.4	&		&	0	&	6		&	$\pm$	0.1	&	6.3	&	6.4	&	58	&		&	3	&	7		&	$\pm$	0.9	&	$<$	&	0	&	7		&				&		&	0	&	6		&	$\pm$	0.2	\\
54		&	72		&		&	$>$	&	11	&	9	&				&		&	6	&	2		&	$\pm$	0.7	&		&	3	&	6		&	$\pm$	0.2	&	5.7	&	6.3	&	56	&		&	20	&	3		&	$\pm$	3.1	&	$<$	&	0	&	5		&				&		&	0	&	3		&	$\pm$	0.1	\\
72		&	80		&		&		&	0	&	4	&	$\pm$	0.02&	$<$	&	0	&	5		&				&	$<$	&	0	&	1		&				&	5.6	&	5.7	&	54	&	$<$	&	0	&	6		&				&	$<$	&	1	&	2		&				&		&\multicolumn{2}{}{}&				\\
\hline
\multicolumn{30}{c}{\bf W51} \\
\hline
-1		&	3		&		&		&	0	&	2	&	$\pm$	0.01&		&\multicolumn{2}{}{}&				&		&\multicolumn{2}{}{}&				&	8.4	&	8.5	&	71	&		&\multicolumn{2}{}{}&				&		&\multicolumn{2}{}{}&				&		&\multicolumn{2}{}{}&				\\
3		&	11		&		&		&	1	&	3	&	$\pm$	0.02&		&\multicolumn{2}{}{}&				&		&\multicolumn{2}{}{}&				&	8.0	&	8.4	&	69	&		&\multicolumn{2}{}{}&				&		&\multicolumn{2}{}{}&				&		&\multicolumn{2}{}{}&				\\
11		&	16		&		&		&	0	&	3	&	$\pm$	0.01&	$<$	&	0	&	5		&				&		&\multicolumn{2}{}{}&				&	7.8	&	8.0	&	68	&		&\multicolumn{2}{}{}&				&	$<$	&	1	&	9		&				&		&\multicolumn{2}{}{}&				\\
16		&	20		&		&		&	0	&	2	&	$\pm$	0.01&	$<$	&	0	&	5		&				&		&\multicolumn{2}{}{}&				&	7.6	&	7.8	&	66	&		&\multicolumn{2}{}{}&				&	$<$	&	3	&	2		&				&		&\multicolumn{2}{}{}&				\\
20		&	32		&		&		&	0	&	6	&	$\pm$	0.02&	$<$	&	0	&	8		&				&	$<$	&	0	&	2		&				&	7.1	&	7.6	&	64	&	$<$	&	1	&	1		&				&	$<$	&	1	&	3		&				&		&\multicolumn{2}{}{}&				\\
36		&	43		&		&		&	0	&	3	&	$\pm$	0.01&		&	1	&	1		&	$\pm$	0.6	&		&	0	&	4		&	$\pm$	0.1	&	6.8	&	7.0	&	61	&		&	2	&	6		&	$\pm$	1.2	&		&	3	&	9		&	$\pm$	2.3	&		&	0	&	4		&	$\pm$	0.3	\\
43		&	50		&		&		&	2	&	5	&	$\pm$	0.05&		&	3	&	0		&	$\pm$	0.7	&		&	0	&	4		&	$\pm$	0.1	&	6.5	&	6.8	&	60	&		&	2	&	5		&	$\pm$	1.1	&		&	1	&	2		&	$\pm$	0.3	&		&	1	&	2		&	$\pm$	0.6	\\
50		&	60		&	5	&		&	4	&	5	&	$\pm$	0.1	&		&	2	&	7		&	$\pm$	0.7	&		&	2	&	0		&	$\pm$	0.2	&	6.3	&	6.5	&	59	&		&	11	&	9		&	$\pm$	2.3	&		&	0	&	6		&	$\pm$	0.2	&		&	0	&	2		&	$\pm$	0.1	\\
60		&	69		&	E	&	$>$	&	5	&	4	&				&	$>$	&	2	&	1		&				&	$>$	&	2	&	5		&				&	6.0	&	6.3	&	57	&	$>$	&	14	&	4		&				&		&\multicolumn{2}{}{}&				&		&\multicolumn{2}{}{}&				\\
69		&	75		&	E	&	$>$	&	0	&	8	&				&		&\multicolumn{2}{}{}&				&	$>$	&	0	&	6		&				&	5.9	&	6.0	&	56	&	$>$	&	3	&	5		&				&		&\multicolumn{2}{}{}&				&		&\multicolumn{2}{}{}&				
\label{TabIntTau}
\end{longtable}
}

\longtab{4}{
\begin{longtable}{r @{\hspace{0.05cm}} r      @{\hspace{0.30cm}} c @{\hspace{0.45cm}} 
                  r @{\hspace{0.05cm}} r@{.}l @{\hspace{0.45cm}}
                  r @{\hspace{0.05cm}} r@{.}l @{\hspace{0.05cm}} l @{\hspace{0.45cm}}
                  r @{\hspace{0.05cm}} r@{.}l @{\hspace{0.05cm}} l @{\hspace{0.45cm}}
                  r @{\hspace{0.05cm}} r@{.}l @{\hspace{0.05cm}} l @{\hspace{0.45cm}}
                  r @{\hspace{0.05cm}} r@{.}l @{\hspace{0.05cm}} l @{\hspace{0.45cm}}
                  r @{\hspace{0.05cm}} r@{.}l @{\hspace{0.05cm}} l @{\hspace{0.45cm}}
                  r @{\hspace{0.05cm}} r@{.}l @{\hspace{0.05cm}} l @{\hspace{0.45cm}}
                  r}
\caption{HI, H$_2$, and total hydrogen column densities per velocity interval. In the last column, the
total hydrogen column densities are estimated over the entire sight line.} \\
\hline
\multicolumn{1}{c}{$\upsilon_{\rm min}$} & \multicolumn{1}{c}{$\upsilon_{\rm max}$} & $*$ & \multicolumn{3}{l}{$N$(H)$^{a}$}       & \multicolumn{4}{l}{$N({\rm CH})$$^{b}$} & \multicolumn{4}{l}{$N({\rm HF})$$^{c}$} & \multicolumn{4}{l}{$N({\rm H}_2)^{d}$} & \multicolumn{4}{l}{$N({\rm H}_2)^{e}$} & \multicolumn{4}{l}{$N_{\rm H}^{d}$}  & \multicolumn{4}{l}{$N_{\rm H}^{e}$}  & $N_{\rm H}^{f}$  \\
\multicolumn{1}{c}{(\kms)}               & \multicolumn{1}{c}{(\kms)}               &     & \multicolumn{3}{l}{(cm$^{-2}$)}        & \multicolumn{4}{l}{(cm$^{-2}$)}         & \multicolumn{4}{l}{(cm$^{-2}$)}         & \multicolumn{4}{l}{(cm$^{-2}$)}        & \multicolumn{4}{l}{(cm$^{-2}$)}        & \multicolumn{4}{l}{(cm$^{-2}$)}      & \multicolumn{4}{l}{(cm$^{-2}$)}      & (cm$^{-2}$)      \\
                                         &                                          &     & \multicolumn{3}{l}{$\times 10^{20}$}   & \multicolumn{4}{l}{$\times 10^{13}$}    & \multicolumn{4}{l}{$\times 10^{12}$}    & \multicolumn{4}{l}{$\times 10^{20}$}   & \multicolumn{4}{l}{$\times 10^{20}$}   & \multicolumn{4}{l}{$\times 10^{20}$} & \multicolumn{4}{l}{$\times 10^{20}$} & $\times 10^{20}$ \\
\hline
\endfirsthead
\caption{continued.} \\
\hline
\multicolumn{1}{c}{$\upsilon_{\rm min}$} & \multicolumn{1}{c}{$\upsilon_{\rm max}$} & $*$ & \multicolumn{3}{l}{$N$(H)$^{a}$}       & \multicolumn{4}{l}{$N({\rm CH})$$^{b}$} & \multicolumn{4}{l}{$N({\rm HF})$$^{c}$} & \multicolumn{4}{l}{$N({\rm H}_2)^{d}$} & \multicolumn{4}{l}{$N({\rm H}_2)^{e}$} & \multicolumn{4}{l}{$N_{\rm H}^{d}$}  & \multicolumn{4}{l}{$N_{\rm H}^{e}$}  & $N_{\rm H}^{f}$  \\
\multicolumn{1}{c}{(\kms)}               & \multicolumn{1}{c}{(\kms)}               &     & \multicolumn{3}{l}{(cm$^{-2}$)}        & \multicolumn{4}{l}{(cm$^{-2}$)}         & \multicolumn{4}{l}{(cm$^{-2}$)}         & \multicolumn{4}{l}{(cm$^{-2}$)}        & \multicolumn{4}{l}{(cm$^{-2}$)}        & \multicolumn{4}{l}{(cm$^{-2}$)}      & \multicolumn{4}{l}{(cm$^{-2}$)}      & (cm$^{-2}$)      \\
                                         &                                          &     & \multicolumn{3}{l}{$\times 10^{20}$}   & \multicolumn{4}{l}{$\times 10^{13}$}    & \multicolumn{4}{l}{$\times 10^{12}$}    & \multicolumn{4}{l}{$\times 10^{20}$}   & \multicolumn{4}{l}{$\times 10^{20}$}   & \multicolumn{4}{l}{$\times 10^{20}$} & \multicolumn{4}{l}{$\times 10^{20}$} & $\times 10^{20}$ \\
\hline
\endhead
\hline
\\
\multicolumn{31}{p{7.2in}}{ $*$ E = absorption line profile observed in the star-forming region; $T_{\rm ex}$ 
may be underestimated, hence the lower limit on $N({\rm HF})$, and $N({\rm CH})$.}\\
\multicolumn{31}{p{7.2in}}{ $^{a}$ From the observations with the VLA interferometer by \citet{Koo1997}, 
\citet{Fish2003}, \citet{Pandian2008}, \citet{Dwarakanath2004}, and \citet{Lang2010} assuming a spin temperature 
of 100 K.}\\
\multicolumn{31}{p{7.2in}}{ $^{b}$ From (\citealt{Gerin2010}; Gerin et al., priv. comm.).}\\
\multicolumn{31}{p{7.2in}}{ $^{c}$ From (\citealt{Neufeld2010}; \citealt{Sonnentrucker2010}; Neufeld et al. priv. comm.; Sonnentrucker et al., priv. comm.).}\\
\multicolumn{31}{p{7.2in}}{ $^{d}$ Estimate from the observation of the CH absorption lines, a mean value of 
the ${\rm HF}/{\rm CH}$ column density ratio of 0.4 (inferred from columns 5 and 6) and an abundance ratio 
$n({\rm HF})/n({\rm H}_2) = 3.6 \times 10^{-8}$, as predicted by UV-dominated chemical models (see Appendix \ref{AppendTracerHH}).}\\
\multicolumn{31}{p{7.2in}}{ $^{e}$ Estimate from the observation of the HF absorption lines and an abundance ratio
$n({\rm HF})/n({\rm H}_2) = 3.6 \times 10^{-8}$, as predicted by UV-dominated chemical models (see Appendix \ref{AppendTracerHH}).}\\
\multicolumn{31}{p{7.2in}}{ $^{f}$ Estimate of $N_{\rm H}$ integrated over the entire line of sight from models of the extinction at 2 $\mu$m by \citet{Marshall2006}.}
\endfoot
\multicolumn{31}{c}{\bf SgrA*+50} \\
\hline
-168	&	-150		&		&		&	1	&	5		&		&\multicolumn{2}{}{}&					&		&\multicolumn{2}{}{}&					&		&\multicolumn{2}{}{}&					&		&\multicolumn{2}{}{}&					&		&\multicolumn{2}{}{}&					&		&\multicolumn{2}{}{}&				&		\\
-150	&	-123		&		&		&	4	&	3		&		&	21	&	4		&	$\pm$	3.3		&		&	35	&	3		&	$\pm$	1.5		&		&	24	&	4		&	$\pm$	3.8		&		&	10	&	1		&	$\pm$	0.4		&		&	53	&	2		&	$\pm$	7.6		&		&	24	&	5		&	$\pm$	0.9	&		\\
-123	&	-95			&		&		&	1	&	9		&		&	8	&	0		&	$\pm$	1.8		&		&	23	&	5		&	$\pm$	1.1		&		&	9	&	2		&	$\pm$	2.1		&		&	6	&	7		&	$\pm$	0.3		&		&	20	&	3		&	$\pm$	4.2		&		&	15	&	3		&	$\pm$	0.6	&		\\
-95		&	-75			&		&		&	1	&	9		&		&	4	&	9		&	$\pm$	0.3		&		&	6	&	9		&	$\pm$	0.7		&		&	5	&	7		&	$\pm$	0.4		&		&	2	&	0		&	$\pm$	0.2		&		&	13	&	2		&	$\pm$	0.7		&		&	5	&	8		&	$\pm$	0.4	&		\\
-75		&	-45			&		&		&	24	&	7		&		&	17	&	4		&	$\pm$	2.0		&	$>$	&	62	&	3		&					&		&	19	&	8		&	$\pm$	2.2		&	$>$	&	17	&	8		&					&		&	64	&	4		&	$\pm$	4.5		&	$>$	&	60	&	3		&				&		\\
-45		&	-15			&		&		&	14	&	3		&		&	18	&	9		&	$\pm$	1.8		&	$>$	&	63	&	9		&					&		&	21	&	6		&	$\pm$	2.1		&	$>$	&	18	&	3		&					&		&	57	&	5		&	$\pm$	4.2		&	$>$	&	50	&	8		&				&	480	\\
-15		&	0			&		&		&	47	&	7		&		&	22	&	8		&	$\pm$	4.7		&	$>$	&	61	&	2		&					&		&	26	&	0		&	$\pm$	5.4		&	$>$	&	17	&	5		&					&		&	99	&	7		&	$\pm$  10.8		&	$>$	&	82	&	7		&				&		\\
0		&	26			&		&		&	61	&	8		&		&	11	&	2		&	$\pm$	2.3		&	$>$	&	26	&	2		&					&		&	12	&	8		&	$\pm$	2.7		&	$>$	&	7	&	5		&					&		&	87	&	5		&	$\pm$	5.3		&	$>$	&	76	&	8		&				&		\\
26		&	49			&	E	&	$>$	&	39	&	0		&		&\multicolumn{2}{}{}&					&	$>$	&	101	&	3		&					&		&\multicolumn{2}{}{}&					&	$>$	&	28	&	9		&					&		&\multicolumn{2}{}{}&					&	$>$	&	96	&	9		&				&		\\
49		&	67			&	E	&	$>$	&	31	&	0		&		&\multicolumn{2}{}{}&					&	$>$	&	35	&	0		&					&		&\multicolumn{2}{}{}&					&	$>$	&	10	&	0		&					&		&\multicolumn{2}{}{}&					&	$>$	&	51	&	0		&				&		\\
67		&	75			&	E	&	$>$	&	8	&	7		&		&\multicolumn{2}{}{}&					&	$>$	&	0	&	9		&					&		&\multicolumn{2}{}{}&					&	$>$	&	0	&	2		&					&		&\multicolumn{2}{}{}&					&	$>$	&	9	&	2		&				&		\\
\hline
\multicolumn{31}{c}{\bf SgrB2(N)} \\
\hline
-135	&	-100		&		&		&	1	&	7		&		&\multicolumn{2}{}{}&					&		&	26	&	0		&	$\pm$	0.8		&		&\multicolumn{2}{}{}&					&		&	7	&	4		&	$\pm$	0.22	&		&\multicolumn{2}{}{}&					&		&	16	&	6	&	$\pm$	0.4		&		\\
-82		&	-62			&		&		&	5	&	4		&		&\multicolumn{2}{}{}&					&		&	46	&	7		&	$\pm$	1.3		&		&\multicolumn{2}{}{}&					&		&	13	&	4		&	$\pm$	0.38	&		&\multicolumn{2}{}{}&					&		&	32	&	1	&	$\pm$	0.8		&		\\
-62		&	-35			&		&		&	17	&	4		&		&\multicolumn{2}{}{}&					&	$>$	&	111	&	8		&					&		&\multicolumn{2}{}{}&					&	$>$	&	32	&	0		&					&		&\multicolumn{2}{}{}&					&	$>$	&	81	&	3	&					&		\\
-35		&	-15			&		&		&	9	&	2		&		&\multicolumn{2}{}{}&					&		&	82	&	1		&	$\pm$	2.8		&		&\multicolumn{2}{}{}&					&		&	23	&	5		&	$\pm$	0.80	&		&\multicolumn{2}{}{}&					&		&	56	&	1	&	$\pm$	1.6		&	450	\\
-15		&	23			&		&		&	55	&	8		&		&\multicolumn{2}{}{}&					&		&	138	&	7		&	$\pm$	3.6		&		&\multicolumn{2}{}{}&					&		&	39	&	6		&	$\pm$	1.01	&		&\multicolumn{2}{}{}&					&		&	135	&	1	&	$\pm$	2.0		&		\\
23		&	38			&		&		&	1	&	0		&		&\multicolumn{2}{}{}&					&		&	19	&	0		&	$\pm$	0.6		&		&\multicolumn{2}{}{}&					&		&	5	&	4		&	$\pm$	0.16	&		&\multicolumn{2}{}{}&					&		&	11	&	9	&	$\pm$	0.3		&		\\
38		&	74			&	E	&	$>$	&	69	&	7		&		&\multicolumn{2}{}{}&					&	$>$	&	143	&	0		&					&		&\multicolumn{2}{}{}&					&	$>$	&	40	&	9		&					&		&\multicolumn{2}{}{}&					&	$>$	&	151	&	4	&					&		\\
74		&	100			&	E	&	$>$	&	14	&	2		&		&\multicolumn{2}{}{}&					&	$>$	&	74	&	6		&					&		&\multicolumn{2}{}{}&					&	$>$	&	21	&	3		&					&		&\multicolumn{2}{}{}&					&	$>$	&	56	&	8	&					&		\\
\hline
\multicolumn{31}{c}{\bf DR21(OH)} \\
\hline
-17		&	-9			&	E	&	$>$	&	36	&	3		&		&\multicolumn{2}{}{}&					&	$>$	&	0	&	8		&					&		&\multicolumn{2}{}{}&					&	$>$	&	0	&	2		&					&		&\multicolumn{2}{}{}&					&	$>$	&	36	&	8		&				&		\\
-8		&	1			&	E	&	$>$	&	40	&	8		&	$>$	&	12	&	8		&					&	$>$	&	30	&	9		&					&	$>$	&	14	&	6		&					&	$>$	&	8	&	8		&					&	$>$	&	70	&	1		&					&	$>$	&	58	&	5		&				&		\\
1		&	7			&		&	$>$	&	24	&	9		&		&	2	&	0		&	$\pm$	0.9		&		&	7	&	0		&	$\pm$	0.2		&		&	2	&	3		&	$\pm$	1.0		&		&	2	&	0		&	$\pm$	0.05	&	$>$	&	29	&	4		&					&	$>$	&	28	&	9		&				&	160	\\
7		&	15			&		&	$>$	&	38	&	5		&		&	11	&	2		&	$\pm$	0.3		&	$>$	&	27	&	4		&					&		&	12	&	9		&	$\pm$	0.3		&	$>$	&	7	&	8		&					&	$>$	&	64	&	2		&					&	$>$	&	54	&	1		&				&		\\
18		&	26			&		&	$>$	&	36	&	3		&	$<$	&	0	&	1		&					&		&	0	&	3		&	$\pm$	0.1		&	$<$	&	0	&	1		&					&		&	0	&	1		&	$\pm$	0.03	&		&\multicolumn{2}{}{}&					&	$>$	&	36	&	5		&				&		\\
\hline
\multicolumn{31}{c}{\bf G34.3+0.1} \\
\hline
-4		&	5			&		&		&	5	&	4		&		&	0	&	4		&	$\pm$	0.3		&		&	1	&	1		&	$\pm$	0.1		&		&	0	&	4		&	$\pm$	0.3		&		&	0	&	3		&	$\pm$	0.04	&		&	6	&	2		&	$\pm$	0.7		&		&	6	&	0		&	$\pm$	0.1	&		\\
5		&	21			&		&		&	20	&	1		&		&	4	&	1		&	$\pm$	0.5		&	$>$	&	19	&	8		&					&		&	4	&	7		&	$\pm$	0.6		&	$>$	&	5	&	7		&					&		&	29	&	4		&	$\pm$	1.2		&	$>$	&	31	&	4		&				&		\\
21		&	34			&		&		&	44	&	2		&		&	4	&	7		&	$\pm$	0.5		&	$>$	&	12	&	9		&					&		&	5	&	3		&	$\pm$	0.6		&	$>$	&	3	&	7		&					&		&	54	&	9		&	$\pm$	1.1		&	$>$	&	51	&	6		&				&	230	\\
36		&	44			&		&		&	15	&	5		&	$<$	&	0	&	2		&					&		&	2	&	7		&	$\pm$	0.1		&	$<$	&	0	&	2		&					&		&	0	&	8		&	$\pm$	0.04	&	$<$	&	15	&	9		&					&		&	17	&	1		&	$\pm$	0.1	&		\\
44		&	58			&	E	&	$>$	&	52	&	8		&	$>$	&	9	&	1		&					&	$>$	&	62	&	9		&					&	$>$	&	10	&	4		&					&	$>$	&	18	&	0		&					&	$>$	&	73	&	7		&					&	$>$	&	88	&	7		&				&		\\
58		&	66			&	E	&	$>$	&	56	&	1		&	$>$	&	15	&	4		&					&	$>$	&	37	&	8		&					&	$>$	&	17	&	6		&					&	$>$	&	10	&	8		&					&	$>$	&	91	&	4		&					&	$>$	&	77	&	7		&				&		\\
\hline
\multicolumn{31}{c}{\bf W31C} \\
\hline
3		&	13			&		&		&	18	&	1		&		&	0	&	9		&	$\pm$	0.4		&		&	9	&	4		&	$\pm$	0.2		&		&	1	&	0		&	$\pm$	0.4		&		&	2	&	7		&	$\pm$	0.1		&		&	20	&	1		&	$\pm$	0.9		&		&	23	&	4		&	$\pm$	0.1	&		\\
13		&	25			&		&		&	35	&	3		&		&	21	&	8		&	$\pm$	0.7		&	$>$	&	59	&	2		&					&		&	24	&	9		&	$\pm$	0.7		&	$>$	&	16	&	9		&					&		&	85	&	1		&	$\pm$	1.5		&	$>$	&	69	&	1		&				&		\\
25		&	31			&		&		&	26	&	1		&		&	20	&	1		&	$\pm$	1.3		&	$>$	&	36	&	2		&					&		&	23	&	0		&	$\pm$	1.4		&	$>$	&	10	&	4		&					&		&	72	&	1		&	$\pm$	2.9		&	$>$	&	46	&	8		&				&		\\
31		&	36			&		&		&	18	&	5		&		&	9	&	2		&	$\pm$	0.4		&	$>$	&	25	&	3		&					&		&	10	&	5		&	$\pm$	0.4		&	$>$	&	7	&	2		&					&		&	39	&	4		&	$\pm$	0.8		&	$>$	&	33	&	0		&				&	220	\\
36		&	44			&		&		&	36	&	9		&		&	17	&	6		&	$\pm$	0.4		&	$>$	&	41	&	0		&					&		&	20	&	1		&	$\pm$	0.4		&	$>$	&	11	&	7		&					&		&	77	&	1		&	$\pm$	0.9		&	$>$	&	60	&	3		&				&		\\
44		&	51			&		&		&	1	&	8		&		&	1	&	7		&	$\pm$	0.3		&	$>$	&	8	&	7		&					&		&	1	&	9		&	$\pm$	0.4		&	$>$	&	2	&	5		&					&		&	5	&	7		&	$\pm$	0.7		&	$>$	&	6	&	8		&				&		\\
51		&	61			&		&		&	2	&	5		&		&	0	&	3		&	$\pm$	0.2		&	$<$	&	0	&	2		&					&		&	0	&	4		&	$\pm$	0.3		&	$<$	&	0	&	05		&					&		&	3	&	3		&	$\pm$	0.5		&	$<$	&	2	&	6		&				&		\\
\hline
\multicolumn{31}{c}{\bf W33A} \\
\hline
14		&	18			&		&		&	1	&	7		&		&	0	&	4		&	$\pm$	0.1		&		&	0	&	3		&	$\pm$	0.2		&		&	0	&	4		&	$\pm$	0.2		&		&	0	&	1		&	$\pm$	0.04	&		&	2	&	5		&	$\pm$	0.3		&		&	1	&	9		&	$\pm$	0.1	&		\\
18		&	25			&		&		&	18	&	0		&		&	2	&	1		&	$\pm$	0.7		&		&	5	&	8		&	$\pm$	0.4		&		&	2	&	4		&	$\pm$	0.8		&		&	1	&	7		&	$\pm$	0.1		&		&	22	&	9		&	$\pm$	1.5		&		&	21	&	3		&	$\pm$	0.2	&		\\
25		&	31			&		&		&	22	&	2		&		&	7	&	0		&	$\pm$	0.4		&	$>$	&	21	&	1		&					&		&	8	&	0		&	$\pm$	0.4		&	$>$	&	6	&	0		&					&		&	38	&	3		&	$\pm$	0.9		&	$>$	&	34	&	3		&				&	100	\\
31		&	35			&		&		&	17	&	9		&		&	8	&	1		&	$\pm$	1.0		&	$>$	&	20	&	9		&					&		&	9	&	2		&	$\pm$	1.2		&	$>$	&	6	&	0		&					&		&	36	&	3		&	$\pm$	2.3		&	$>$	&	29	&	8		&				&		\\
35		&	39			&	E	&	$>$	&	7	&	7		&	$>$	&	9	&	1		&					&	$>$	&	20	&	9		&					&	$>$	&	10	&	4		&					&	$>$	&	6	&	0		&					&	$>$	&	28	&	5		&					&	$>$	&	19	&	6		&				&		\\
39		&	47			&	E	&	$>$	&	9	&	8		&	$>$	&	5	&	5		&					&	$>$	&	17	&	9		&					&	$>$	&	6	&	3		&					&	$>$	&	5	&	1		&					&	$>$	&	22	&	4		&					&	$>$	&	20	&	1		&				&		\\
\hline
\multicolumn{31}{c}{\bf W49N} \\
\hline
5		&	20			&	E	&	$>$	&	89	&	1		&	$>$	&	7	&	2		&					&	$>$	&	75	&	6		&					&	$>$	&	8	&	2		&					&	$>$	&	21	&	6		&					&	$>$	&	105	&	5		&					&	$>$	&	132	&	3		&				&		\\
20		&	30			&		&		&	19	&	3		&		&	1	&	6		&	$\pm$	0.4		&		&	6	&	1		&	$\pm$	0.1		&		&	1	&	8		&	$\pm$	0.5		&		&	1	&	7		&	$\pm$	0.04	&		&	23	&	0		&	$\pm$	1.0		&		&	22	&	8		&	$\pm$	0.1	&		\\
30		&	37			&		&		&	19	&	6		&		&	5	&	2		&	$\pm$	1.7		&	$>$	&	23	&	7		&					&		&	5	&	9		&	$\pm$	1.9		&	$>$	&	6	&	8		&					&		&	31	&	4		&	$\pm$	3.8		&	$>$	&	33	&	1		&				&		\\
37		&	44			&		&		&	30	&	9		&		&	5	&	8		&	$\pm$	1.7		&	$>$	&	24	&	4		&					&		&	6	&	6		&	$\pm$	2.0		&	$>$	&	7	&	0		&					&		&	44	&	1		&	$\pm$	4.0		&	$>$	&	44	&	8		&				&	460	\\
44		&	49			&		&		&	9	&	0		&		&	1	&	0		&	$\pm$	0.3		&		&	3	&	3		&	$\pm$	0.1		&		&	1	&	1		&	$\pm$	1.4		&		&	0	&	9		&	$\pm$	0.03	&		&	11	&	2		&	$\pm$	2.8		&		&	10	&	8		&	$\pm$	0.1	&		\\
49		&	54			&		&		&	11	&	5		&		&	3	&	2		&	$\pm$	1.4		&		&	11	&	3		&	$\pm$	0.2		&		&	3	&	7		&	$\pm$	1.7		&		&	3	&	2		&	$\pm$	0.1		&		&	18	&	8		&	$\pm$	3.3		&		&	18	&	0		&	$\pm$	0.1	&		\\
54		&	72			&		&		&	52	&	6		&		&	19	&	8		&	$\pm$	3.0		&	$>$	&	56	&	0		&					&		&	22	&	6		&	$\pm$	3.4		&	$>$	&	16	&	0		&					&		&	97	&	8		&	$\pm$	6.9		&	$>$	&	84	&	6		&				&		\\
72		&	80			&		&		&	21	&	7		&		&	0	&	8		&	$\pm$	0.4		&		&	0	&	4		&	$\pm$	0.1		&		&	0	&	9		&	$\pm$	0.4		&		&	0	&	1		&	$\pm$	0.03	&		&	23	&	5		&	$\pm$	0.8		&		&	21	&	9		&	$\pm$	0.1	&		\\
\hline
\multicolumn{31}{c}{\bf W51} \\
\hline
3		&	11			&		&		&	14	&	8		&		&	3	&	1		&	$\pm$	1.0		&		&	15	&	2		&	$\pm$	0.4		&		&	3	&	5		&	$\pm$	1.1		&		&	4	&	3		&	$\pm$	0.1		&		&\multicolumn{2}{}{}&					&		&	23	&	5		&	$\pm$	0.2	&		\\
11		&	16			&		&		&	4	&	3		&		&	0	&	7		&	$\pm$	0.7		&		&	2	&	1		&	$\pm$	0.1		&		&	0	&	8		&	$\pm$	0.8		&		&	0	&	6		&	$\pm$	0.03	&		&\multicolumn{2}{}{}&					&		&	5	&	5		&	$\pm$	0.1	&		\\
16		&	20			&		&		&	2	&	3		&		&\multicolumn{2}{}{}&					&		&	0	&	4		&	$\pm$	0.1		&		&\multicolumn{2}{}{}&					&		&	0	&	1		&	$\pm$	0.03	&		&\multicolumn{2}{}{}&					&		&	2	&	5		&	$\pm$	0.0	&		\\
20		&	32			&		&		&	8	&	6		&		&\multicolumn{2}{}{}&					&		&	1	&	4		&	$\pm$	0.1		&		&\multicolumn{2}{}{}&					&		&	0	&	4		&	$\pm$	0.04	&		&\multicolumn{2}{}{}&					&		&	9	&	4		&	$\pm$	0.1	&		\\
36		&	43			&		&		&	2	&	9		&		&\multicolumn{2}{}{}&					&		&	0	&	1		&	$\pm$	0.1		&		&\multicolumn{2}{}{}&					&		&	0	&	04		&	$\pm$	0.04	&		&\multicolumn{2}{}{}&					&		&	3	&	0		&	$\pm$	0.1	&	250	\\
43		&	50			&		&		&	18	&	7		&		&	2	&	3		&	$\pm$	0.7		&	$>$	&	15	&	5		&					&		&	2	&	6		&	$\pm$	0.8		&	$>$	&	4	&	4		&					&		&\multicolumn{2}{}{}&					&	$>$	&	27	&	6		&				&		\\
50		&	60			&		&		&	50	&	9		&		&\multicolumn{2}{}{}&					&	$>$	&	41	&	5		&					&		&\multicolumn{2}{}{}&					&	$>$	&	11	&	8		&					&		&\multicolumn{2}{}{}&					&	$>$	&	74	&	6		&				&		\\
60		&	69			&	E	&	$>$	&	64	&	2		&	$>$	&	12	&	5		&					&	$>$	&	53	&	1		&					&	$>$	&	14	&	3		&					&	$>$	&	15	&	2		&					&	$>$	&	92	&	8		&					&	$>$	&	94	&	5		&				&		\\
69		&	75			&	E	&	$>$	&	10	&	4		&	$>$	&	0	&	8		&					&	$>$	&	12	&	0		&					&	$>$	&	0	&	9		&					&	$>$	&	3	&	4		&					&	$>$	&	12	&	2		&					&	$>$	&	17	&	2		&				&		
\label{TabHI}
\end{longtable}
}

\appendix

\section{Gaussian decomposition and calculation of column densities} \label{AppendGauss}

\begin{figure}[!h]
\begin{center}
\includegraphics[width=7.8cm,angle=0]{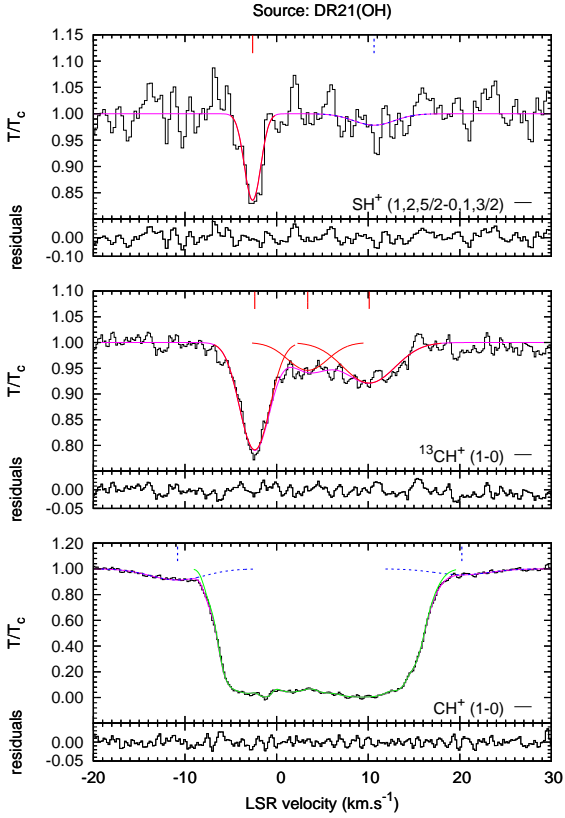}
\caption{Observational data (black lines) compared to the multi-Gaussian decomposition (purple lines) of 
the \SHp\ $(1,2,5/2 \gets 0,1,3/2)$ (top), \thCHp $(1 \gets 0)$ (middle), and \CHp\ $(1 \gets 0)$ (bottom) 
absorption spectra observed towards DR21(OH). The solid red lines and the dashed blue lines correspond
to the confirmed (C) and uncertain (U) Gaussian components, respectively (see main text). The solid green 
lines are the empirical modelling of the \CHp saturated line profiles.}
\label{FigGaussDR21}
\end{center}
\end{figure}

\begin{figure}[!h]
\begin{center}
\includegraphics[width=7.8cm,angle=0]{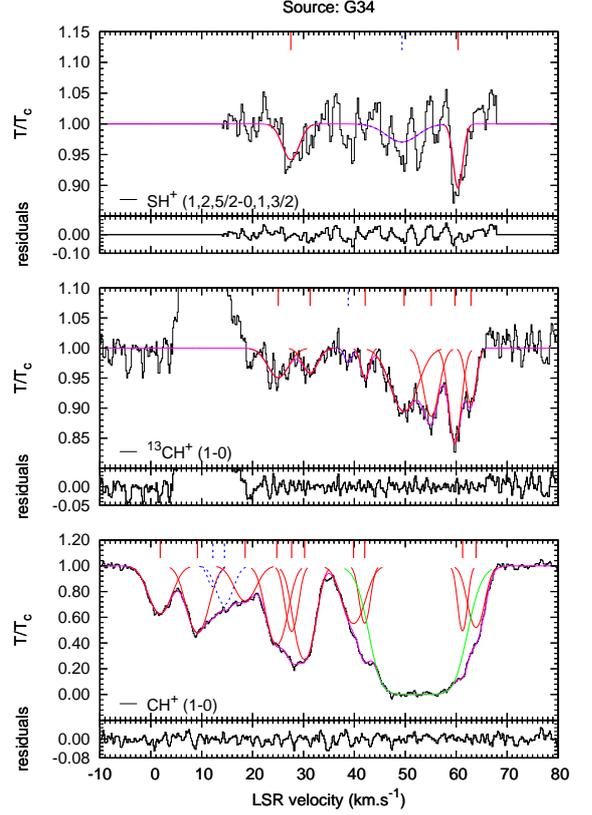}
\caption{Same as Fig. \ref{FigGaussDR21} for the observations towards G34.3+0.1.}
\label{FigGaussG34}
\end{center}
\end{figure}

\begin{figure}[!h]
\begin{center}
\includegraphics[width=7.8cm,angle=0]{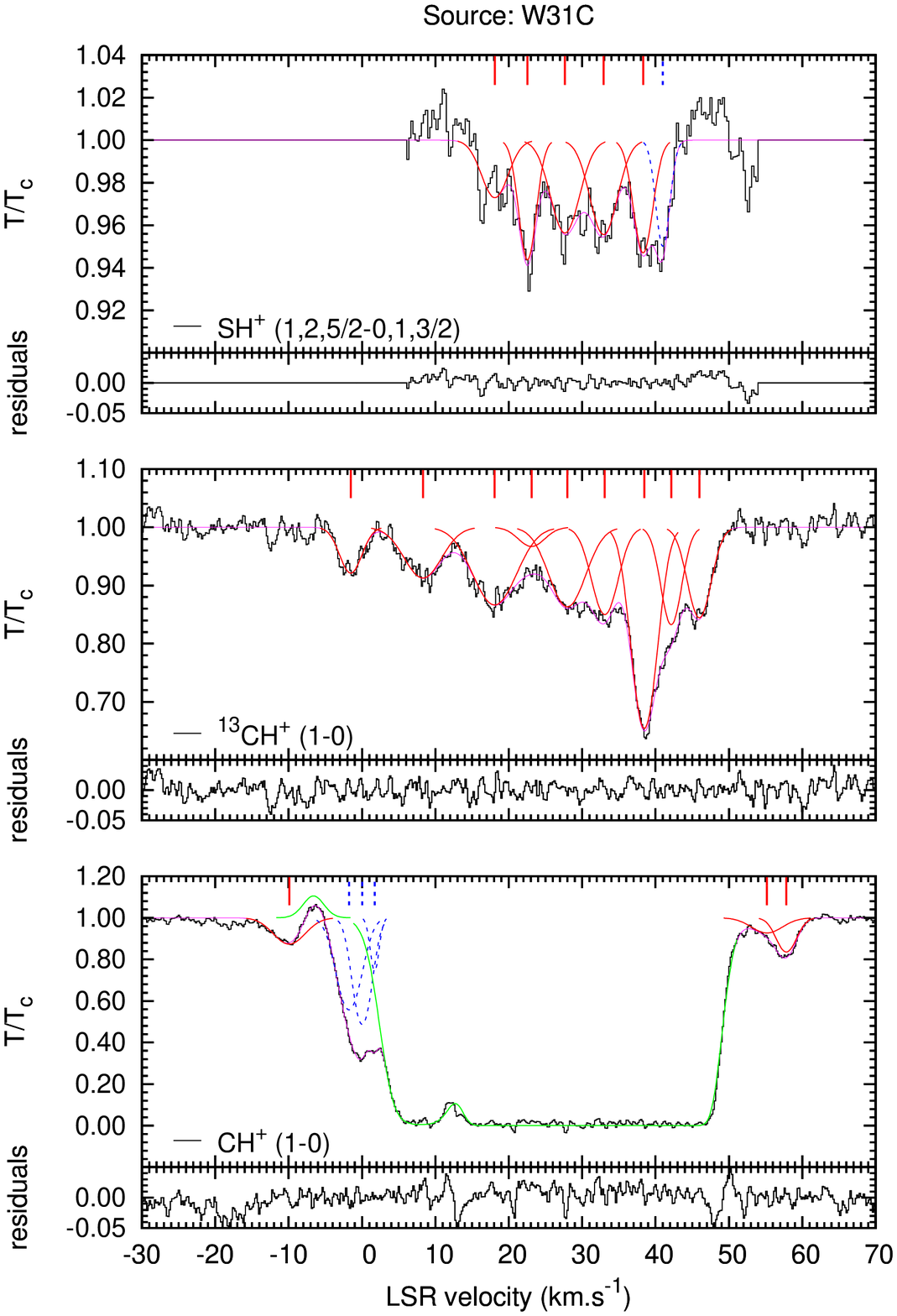}
\caption{Same as Fig. \ref{FigGaussDR21} for the observations towards W31C.}
\label{FigGaussW31C}
\end{center}
\end{figure}

\begin{figure}[!h]
\begin{center}
\includegraphics[width=7.8cm,angle=0]{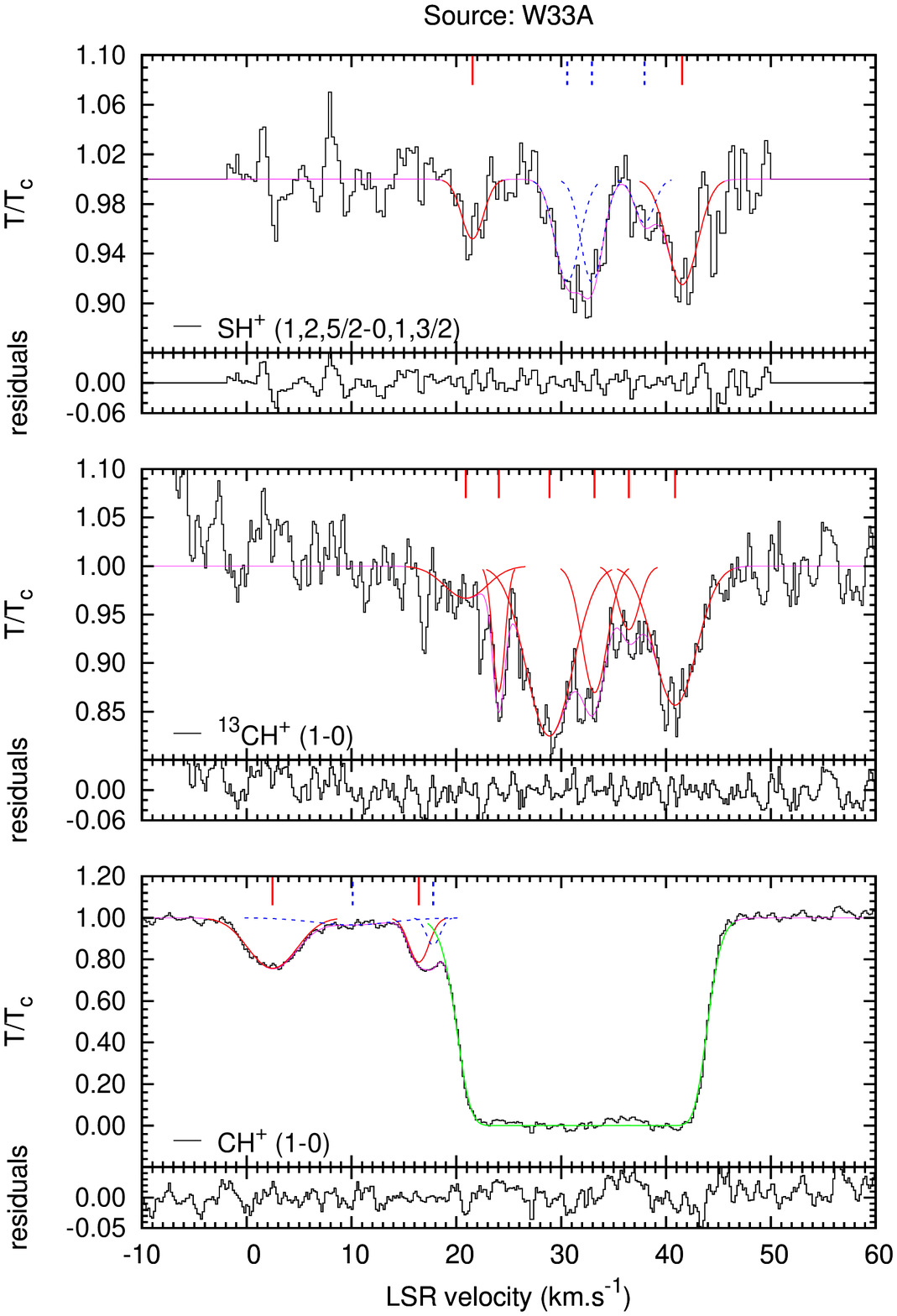}
\caption{Same as Fig. \ref{FigGaussDR21} for the observations towards W33A.}
\label{FigGaussW33A}
\end{center}
\end{figure}

\begin{figure}[!h]
\begin{center}
\includegraphics[width=7.8cm,angle=0]{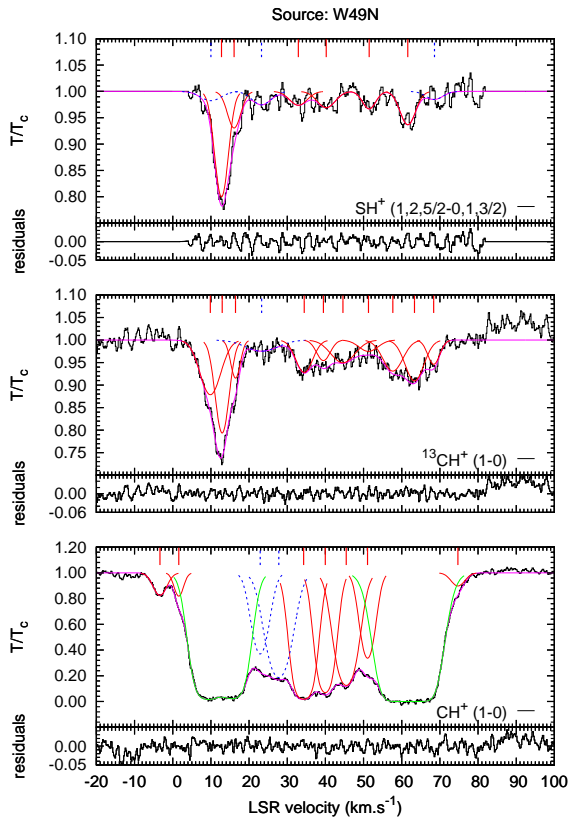}
\caption{Same as Fig. \ref{FigGaussDR21} for the observations towards W49N.}
\label{FigGaussW49N}
\end{center}
\end{figure}

\begin{figure}[!h]
\begin{center}
\includegraphics[width=7.8cm,angle=0]{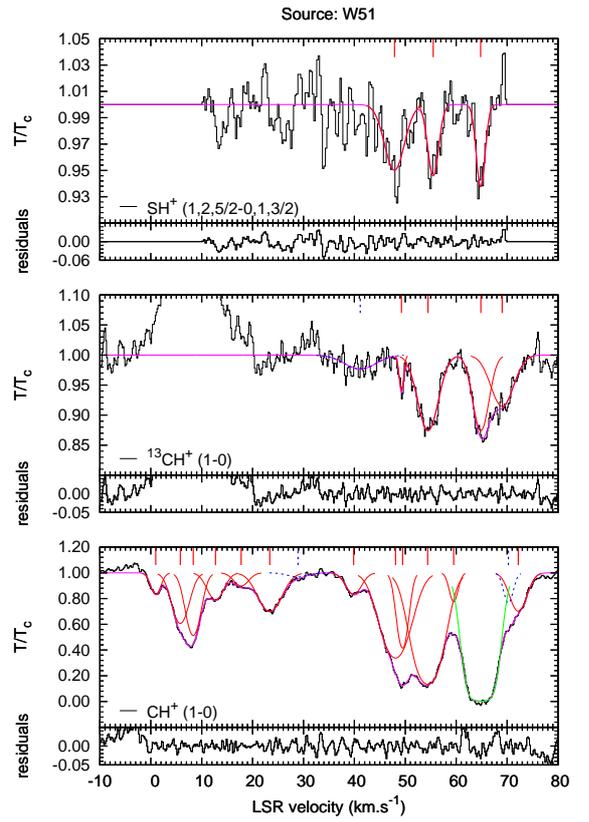}
\caption{Same as Fig. \ref{FigGaussDR21} for the observations towards W51.}
\label{FigGaussW51}
\end{center}
\end{figure}

\begin{figure}[!h]
\begin{center}
\includegraphics[width=7.8cm,angle=0]{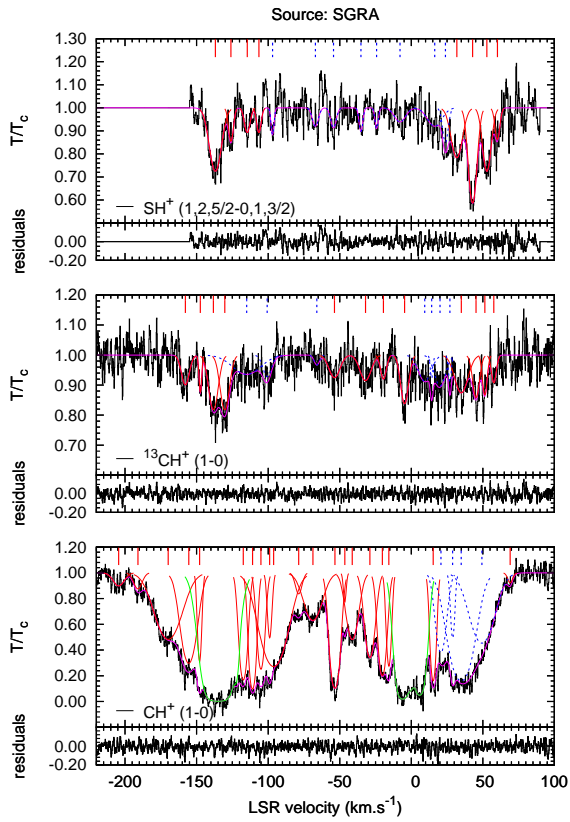}
\caption{Same as Fig. \ref{FigGaussDR21} for the observations towards SgrA*+50.}
\label{FigGaussSGRA}
\end{center}
\end{figure}

\begin{figure}[!h]
\begin{center}
\includegraphics[width=7.8cm,angle=0]{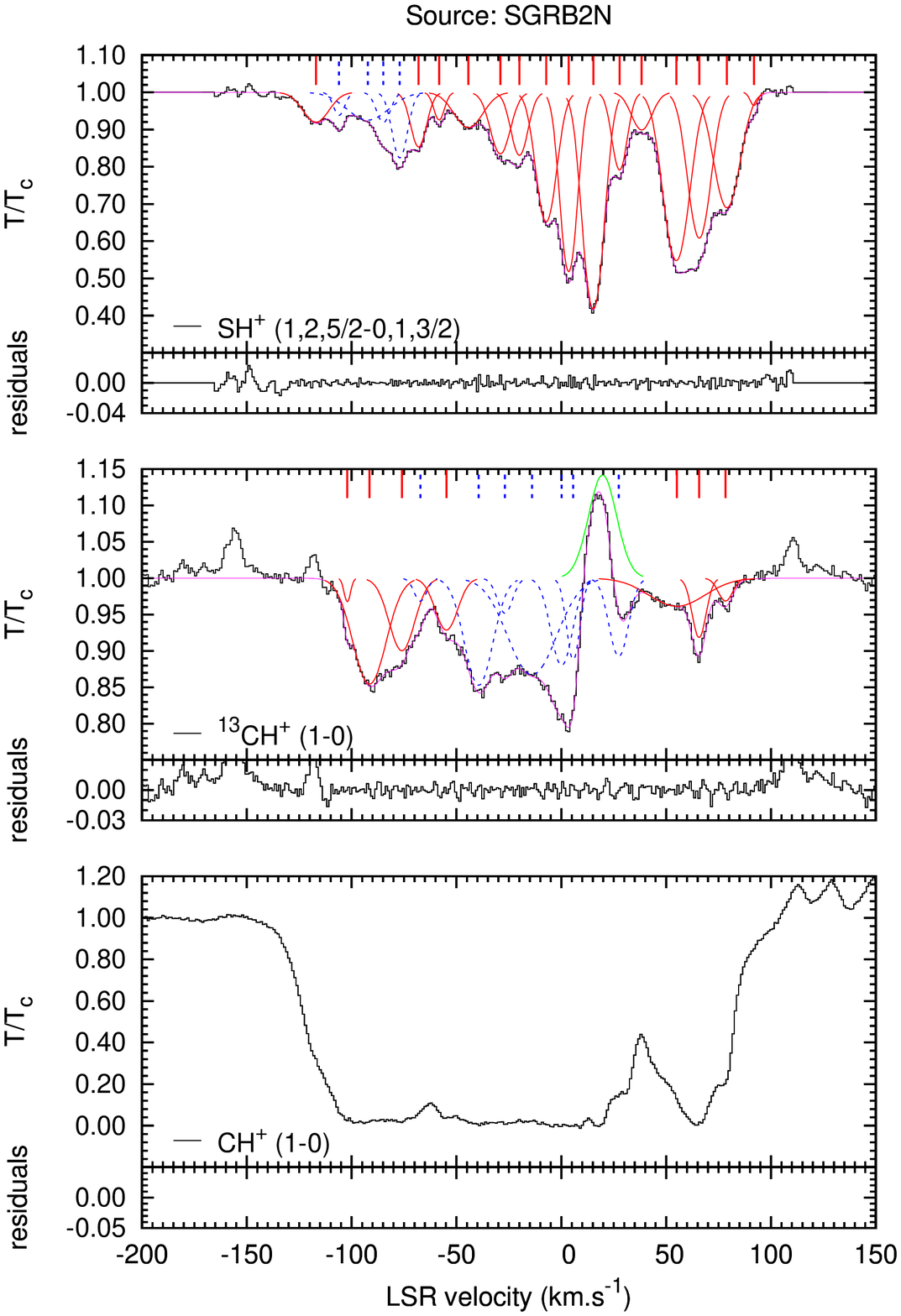}
\caption{Same as Fig. \ref{FigGaussDR21} for the observations towards SgrB2(N).}
\label{FigGaussSGRB2N}
\end{center}
\end{figure}

The results of the Gaussian decomposition procedure applied to the spectra are 
given in Tables \ref{TabFit}. In turn, Figs. \ref{FigGaussDR21} - \ref{FigGaussSGRB2N} 
display the comparison between the result of the fit and the original data, 
along with the associated residuals.
The uncertainties given in the Table \ref{TabFit}
are the formal 1-$\sigma$ errors derived from the diagonal elements of the 
covariance matrix and do not take into account the systematic errors introduced 
by (1) the finite velocity resolution, (2) the uncertainty on the continuum
level, and (3) the error introduced by the decomposition procedure, which depends 
on its convergence criteria.

According to \citet{Godard2010} (Appendix B), the finite velocity resolution 
of the spectra introduces an error on the linewidth determination smaller than 
8 \% for the \CHp\ and \thCHp\ components, and smaller than 14 \% for the \SHp\ 
components, if $\Delta \upsilon > 1.3$ \kms\ (the smallest observed linewidth).
Furthermore, a relative uncertainty $\epsilon$ on the continuum temperature 
leads to an error $\delta \tau \sim \epsilon - {\rm ln}(1 + \epsilon e^{\tau})$
on the opacity that ranges between 6\% and 30 \% when $\tau$ varies 
between 0.03 and 2.7 (the lowest and highest observed central opacities) 
and using $\epsilon = 10$ \% (value corresponding to the error on the beam
efficiency and the side band ratio, \citealt{Roelfsema2012}).

Unlike the other numerical fitting methods\footnote{such as the
  grid-Powell method, the Montecarlo-Powell search or the simulated
  annealing method} that are devised to find the global minimum of
$\chi^{2}$, the Levenberg Marquardt algorithm is a so-called single-shot 
method: used when the initial conditions are easily fixed, this
algorithm takes advantage of the information on the $\chi^{2}$
derivatives with respect to each parameter to converge on the nearest
local minimum. However, since the minimum only corresponds to a
statistical estimate of the fit parameters, the code can converge on
different solutions close to each other depending on the initial
condition, the $\chi^{2}$ valley topography, and the criteria to stop
the optimization. To estimate the associated error, we
analysed the dispersion of the best-fit solutions when varying the
initial conditions: for each spectrum, we assumed an uncertainty of 2
\kms\ on the initial guess of the positions and the linewidths of the
Gaussians. The resulting 4200 fits are found to bracket the optimum
solutions given in Table \ref{TabFit} with a standard deviation of 
10 \% on the central optical depths and the Gaussian linewidths. 

When combined, the robustness of the decomposition procedure, the finite 
resolution, and the uncertainty on the continuum temperature correspond 
to maximal errors on the calculation of $\Delta \upsilon$ 
and $\tau$ of 20 \% and 30 \%, respectively.

\longtab{1}{
\begin{longtable}{r @{\hspace{0.1cm}} c @{\hspace{0.1cm}} c @{\hspace{0.1cm}} c @{\hspace{0.3cm}} 
                  r @{\hspace{0.1cm}} c @{\hspace{0.1cm}} c @{\hspace{0.1cm}} c @{\hspace{0.3cm}}
                  r @{\hspace{0.1cm}} c @{\hspace{0.1cm}} c @{\hspace{0.1cm}} c}
\caption{\CHp\ $(1 \gets 0)$, \thCHp\ $(1 \gets 0)$, and \SHp\ $(1 \gets 0)$ absorption line analysis results.} \\
\multicolumn{4}{c}{\CHp\ $(1 \gets 0)$} & \multicolumn{4}{c}{\thCHp\ $(1 \gets 0)$} & \multicolumn{4}{c}{\SHp\ $(1,2,5/2 \gets 0,1,3/2)$} \\
\hline
\multicolumn{1}{c}{$\upsilon_0$} & $\Delta \upsilon \pm  \sigma(\Delta \upsilon)$ & $\tau_0 \pm \sigma(\tau_0)$ & V$^{a}$ & \multicolumn{1}{c}{$\upsilon_0$} & $\Delta \upsilon \pm  \sigma(\Delta \upsilon)$ & $\tau_0 \pm \sigma(\tau_0)$ & V$^{a}$ & \multicolumn{1}{c}{$\upsilon_0$} & $\Delta \upsilon \pm  \sigma(\Delta \upsilon)$ & $\tau_0 \pm \sigma(\tau_0)$ & V$^{a}$ \\
\multicolumn{1}{c}{(km s$^{-1}$)}& (km s$^{-1}$)                                  &                             &         & \multicolumn{1}{c}{(km s$^{-1}$)}& (km s$^{-1}$)                                  &                             &         & \multicolumn{1}{c}{(km s$^{-1}$)}& (km s$^{-1}$)                                  &                             &         \\
\hline
\endfirsthead
\caption{continued.} \\
\multicolumn{4}{c}{\CHp\ $(1 \gets 0)$} & \multicolumn{4}{c}{\thCHp\ $(1 \gets 0)$} & \multicolumn{4}{c}{\SHp\ $(1,2,5/2 \gets 0,1,3/2)$} \\
\hline
$\upsilon_0$ & $\Delta \upsilon \pm  \sigma(\Delta \upsilon)$ & $\tau_0 \pm \sigma(\tau_0)$ & V$^{a}$ & $\upsilon_0$ & $\Delta \upsilon \pm  \sigma(\Delta \upsilon)$ & $\tau_0 \pm \sigma(\tau_0)$ & V$^{a}$ & $\upsilon_0$ & $\Delta \upsilon \pm  \sigma(\Delta \upsilon)$ & $\tau_0 \pm \sigma(\tau_0)$ & V$^{a}$ \\
(km s$^{-1}$)& (km s$^{-1}$)                                  &                             &         & (km s$^{-1}$)& (km s$^{-1}$)                                  &                             &         & (km s$^{-1}$)& (km s$^{-1}$)                                  &                             &         \\
\hline
\endhead
\hline
\\
\multicolumn{12}{l}{$^{a}$ Validity of the Gaussian detection: C = confirmed if $\Delta \upsilon > 3 \sigma(\Delta \upsilon)$ and $\tau_0 > 2.5 \sigma(\tau_0)$; 
U = uncertain otherwise} \\
\endfoot
\multicolumn{12}{c}{{\bf DR21(OH)}} \\
\hline
-10.8	$\pm$	0.3	&	6.9	$\pm$	0.5	&	9.1 (-2)	$\pm$	4 (-3)	&	U	&	-2.4	$\pm$	0.1	&	3.7	$\pm$	0.1	&	2.3 (-1)	$\pm$	6 (-3)	&	C	&	-2.6	$\pm$	0.1	&	2.0	$\pm$	0.4	&	1.8 (-1)	$\pm$	3 (-2)	&	C	\\
20.2	$\pm$	1.8	&	7.0	$\pm$	2.4	&	4.7 (-2)	$\pm$	1 (-2)	&	U	&	3.4		$\pm$	0.2	&	5.1	$\pm$	0.4	&	5.6 (-2)	$\pm$	4 (-3)	&	C	&	10.6	$\pm$	1.2	&	5.2	$\pm$	2.7	&	2.2 (-2)	$\pm$	2 (-2)	&	U	\\
					&					&								&		&	10.1	$\pm$	0.4	&	6.6	$\pm$	0.7	&	8.2 (-2)	$\pm$	4 (-3)	&	C	&						&					&								&		\\
\hline
\multicolumn{12}{c}{{\bf G34}} \\
\hline
1.8		$\pm$	0.1	&	5.0	$\pm$	0.1	&	4.6 (-1)	$\pm$	9 (-3)	&	C	&	25.0	$\pm$	0.3	&	4.8	$\pm$	0.7	&	5.1 (-2)	$\pm$	6 (-3)	&	C	&	27.5	$\pm$	0.3	&	3.8	$\pm$	0.6	&	6.2 (-2)	$\pm$	2 (-2)	&	C	\\
9.2		$\pm$	0.1	&	4.6	$\pm$	0.2	&	7.3 (-1)	$\pm$	1 (-2)	&	C	&	31.3	$\pm$	0.3	&	3.5	$\pm$	0.7	&	4.2 (-2)	$\pm$	7 (-3)	&	C	&	49.4	$\pm$	0.8	&	7.3	$\pm$	2.4	&	3.3 (-2)	$\pm$	2 (-2)	&	U	\\
12.2	$\pm$	0.3	&	2.2	$\pm$	0.9	&	1.8 (-1)	$\pm$	2 (-1)	&	U	&	38.8	$\pm$	0.4	&	1.7	$\pm$	0.8	&	2.4 (-2)	$\pm$	1 (-2)	&	U	&	60.4	$\pm$	0.1	&	2.2	$\pm$	0.3	&	1.1 (-1)	$\pm$	3 (-2)	&	C	\\
14.5	$\pm$	0.5	&	3.7	$\pm$	1.6	&	3.6 (-1)	$\pm$	9 (-2)	&	U	&	42.1	$\pm$	0.2	&	1.9	$\pm$	0.4	&	5.2 (-2)	$\pm$	9 (-3)	&	C	&						&					&								&		\\
18.5	$\pm$	0.6	&	4.7	$\pm$	0.9	&	3.2 (-1)	$\pm$	4 (-2)	&	C	&	49.8	$\pm$	0.2	&	6.2	$\pm$	0.4	&	1.1 (-1)	$\pm$	6 (-3)	&	C	&						&					&								&		\\
24.8	$\pm$	0.2	&	4.2	$\pm$	0.5	&	9.5 (-1)	$\pm$	5 (-2)	&	C	&	55.1	$\pm$	0.1	&	3.5	$\pm$	0.3	&	1.2 (-1)	$\pm$	8 (-3)	&	C	&						&					&								&		\\
27.7	$\pm$	0.2	&	2.7	$\pm$	0.4	&	7.1 (-1)	$\pm$	2 (-1)	&	C	&	59.8	$\pm$	0.1	&	2.8	$\pm$	0.2	&	1.7 (-1)	$\pm$	9 (-3)	&	C	&						&					&								&		\\
30.2	$\pm$	0.2	&	4.0	$\pm$	0.2	&	1.3 (+0)	$\pm$	8 (-2)	&	C	&	62.9	$\pm$	0.1	&	2.3	$\pm$	0.2	&	9.9 (-2)	$\pm$	9 (-3)	&	C	&						&					&								&		\\
39.8	$\pm$	0.3	&	4.8	$\pm$	0.3	&	6.0 (-1)	$\pm$	4 (-2)	&	C	&						&					&								&		&						&					&								&		\\
42.0	$\pm$	0.0	&	2.3	$\pm$	0.2	&	5.9 (-1)	$\pm$	8 (-2)	&	C	&						&					&								&		&						&					&								&		\\
61.3	$\pm$	0.0	&	1.9	$\pm$	0.1	&	7.1 (-1)	$\pm$	3 (-2)	&	C	&						&					&								&		&						&					&								&		\\
63.9	$\pm$	0.1	&	3.5	$\pm$	0.1	&	6.5 (-1)	$\pm$	1 (-2)	&	C	&						&					&								&		&						&					&								&		\\
\hline
\multicolumn{12}{c}{{\bf W31C}} \\
\hline
-9.9	$\pm$	0.1	&	5.0	$\pm$	0.2	&	1.4 (-1)	$\pm$	4 (-3)	&	C	&	-1.5	$\pm$	0.1	&	3.5	$\pm$	0.2	&	8.2 (-2)	$\pm$	5 (-3)	&	C	&	18.1	$\pm$	0.4	&	4.2	$\pm$	1.0	&	2.7 (-2)	$\pm$	7 (-3)	&	C	\\
-1.8	$\pm$	0.8	&	3.7	$\pm$	0.6	&	5.8 (-1)	$\pm$	3 (-1)	&	U	&	8.3		$\pm$	0.2	&	5.9	$\pm$	0.4	&	9.1 (-2)	$\pm$	4 (-3)	&	C	&	22.6	$\pm$	0.2	&	2.8	$\pm$	0.4	&	5.8 (-2)	$\pm$	1 (-2)	&	C	\\
0.0		$\pm$	0.2	&	2.8	$\pm$	0.5	&	7.2 (-1)	$\pm$	4 (-1)	&	U	&	18.1	$\pm$	0.2	&	6.8	$\pm$	0.5	&	1.4 (-1)	$\pm$	5 (-3)	&	C	&	27.7	$\pm$	0.5	&	4.6	$\pm$	1.3	&	4.5 (-2)	$\pm$	8 (-3)	&	C	\\
1.7		$\pm$	0.1	&	1.3	$\pm$	0.2	&	2.9 (-1)	$\pm$	8 (-2)	&	U	&	23.1	$\pm$	0.4	&	4.2	$\pm$	0.8	&	3.4 (-2)	$\pm$	7 (-3)	&	C	&	32.9	$\pm$	0.4	&	4.4	$\pm$	1.3	&	4.5 (-2)	$\pm$	8 (-3)	&	C	\\
55.2	$\pm$	1.6	&	4.9	$\pm$	1.6	&	7.6 (-2)	$\pm$	3 (-2)	&	C	&	28.0	$\pm$	0.2	&	5.7	$\pm$	0.4	&	1.5 (-1)	$\pm$	5 (-3)	&	C	&	38.3	$\pm$	0.4	&	3.1	$\pm$	1.0	&	5.4 (-2)	$\pm$	1 (-2)	&	C	\\
57.8	$\pm$	0.2	&	3.1	$\pm$	0.4	&	1.8 (-1)	$\pm$	6 (-2)	&	C	&	33.1	$\pm$	0.1	&	4.1	$\pm$	0.2	&	1.6 (-1)	$\pm$	7 (-3)	&	C	&	41.0	$\pm$	0.3	&	2.2	$\pm$	0.5	&	5.1 (-2)	$\pm$	3 (-2)	&	U	\\
					&					&								&		&	38.5	$\pm$	0.1	&	3.9	$\pm$	0.3	&	4.2 (-1)	$\pm$	1 (-2)	&	C	&						&					&								&		\\
					&					&								&		&	42.1	$\pm$	0.2	&	3.2	$\pm$	0.6	&	1.8 (-1)	$\pm$	2 (-2)	&	C	&						&					&								&		\\
					&					&								&		&	46.0	$\pm$	0.2	&	3.7	$\pm$	0.3	&	1.7 (-1)	$\pm$	9 (-3)	&	C	&						&					&								&		\\
\hline
\multicolumn{12}{c}{{\bf W33A}} \\
\hline
2.5		$\pm$	0.1	&	5.2	$\pm$	0.2	&	2.8 (-1)	$\pm$	1 (-2)	&	C	&	20.9	$\pm$	0.8	&	4.7	$\pm$	1.3 &	4.2 (-2)	$\pm$	1 (-2)	&	C	&	21.5	$\pm$	0.2	&	2.4	$\pm$	0.5	&	4.9 (-2)	$\pm$	1 (-2)	&	C	\\
10.1	$\pm$	1.2	&	8.6	$\pm$	4.9	&	3.5 (-2)	$\pm$	5 (-3)	&	U	&	24.1	$\pm$	0.1	&	1.3	$\pm$	0.1 &	1.4 (-1)	$\pm$	1 (-2)	&	C	&	30.6	$\pm$	1.0	&	2.7	$\pm$	1.4	&	8.6 (-2)	$\pm$	7 (-2)	&	U	\\
16.4	$\pm$	0.5	&	2.1	$\pm$	0.6	&	2.4 (-1)	$\pm$	6 (-2)	&	C	&	28.9	$\pm$	0.2	&	5.0	$\pm$	0.4 &	1.9 (-1)	$\pm$	7 (-3)	&	C	&	32.9	$\pm$	0.9	&	2.4	$\pm$	1.1	&	8.7 (-2)	$\pm$	9 (-2)	&	U	\\
17.8	$\pm$	0.3	&	1.5	$\pm$	0.4	&	1.4 (-1)	$\pm$	1 (-1)	&	U	&	33.2	$\pm$	0.2	&	2.7	$\pm$	0.5 &	1.4 (-1)	$\pm$	1 (-2)	&	C	&	38.0	$\pm$	0.3	&	2.1	$\pm$	0.7	&	3.6 (-2)	$\pm$	2 (-2)	&	U	\\
					&					&								&		&	36.5	$\pm$	0.2	&	2.3	$\pm$	0.6 &	7.0 (-2)	$\pm$	1 (-2)	&	C	&	41.6	$\pm$	0.2	&	3.4	$\pm$	0.3	&	8.9 (-2)	$\pm$	1 (-2)	&	C	\\
					&					&								&		&	40.9	$\pm$	0.1	&	4.6	$\pm$	0.3 &	1.6 (-1)	$\pm$	6 (-3)	&	C	&						&					&								&		\\
\hline
\multicolumn{12}{c}{{\bf W49N}} \\
\hline
-3.4	$\pm$	0.1	&	4.1	$\pm$	0.2	&	1.9 (-1)	$\pm$	6 (-3)	&	C	&	9.9		$\pm$	0.1	&	6.2	$\pm$	0.3	&	1.3 (-1)	$\pm$	7 (-3)	&	C	&	10.0	$\pm$	1.0	&	6.3	$\pm$	2.0	&	1.8 (-2)	$\pm$	2 (-2)	&	U	\\
1.6		$\pm$	0.1	&	2.8	$\pm$	0.2	&	2.0 (-1)	$\pm$	9 (-3)	&	C	&	13.0	$\pm$	0.1	&	4.1	$\pm$	0.2	&	2.3 (-1)	$\pm$	7 (-3)	&	C	&	12.8	$\pm$	0.1	&	4.0	$\pm$	0.2	&	2.2 (-1)	$\pm$	1 (-2)	&	C	\\
22.9	$\pm$	0.8	&	4.8	$\pm$	0.7	&	9.9 (-1)	$\pm$	8 (-1)	&	U	&	16.5	$\pm$	0.1	&	2.9	$\pm$	0.2	&	8.8 (-2)	$\pm$	1 (-2)	&	C	&	16.1	$\pm$	0.2	&	4.0	$\pm$	0.5	&	7.2 (-2)	$\pm$	2 (-2)	&	C	\\
27.8	$\pm$	0.5	&	6.5	$\pm$	4.1	&	1.7 (+0)	$\pm$	3 (-1)	&	U	&	23.3	$\pm$	1.2	&	9.9	$\pm$	2.3	&	2.5 (-2)	$\pm$	3 (-3)	&	U	&	23.3	$\pm$	0.6	&	5.8	$\pm$	1.3	&	2.6 (-2)	$\pm$	7 (-3)	&	U	\\
34.3	$\pm$	0.3	&	5.2	$\pm$	1.6	&	4.0 (+0)	$\pm$	4 (-1)	&	U	&	34.5	$\pm$	0.2	&	5.1	$\pm$	0.4	&	7.5 (-2)	$\pm$	7 (-3)	&	C	&	32.9	$\pm$	0.5	&	5.4	$\pm$	1.0	&	2.7 (-2)	$\pm$	7 (-3)	&	C	\\
40.0	$\pm$	0.4	&	4.6	$\pm$	1.0	&	2.7 (+0)	$\pm$	4 (-1)	&	C	&	39.5	$\pm$	0.2	&	4.1	$\pm$	0.4	&	4.7 (-2)	$\pm$	4 (-3)	&	C	&	40.2	$\pm$	0.4	&	5.5	$\pm$	0.8	&	3.1 (-2)	$\pm$	7 (-3)	&	C	\\
45.5	$\pm$	0.3	&	5.8	$\pm$	0.9	&	2.1 (+0)	$\pm$	9 (-2)	&	C	&	44.6	$\pm$	0.9	&	7.3	$\pm$	1.8	&	5.2 (-2)	$\pm$	4 (-3)	&	C	&	51.5	$\pm$	0.3	&	4.0	$\pm$	0.6	&	3.4 (-2)	$\pm$	8 (-3)	&	C	\\
51.1	$\pm$	0.2	&	4.1	$\pm$	0.2	&	1.1 (+0)	$\pm$	1 (-1)	&	C	&	51.3	$\pm$	0.4	&	5.8	$\pm$	0.9	&	2.6 (-2)	$\pm$	4 (-3)	&	C	&	61.6	$\pm$	0.2	&	4.7	$\pm$	0.3	&	6.5 (-2)	$\pm$	8 (-3)	&	C	\\
74.7	$\pm$	0.1	&	4.1	$\pm$	0.2	&	1.1 (-1)	$\pm$	6 (-3)	&	C	&	57.7	$\pm$	0.3	&	6.0	$\pm$	0.6	&	7.2 (-2)	$\pm$	9 (-3)	&	C	&	68.6	$\pm$	0.8	&	5.2	$\pm$	1.6	&	1.5 (-2)	$\pm$	7 (-3)	&	U	\\
					&					&								&		&	63.4	$\pm$	0.3	&	5.7	$\pm$	0.5	&	9.5 (-2)	$\pm$	5 (-3)	&	C	&						&					&								&		\\
					&					&								&		&	68.4	$\pm$	0.2	&	3.6	$\pm$	0.4	&	5.5 (-2)	$\pm$	6 (-3)	&	C	&						&					&								&		\\
\hline
\multicolumn{12}{c}{{\bf W51}} \\
\hline
1.0		$\pm$	0.1	&	2.3	$\pm$	0.1	&	1.8 (-1)	$\pm$	9 (-3)	&	C	&	41.1	$\pm$	0.7	&	7.2	$\pm$	1.6	&	2.4 (-2)	$\pm$	5 (-3)	&	U	&	47.9	$\pm$	0.2	&	4.5	$\pm$	0.6	&	5.1 (-2)	$\pm$	1 (-2)	&	C	\\
5.8		$\pm$	0.8	&	3.7	$\pm$	0.7	&	5.0 (-1)	$\pm$	2 (-1)	&	C	&	49.3	$\pm$	0.1	&	1.0	$\pm$	0.2	&	5.9 (-2)	$\pm$	1 (-2)	&	C	&	55.5	$\pm$	0.2	&	2.5	$\pm$	0.4	&	5.6 (-2)	$\pm$	2 (-2)	&	C	\\
8.3		$\pm$	0.4	&	3.2	$\pm$	0.4	&	6.7 (-1)	$\pm$	2 (-1)	&	C	&	54.5	$\pm$	0.1	&	4.6	$\pm$	0.3	&	1.3 (-1)	$\pm$	6 (-3)	&	C	&	64.8	$\pm$	0.1	&	2.0	$\pm$	0.3	&	6.4 (-2)	$\pm$	2 (-2)	&	C	\\
12.7	$\pm$	0.1	&	3.7	$\pm$	0.4	&	2.4 (-1)	$\pm$	8 (-3)	&	C	&	64.9	$\pm$	0.3	&	3.7	$\pm$	0.4	&	1.3 (-1)	$\pm$	2 (-2)	&	C	&						&					&								&		\\
17.8	$\pm$	0.2	&	3.3	$\pm$	0.5	&	1.1 (-1)	$\pm$	8 (-3)	&	C	&	69.0	$\pm$	0.6	&	5.2	$\pm$	1.0	&	8.9 (-2)	$\pm$	9 (-3)	&	C	&						&					&								&		\\
23.4	$\pm$	0.1	&	5.2	$\pm$	0.4	&	3.6 (-1)	$\pm$	7 (-3)	&	C	&						&					&								&		&						&					&								&		\\
28.9	$\pm$	1.2	&	4.7	$\pm$	1.9	&	3.3 (-2)	$\pm$	1 (-2)	&	U	&						&					&								&		&						&					&								&		\\
39.9	$\pm$	0.1	&	3.5	$\pm$	0.2	&	1.8 (-1)	$\pm$	7 (-3)	&	C	&						&					&								&		&						&					&								&		\\
48.1	$\pm$	0.7	&	6.2	$\pm$	0.7	&	1.1 (+0)	$\pm$	2 (-1)	&	C	&						&					&								&		&						&					&								&		\\
49.5	$\pm$	0.1	&	2.4	$\pm$	0.3	&	8.8 (-1)	$\pm$	2 (-1)	&	C	&						&					&								&		&						&					&								&		\\
54.4	$\pm$	0.2	&	5.8	$\pm$	0.4	&	2.0 (+0)	$\pm$	1 (-1)	&	C	&						&					&								&		&						&					&								&		\\
59.5	$\pm$	0.1	&	1.9	$\pm$	0.2	&	2.5 (-1)	$\pm$	3 (-2)	&	C	&						&					&								&		&						&					&								&		\\
70.3	$\pm$	0.1	&	2.1	$\pm$	0.4	&	2.7 (-1)	$\pm$	1 (-1)	&	U	&						&					&								&		&						&					&								&		\\
72.2	$\pm$	0.5	&	3.7	$\pm$	0.6	&	3.5 (-1)	$\pm$	5 (-2)	&	C	&						&					&								&		&						&					&								&		\\
\hline
\multicolumn{12}{c}{{\bf SgrA*+50}} \\
\hline
-204.4	$\pm$	0.5	&	8.9	$\pm$	1.1	&	1.1 (-1)	$\pm$	1 (-2)	&	C	&	-157.9	$\pm$	0.4	&	6.4	$\pm$	0.9	&	1.0 (-1)	$\pm$	1 (-2)	&	C	&	-136.9	$\pm$	0.2	&	9.4	$\pm$	0.5	&	3.2 (-1)	$\pm$	3 (-2)	&	C	\\
-191.0	$\pm$	0.4	&	6.7	$\pm$	1.0	&	1.3 (-1)	$\pm$	2 (-2)	&	C	&	-147.4	$\pm$	0.2	&	2.1	$\pm$	0.5	&	1.3 (-1)	$\pm$	3 (-2)	&	C	&	-126.0	$\pm$	0.3	&	3.8	$\pm$	0.6	&	1.6 (-1)	$\pm$	5 (-2)	&	C	\\
-169.8	$\pm$	0.9	&  20.1	$\pm$	1.6	&	7.3 (-1)	$\pm$	2 (-2)	&	C	&	-138.3	$\pm$	0.9	&	7.5	$\pm$	1.5	&	2.0 (-1)	$\pm$	2 (-2)	&	C	&	-114.6	$\pm$	0.4	&	4.5	$\pm$	0.9	&	1.1 (-1)	$\pm$	4 (-2)	&	C	\\
-155.5	$\pm$	0.3	&  11.4	$\pm$	0.8	&	1.2 (+0)	$\pm$	7 (-2)	&	C	&	-130.3	$\pm$	0.8	&	7.3	$\pm$	1.7	&	2.0 (-1)	$\pm$	3 (-2)	&	C	&	-106.5	$\pm$	0.4	&	3.6	$\pm$	0.8	&	1.1 (-1)	$\pm$	4 (-2)	&	C	\\
-147.9	$\pm$	0.1	&	3.0	$\pm$	0.2	&	1.2 (+0)	$\pm$	8 (-2)	&	C	&	-115.0	$\pm$	2.7	&  22.7	$\pm$  10.5	&	6.7 (-2)	$\pm$	8 (-3)	&	U	&	-97.0	$\pm$	0.3	&	2.4	$\pm$	0.6	&	1.2 (-1)	$\pm$	6 (-2)	&	U	\\
-117.4	$\pm$	0.4	&	5.8	$\pm$	0.7	&	1.8 (+0)	$\pm$	1 (-1)	&	C	&	-100.7	$\pm$	0.7	&	6.9	$\pm$	2.4	&	7.4 (-2)	$\pm$	2 (-2)	&	U	&	-67.0	$\pm$	0.6	&	4.8	$\pm$	1.4	&	8.4 (-2)	$\pm$	4 (-2)	&	U	\\
-110.9	$\pm$	0.5	&	5.0	$\pm$	1.2	&	2.3 (+0)	$\pm$	3 (-1)	&	C	&	-66.0	$\pm$	1.1	&	4.4	$\pm$	2.5	&	3.3 (-2)	$\pm$	2 (-2)	&	U	&	-54.4	$\pm$	0.5	&	4.5	$\pm$	1.1	&	9.1 (-2)	$\pm$	4 (-2)	&	U	\\
-105.1	$\pm$	0.8	&	5.5	$\pm$	1.8	&	1.4 (+0)	$\pm$	4 (-1)	&	C	&	-53.7	$\pm$	0.7	&	8.8	$\pm$	1.6	&	7.8 (-2)	$\pm$	1 (-2)	&	C	&	-35.2	$\pm$	0.3	&	3.0	$\pm$	0.7	&	1.1 (-1)	$\pm$	5 (-2)	&	U	\\
-98.9	$\pm$	0.5	&	3.3	$\pm$	1.1	&	7.0 (-1)	$\pm$	2 (-1)	&	C	&	-32.0	$\pm$	0.6	&	9.0	$\pm$	1.4	&	8.9 (-2)	$\pm$	1 (-2)	&	C	&	-24.2	$\pm$	0.3	&	2.6	$\pm$	0.8	&	9.5 (-2)	$\pm$	5 (-2)	&	U	\\
-96.2	$\pm$	2.1	&  19.5	$\pm$	3.1	&	1.3 (+0)	$\pm$	1 (-1)	&	C	&	-19.4	$\pm$	0.5	&	4.7	$\pm$	1.0	&	8.3 (-2)	$\pm$	2 (-2)	&	C	&	-7.9	$\pm$	1.0	&	9.0	$\pm$	2.2	&	6.3 (-2)	$\pm$	3 (-2)	&	U	\\
-78.7	$\pm$	0.4	&	5.2	$\pm$	1.3	&	1.8 (-1)	$\pm$	5 (-2)	&	C	&	-4.7	$\pm$	0.3	&	6.3	$\pm$	0.7	&	1.8 (-1)	$\pm$	2 (-2)	&	C	&	16.4	$\pm$	2.1	&  11.7	$\pm$	4.2	&	8.3 (-2)	$\pm$	3 (-2)	&	U	\\
-68.7	$\pm$	0.4	&  14.0	$\pm$	1.3	&	4.6 (-1)	$\pm$	2 (-2)	&	C	&	9.4		$\pm$	2.5	&  10.4	$\pm$	4.9	&	9.1 (-2)	$\pm$	2 (-2)	&	U	&	23.9	$\pm$	0.5	&	4.5	$\pm$	1.3	&	1.6 (-1)	$\pm$	9 (-2)	&	U	\\
-53.4	$\pm$	0.1	&	6.8	$\pm$	0.2	&	2.3 (+0)	$\pm$	9 (-2)	&	C	&	14.2	$\pm$	0.4	&	2.1	$\pm$	1.1	&	8.1 (-2)	$\pm$	4 (-2)	&	U	&	31.8	$\pm$	0.5	&	8.9	$\pm$	1.7	&	2.4 (-1)	$\pm$	4 (-2)	&	C	\\
-46.4	$\pm$	0.2	&	2.2	$\pm$	0.6	&	2.5 (-1)	$\pm$	7 (-2)	&	C	&	20.1	$\pm$	1.6	&	9.0	$\pm$	4.5	&	1.1 (-1)	$\pm$	2 (-2)	&	U	&	42.9	$\pm$	0.2	&	6.5	$\pm$	0.5	&	5.3 (-1)	$\pm$	5 (-2)	&	C	\\
-41.2	$\pm$	0.3	&	9.3	$\pm$	1.0	&	7.3 (-1)	$\pm$	2 (-2)	&	C	&	27.0	$\pm$	0.3	&	2.3	$\pm$	0.9	&	9.4 (-2)	$\pm$	3 (-2)	&	U	&	52.8	$\pm$	0.2	&	6.7	$\pm$	0.8	&	3.2 (-1)	$\pm$	4 (-2)	&	C	\\
-28.9	$\pm$	0.2	&	7.5	$\pm$	0.6	&	1.1 (+0)	$\pm$	3 (-2)	&	C	&	34.9	$\pm$	0.6	&  10.6	$\pm$	2.6	&	1.3 (-1)	$\pm$	1 (-2)	&	C	&	60.2	$\pm$	0.4	&	3.9	$\pm$	0.9	&	1.5 (-1)	$\pm$	5 (-2)	&	C	\\
-20.2	$\pm$	0.4	&	6.1	$\pm$	0.9	&	1.6 (+0)	$\pm$	7 (-2)	&	C	&	45.3	$\pm$	0.4	&	5.5	$\pm$	1.4	&	1.5 (-1)	$\pm$	2 (-2)	&	C	&						&					&								&		\\
-15.7	$\pm$	0.3	&	3.6	$\pm$	0.3	&	1.3 (+0)	$\pm$	3 (-1)	&	C	&	51.4	$\pm$	0.4	&	3.7	$\pm$	0.9	&	1.4 (-1)	$\pm$	2 (-2)	&	C	&						&					&								&		\\
15.3	$\pm$	0.2	&	4.1	$\pm$	0.3	&	1.8 (+0)	$\pm$	3 (-1)	&	C	&	57.7	$\pm$	0.4	&	3.9	$\pm$	0.9	&	9.7 (-2)	$\pm$	2 (-2)	&	C	&						&					&								&		\\
20.7	$\pm$	0.8	&	7.9	$\pm$	3.2	&	9.7 (-1)	$\pm$	4 (-1)	&	U	&						&					&								&		&						&					&								&		\\
28.9	$\pm$	0.3	&	3.1	$\pm$	1.1	&	7.1 (-1)	$\pm$	3 (-1)	&	U	&						&					&								&		&						&					&								&		\\
34.9	$\pm$	1.6	&  17.2	$\pm$	6.0	&	1.8 (+0)	$\pm$	4 (-1)	&	U	&						&					&								&		&						&					&								&		\\
49.4	$\pm$	5.6	&  18.2	$\pm$	4.4	&	7.9 (-1)	$\pm$	5 (-1)	&	U	&						&					&								&		&						&					&								&		\\
69.1	$\pm$	0.5	&	3.9	$\pm$	1.2	&	8.0 (-2)	$\pm$	2 (-2)	&	C	&						&					&								&		&						&					&								&		\\
\hline
\multicolumn{12}{c}{{\bf SgrB2(N)}} \\
\hline
					&					&								&		&	-102.0	$\pm$	0.3	&	3.6	$\pm$	0.7	&	3.3 (-2)	$\pm$	5 (-3)	&	C	&	-117.0	$\pm$	1.4	&  14.4	$\pm$	1.4	&	8.4 (-2)	$\pm$	3 (-2)	&	C	\\
					&					&								&		&	-91.5	$\pm$	0.9	&  18.0	$\pm$	1.3	&	1.6 (-1)	$\pm$	8 (-3)	&	C	&	-106.0	$\pm$	0.3	&	6.3	$\pm$	1.2	&	6.4 (-2)	$\pm$	4 (-2)	&	U	\\
					&					&								&		&	-76.0	$\pm$	0.8	&  14.1	$\pm$	3.4	&	1.1 (-1)	$\pm$	1 (-2)	&	C	&	-92.2	$\pm$  10.6	&  23.0	$\pm$  29.7	&	7.8 (-2)	$\pm$	4 (-2)	&	U	\\
					&					&								&		&	-67.2	$\pm$	0.8	&	7.0	$\pm$	2.6	&	3.2 (-2)	$\pm$	2 (-2)	&	U	&	-84.9	$\pm$	1.3	&	7.6	$\pm$	3.2	&	8.5 (-2)	$\pm$	2 (-1)	&	U	\\
					&					&								&		&	-54.8	$\pm$	1.1	&  12.4	$\pm$	2.3	&	7.4 (-2)	$\pm$	1 (-2)	&	C	&	-77.0	$\pm$	0.4	&	8.4	$\pm$	1.6	&	1.9 (-1)	$\pm$	2 (-1)	&	U	\\
					&					&								&		&	-39.4	$\pm$	2.9	&  15.5	$\pm$	4.7	&	1.6 (-1)	$\pm$	1 (-1)	&	U	&	-68.1	$\pm$	0.7	&	8.6	$\pm$	1.3	&	1.6 (-1)	$\pm$	6 (-2)	&	C	\\
					&					&								&		&	-26.9	$\pm$	0.8	&	8.4	$\pm$	4.6	&	4.9 (-2)	$\pm$	8 (-2)	&	U	&	-58.3	$\pm$	0.2	&	5.8	$\pm$	0.6	&	7.7 (-2)	$\pm$	1 (-2)	&	C	\\
					&					&								&		&	-13.9	$\pm$  10.5	&  26.5	$\pm$  56.9	&	1.4 (-1)	$\pm$	3 (-2)	&	U	&	-44.3	$\pm$	0.6	&  15.8	$\pm$	1.7	&	1.0 (-1)	$\pm$	5 (-3)	&	C	\\
					&					&								&		&	0.2		$\pm$	6.7	&  11.9	$\pm$  16.8	&	1.3 (-1)	$\pm$	4 (-1)	&	U	&	-29.1	$\pm$	0.7	&  10.8	$\pm$	1.5	&	1.8 (-1)	$\pm$	3 (-2)	&	C	\\
					&					&								&		&	5.7		$\pm$	0.8	&	7.3	$\pm$	2.7	&	1.2 (-1)	$\pm$	2 (-1)	&	U	&	-20.0	$\pm$	0.5	&	9.2	$\pm$	1.2	&	1.9 (-1)	$\pm$	5 (-2)	&	C	\\
					&					&								&		&	27.4	$\pm$	1.0	&  10.6	$\pm$	2.9	&	1.1 (-1)	$\pm$	1 (-1)	&	U	&	-7.2	$\pm$	0.2	&  10.8	$\pm$	0.6	&	4.3 (-1)	$\pm$	1 (-2)	&	C	\\
					&					&								&		&	55.1	$\pm$	2.4	&  31.1	$\pm$	6.1	&	3.9 (-2)	$\pm$	2 (-3)	&	C	&	3.5		$\pm$	0.1	&	8.8	$\pm$	0.3	&	6.6 (-1)	$\pm$	3 (-2)	&	C	\\
					&					&								&		&	65.7	$\pm$	0.2	&	7.7	$\pm$	0.6	&	8.5 (-2)	$\pm$	8 (-3)	&	C	&	15.3	$\pm$	0.1	&  10.6	$\pm$	0.2	&	8.7 (-1)	$\pm$	1 (-2)	&	C	\\
					&					&								&		&	78.3	$\pm$	0.4	&	8.1	$\pm$	1.6	&	3.2 (-2)	$\pm$	7 (-3)	&	C	&	27.8	$\pm$	0.2	&	8.2	$\pm$	0.5	&	2.3 (-1)	$\pm$	2 (-2)	&	C	\\
					&					&								&		&						&					&								&		&	38.3	$\pm$	0.4	&  11.1	$\pm$	1.9	&	1.1 (-1)	$\pm$	9 (-3)	&	C	\\
					&					&								&		&						&					&								&		&	54.8	$\pm$	0.6	&  13.3	$\pm$	1.0	&	6.0 (-1)	$\pm$	7 (-2)	&	C	\\
					&					&								&		&						&					&								&		&	65.8	$\pm$	0.5	&  11.8	$\pm$	1.1	&	5.0 (-1)	$\pm$	1 (-1)	&	C	\\
					&					&								&		&						&					&								&		&	79.0	$\pm$	0.4	&  13.8	$\pm$	0.6	&	3.7 (-1)	$\pm$	2 (-2)	&	C	\\
					&					&								&		&						&					&								&		&	91.9	$\pm$	0.2	&	4.3	$\pm$	0.7	&	3.6 (-2)	$\pm$	9 (-3)	&	C	\\
\label{TabFit}
\end{longtable}
}

\section{Determination of the excitation temperatures and of the subsequent column densities} \label{AppendCD}

The column densities given in Table \ref{TabIntTau} are derived assuming a single 
excitation temperature $T_{\rm ex}$ for all levels of a given molecule as
\begin{equation} \label{EqDcol}
N = Q(T_{\rm ex}) \frac{8 \pi \nu_0^{3}}{c^{3}} \frac{1}{g_u} \frac{1}{A_{ul}}
\left[1-e^{-h\nu_0/k T_{\rm ex}} \right]^{-1} \int \tau \, d\upsilon 
\end{equation}
where $\nu_0$, $g_u$, $g_l$ and $A_{ul}$ are the rest frequency, the upper and
lower level degeneracies and the Einstein's coefficients of the observed
transition, $Q(T_{\rm ex})$ is the partition funtion, and $c$ is the speed of 
light.

To derive the \CHp\ and \SHp\ $J=1 \gets 0$ line excitation
temperatures $T_{\rm ex}$, we adopt the recent computation of the \CHp\ 
de-excitation rate coefficients by collision with He and $e^-$ \citep{Hammami2009,Lim1999}, 
and assume a generic \SHp\ de-excitation rate coefficient by collision with H and \HH\ 
of $10^{-10}$ cm$^{3}$ s$^{-1}$. Because the corresponding critical densities 
at 100 K, $n_{\rm crit,He }(\CHp) \sim 6 \times 10^{7}$ \cc, $n_{{\rm
    crit},e^-}(\CHp) \sim 4 \times 10^{4}$ \cc, and $n_{\rm crit,\HH
}(\SHp) \sim 7 \times 10^{6}$ \cc, are considerably higher than those
of the diffuse ISM, collisions are irrelevant in the excitation of
\CHp\ and \SHp, all the more so collisions of \CHp\ with \HH\ are its 
destruction pathway. Therefore, the level populations of these two ions 
only result from radiative excitation and possible additional excitation 
processes during their chemical formation. Neglecting the latter, we derive 
the \CHp\ and \SHp\ $J=1 \gets 0$ line excitation temperatures $T_{\rm ex}$, 
from statistical equilibrium, using the submillimetre interstellar radiation 
field measured at $l=45^{\circ}$ and $b=0^{\circ}$ with the {\it Cosmic
  Background Explorer} \citep{Reach1995}. We obtain low excitation temperatures
 $T_{\rm ex}(835\,{\rm  GHz}) = 4.4$ K, and $T_{\rm ex}(526\,{\rm GHz}) = 3.0$ K
that agree with the lack of \CHp\ $(2 \gets 1)$ 
and \SHp\ $(21 \gets 11)$, $(22 \gets 11)$, $(21 \gets 10)$, and $(22 \gets 12)$
detection in absorption at the velocities of the diffuse medium along the same 
lines of sight (PRISMAS and HEXOS observations, to be published). 
Note that an error of about 1 K 
corresponds to an error on the \CHp\ and \SHp\ column densities smaller 
than 1 \%.  While these values of $T_{\rm ex}$ are valid for the velocity 
components associated with the diffuse interstellar gas, they clearly correspond 
to lower limits for those associated with the SFRs molecular environments. In 
the latter case, since the PRISMAS lines of sight target bright IR and submm
emitting sources, a complete description of the IR radiative transfer
accross the molecular gas surrounding the sources is needed (De Luca
et al., in prep.; Vastel et al., in prep.).

For excitation temperatures of $^{12,13}\CHp$ and \SHp\ of 4.4 K and 
3.0 K, respectively, Eq. \ref{EqDcol} finally becomes
\begin{equation}
N(\CHp)   = 3.11 \times 10^{12} \int \tau \, d\upsilon \quad {\rm cm}^{-2},
\end{equation}
\begin{equation}
N(\thCHp) = 3.05 \times 10^{12} \int \tau \, d\upsilon \quad {\rm cm}^{-2},
\end{equation}
and
\begin{equation}
N(\SHp)   = 1.42 \times 10^{13} \int \tau \, d\upsilon \quad {\rm cm}^{-2},
\end{equation}
using the \CHp\ $J=1 \gets 0$, \thCHp\ $J=1 \gets 0$, and the \SHp\ 
$N,J,F=1,2,5/2 \gets 0,1,3/2$ transitions, respectively.

The \CHp\ line opacity for $N(\CHp)=10^{14}$ \cq\ and a linewidth of 4 \kms\
is of the order of 15 and the expected line emission is about 4 mK (below
the noise level, see Tables \ref{TabSources} and \ref{TabCondObs}). The intensities
of the \thCHp\ and \SHp\ lines that are optically thin are even lower.

\section{Tracers of molecular hydrogen in the diffuse ISM} \label{AppendTracerHH}

The methylidyne radical CH has often been used as a probe of molecular hydrogen 
because of its linear relation with \HH\ observed in the local diffuse medium (Federman 1982) and in 
dark clouds \citep{Mattila1986}. Compiling the UV and visible data obtained on 48 lines of sight
\citep{Savage1977,Rachford2002,Crane1995,Crawford1995,Allen1994,Gredel1993,Federman1994,Danks1984,
Jenniskens1992}, \citet{Liszt2002} derived a mean ratio $\NCH / \NH2 \sim 4.3 \times 10^{-8}$, while 
\citet{Sheffer2008} found $\NCH / \NH2 \sim 3.5 \times 10^{-8}$, using their observations of the absorption 
profiles of \HH\ and CH on a sample of 90 diffuse clouds. 
Although the latter could be used in this work to estimate \NH2, based on the
Herschel observations of the six hyperfine components of the CH $J=1/2-3/2$ absorption line 
\citep{Gerin2010}, it raises several problems that have to be taken into account.
\begin{itemize}

\item[$\bullet$] As quoted by \citet{Liszt2002} and \citet{Sheffer2008}, there is a significant scatter 
(a dispersion of about a factor of 3) on the $\NCH - \NH2$ relation observed in the local diffuse 
medium, up to $\NH2 = 10^{21}$ cm$^{-2}$.
\item[$\bullet$] Despite the similarity of the CH absorption profiles with those of other tracers of 
molecular hydrogen (PRISMAS observations, to be published), a $\NCH - \NH2$ correlation has never been 
directly measured in the inner Galactic ISM.
\item[$\bullet$] \citet{Crane1995} and \citet{Lambert1990} found that the linewidths
of CH observed in absorption are highly variable from one diffuse cloud component to another. They reported 
the detection of CH line profiles either similar to or more Gaussian and less broad than those of \CHp.
This result was then confirmed by \citet{Pan2005}, who divided their observed CH absorption features
into two categories: those with CN-like and those with \CHp-like line profiles. This implies that the
production of CH in the interstellar medium can sometimes be linked to the production of \CHp and 
depends only slightly on the molecular hydrogen abundance (Godard et al., in prep.). 
\end{itemize}

For these reasons, we derive the \HH\ column densities from observations 
of the $J=1 \gets 0$ ground-state rotational transition of hydrogen fluoride (HF)
performed with the {\it Herschel}/HIFI instrument. 
If HF is saturated or not available, \HH\ is deduced from CH, assuming a mean HF/CH 
abundance ratio of 0.4 (see columns 5 and 6 of Table \ref{TabHI}).
The absorption line analysis performed by 
\citet{Neufeld2010} and \citet{Sonnentrucker2010} showed that (1) the HF profiles are similar
to those of \HdO, and (2) HF accounts for more than 30 \% of the total fluorine nuclei 
in the gas-phase along these sight-lines. Since both results agree with the chemical models
that predict that the production of HF does not depend on the dynamics of the gas and that 
HF is the main F-bearing species, we deduce the $\NHF - \NH2$ relation from an analytical study of 
the fluorine chemistry. For the range of parameters defining the diffuse and transluscent gas 
\citep{Snow2006}, $10 < \dens < 5000 \cc$, $0 < A_V < 5$, and $\nCO / [{\rm C}] < 0.9$, the abundance 
of HF is driven by only four chemical reactions \citep{Neufeld2009} :
\begin{equation}
\begin{array}{l l l l l l l l}
{\rm F}      & + & {\rm H}_2   & \rightarrow & {\rm HF}     & + & {\rm H} & \qquad (k_1) , \\
{\rm HF}     & + & \gamma      & \rightarrow & {\rm F}      & + & {\rm H} & \qquad (\zeta_{\rm HF}) , \\
{\rm HF}     & + & {\rm C}^{+} & \rightarrow & {\rm CF}^{+} & + & {\rm H} & \qquad (k_2) , \,\, {\rm and} \\
{\rm CF}^{+} & + & {\rm e}^{-} & \rightarrow & {\rm F}      & + & {\rm C} & \qquad (k_3), \\
\end{array}
\end{equation}
where
\begin{equation}
\begin{array}{l l l}
k_1 & = & 1.00 \times 10^{-10} \,\, [{\rm exp}(-450 {\rm K}/T) + 0.078 \, {\rm exp}(-80 {\rm K}/T) \\
    &   & + 0.0155 \, {\rm exp} (-10 {\rm K}/T ) ] \,\, {\rm cm}^3{\rm s}^{-1} , \\
\zeta_{\rm HF} & = & 1.17 \times 10^{-10} \,\, {\rm exp}(-2.2 A_V) \,\, {\rm s}^{-1} , \\
k_2 & = & 7.20 \times 10^{-09} \,\, (T/300K)^{-0.15} \,\, {\rm cm}^3{\rm s}^{-1} , \,\, {\rm and} \\
k_3 & = & 5.20 \times 10^{-08} \,\, (T/300K)^{-0.80} \,\, {\rm cm}^3{\rm s}^{-1}
\end{array}
\end{equation}
are their respective reaction rate coefficients from the calculations and laboratory 
measurements \citep{Zhu2002,Brown2000,Neufeld2005,Novotny2005}.
Therefore, assuming that there is no depletion of F on interstellar dust, and that the carbon is 
fully ionized and the sole source of electrons, the HF and \HH\ abundances are related by 
\begin{equation} \label{eqHFHH}
\frac{n({\rm HF})}{n({\rm H}_2)} = \frac{ \left[{\rm F}\right] }{ \big( \left[{\rm C}\right]k_2 
+ \zeta_{\rm HF}/n_{\rm H} \big) / k_1 + 1/2 f_{{\rm H}_2} \big( 1+k_2/k_3 \big)},
\end{equation}
where $\left[ {\rm F} \right]$ and $\left[ {\rm C} \right]$ are the fluorine and carbon elemental 
gas phase abundances relative to H nuclei
(assumed to equal those observed in the solar neighbourhood, $\sim 1.8 \times 10^{-8}$ and 
$\sim 1.4 \times 10^{-4}$ respectively, \citealt{Savage1996,Sofia2001,Snow2007}), 
$\dens$ is the gas density, and $f_{{\rm H}_2}$ is the molecular 
fraction defined as $f_{{\rm H}_2} = 2n({\rm H}_2)/n_{\rm H}$.
For $\dens > 30$ \cc\ and $A_V > 0.2$, because F reacts rapidly with \HH, the left-hand side of 
the denominator of Eq. \ref{eqHFHH} can be neglected, more than 90 \% of the interstellar 
fluorine being in HF and less than 10 \% in \CFp. We thus assume a $\nHF / \nH2$ abundance 
ratio of $3.6 \times 10^{-8}$.

Although HF is preferred to CH here as a tracer of molecular hydrogen, relation \ref{eqHFHH} has
to be used cautiously because: (1) it has never been proven observationally, (2) this abundance
ratio linearly depends on the fluorine elemental abundance, which can vary with the Galactocentric 
distance \citep{Daflon2004,Rudolph2006,Abia2010}, and (3) it also depends on the depletion of 
fluorine onto grains, which can become important in the molecular gas.
We note that the \HH\ molecular fraction computed as $2 N(\HH)/\NHt$ on the velocity intervals 
given in Table \ref{TabHI} ranges between 0.05 and 1 with an average value of $\sim 0.4$ 
(Neufeld et al., in prep.). While this value suggests that the diffuse medium sampled by \CHp, \thCHp, 
and \SHp\ is on average weakly molecular, it does not preclude the existence within this gas of 
regions where $f_{\HH} = 1$. A complete analysis of HF and CH as tracers of molecular hydrogen will
be performed by Godard et al. (in prep).

\end{document}